\documentclass[usenatbib]{mnras}

\usepackage{multicol}
\usepackage{aas_macros,times,multirow,amsmath,amssymb,longtable,breqn}

\usepackage[T1]{fontenc}
\usepackage[varg]{txfonts}

\usepackage{grffile} 
\usepackage{graphicx}

\usepackage{color}
\usepackage{xcolor}

\def\ltsima{$\; \buildrel < \over \sim \;$}
\def\lsim{\lower.5ex\hbox{\ltsima}}
\def\gtsima{$\; \buildrel > \over \sim \;$}
\def\gsim{\lower.5ex\hbox{\gtsima}}

\newcommand{\be}{\begin{equation}}
\newcommand{\en}{\end{equation}}

\def\flux {\mbox{erg cm$^{-2}$ s$^{-1}$}}
\def\lum {\mbox{erg s$^{-1}$}}

\newcommand\cxo{{\em Chandra}}

\newcommand\xmm{{\em XMM--Newton}}
\newcommand{\nustar}{{\em NuSTAR}}

\newcommand{\suzaku}{{\em Suzaku}}
\newcommand{\rxte}{{\em RXTE}}
\newcommand{\swift}{{\em Swift}}
\newcommand{\fermi}{{\em Fermi}}
\newcommand{\integral}{{\em INTEGRAL}}

\begin{document}
\label{firstpage}
\pagerange{\pageref{firstpage}--\pageref{lastpage}}

\title[Non-thermal X-ray emission from gamma-ray pulsars]{Spectral characterization of the non-thermal X-ray emission of gamma-ray pulsars}
\author[F. Coti Zelati et al.]
{Francesco Coti Zelati,$^{1,2}$ Diego F. Torres,$^{3,1,2}$ Jian Li,$^{4}$ and Daniele Vigan\`o$^{2,5,6}$\\
$^{1}$ Institute of Space Sciences (ICE, CSIC), Campus UAB, Carrer de Can Magrans s/n, 08193 Barcelona, Spain \\
$^{2}$ Institut d'Estudis Espacials de Catalunya (IEEC), Gran Capit\`a 2-4, 08034 Barcelona, Spain \\
$^{3}$ Instituci\'o Catalana de Recerca i Estudis Avan\c cats (ICREA), 08010 Barcelona, Spain \\
$^{4}$ Deutsches Elektronen Synchrotron DESY, D-15738 Zeuthen, Germany \\
$^{5}$ Departament de F\'isica, Universitat de les Illes Balears, Palma de Mallorca, Baleares E-07122, Spain \\
$^{6}$ Institut Aplicacions Computationals (IAC3) Universitat de les Illes Balears, Palma de Mallorca, Baleares E-07122, Spain
}

\date{Accepted 2019 December 5. Received 2019 October 31; in original form 2019 June 17}
\maketitle

\pubyear{2019}

\begin{abstract}
We report on a detailed spectral characterization of the non-thermal X-ray emission for a large sample of gamma-ray pulsars in the 
second \fermi\ LAT catalogue. 
We outline the criteria adopted for the selection of our sample, its completeness, and critically describe different approaches
to estimate the spectral shape and flux of pulsars. We perform a systematic modelling of the pulsars' X-ray spectra using archival 
observations with \xmm, \cxo\ and \nustar\ and extract the corresponding non-thermal X-ray spectral distributions. This set of data 
is made available online and is useful to confront with predictions of theoretical models.

\end{abstract}


\section{Introduction}
\label{sec:intro}

The second \fermi\ Large Area Telescope (LAT) catalogue of pulsars (henceforth 2PC) lists 117 pulsars
with significant detections in the GeV energy domain \citep{2fpc}. 
Most of these sources have been the subject of detailed follow-up observing campaigns in the X-ray band 
over the past few years.
These observations were aimed at searching for their X-ray counterparts and characterising their 
spectral and timing properties. 
Particular mention is warranted to the PhD thesis by \cite{Marelli2012}, who studied global properties of the 
X-ray emission of pulsars that have been detected by \fermi\ LAT (see also \citealt{marelli11}). 
However, to our knowledge, essentially none of the related studies in the literature report the data of the 
spectral energy distribution (SED) of the non-thermal component of these pulsars.
Notable exceptions are the soft gamma-ray catalog by \cite{kuiper15}, which provides the SEDs for the non-thermal 
pulsed emission of pulsars known up to 2015, and our recent study by \cite{li18}, which provides the SEDs for 
J1747$-$2958, J1826$-$1256 and J2021$+$3651.
All in all, non-thermal X-ray SEDs are available only for a limited number of sources.
and for even fewer if we were to consider only the non-thermal X-ray sources having a gamma-ray detection.

Broad-band SED data are needed to test any theoretical prediction for the spectral 
shape of pulsars' emission against observations.
For instance, they are of particular interest to probe the 
models developed by \citet{takata08,takata17,hirotani15,harding18,paper0,torres18}, to name a few.
In particular, we have recently developed a model that was successfully applied to the broad-band SEDs of some 
gamma-ray pulsars with detected non-thermal X-ray emission \citep{torres18}.
Testing this model against an extended sample of X-ray and gamma-ray pulsars is needed to gain insight into 
the physics of pulsar emission, e.g., to corroborate the correlations between model parameters, unveil new trends 
between these parameters, and test model limitations.

This paper presents a collection of non-thermal X-ray SEDs for a large sample of X-ray and gamma-ray pulsars,
and is understood to be complete as of April 2019.
The manuscript is structured as follows: we describe the selection of our sample in Section~\ref{sec:selection}. 
We present the different methods that can be adopted to characterize the pulsars spectra and estimate their fluxes in Section~\ref{sec:approx}.
We report on the data analysis in Section~\ref{sec:analysis}. We describe the spectral modelling and the extraction of the non-thermal
X-ray SEDs in Section~\ref{sec:xspec}.
Discussion of our results follows in Section~\ref{sec:discuss}.
%

\section{Selection of the sample}
\label{sec:selection}

We shall dub here `normal pulsars', or just PSRs, those pulsars having spin periods $P>10$ ms, and millisecond pulsars (MSPs) otherwise. 
The treatment of the observational data is however the same for both classes.

We consider here the pulsars with GeV emission reported in the 2PC catalogue and in subsequent publications (e.g., \citealt{guillemot12,Xing2016,kuiper18}).
All of them have a publicly available phase-averaged gamma-ray SED.
Out of these pulsars, we consider only those with a clearly detected non-thermal component in X-rays, as reported in the literature (see e.g. the sources labelled as "2" in Tables 
15 and 16 by \citealt{2fpc}), and typically well described by a power-law fit.

We also include in our sample the pulsars J1846$-$0258 and J2022$+$3842, which were detected as gamma-ray pulsars only after the release of the 2PC catalogue (e.g. 
\citealt{kuiper18}), as well as the three pulsars B1821$-$24, B1937+21 and J0218+4232, as analysed by \cite{Gotthelf2017}. 
As mentioned above, the SEDs of the pulsars J1747$-$2958, J1826$-$1256 and J2021$+$3651, have been already
extracted by us before \citep{li18} using the same procedure adopted in this paper (see Section~\ref{sec:nthseds}).
These results are also shown here for completeness. 
This sample of pulsars with known non-thermal emission in X-rays and gamma-rays includes 42 PSRs and 13 MSPs, and is understood to be complete 
as of 2019 April. 
Out of this list, we further reduce the number of targets for our analysis based on statistics (see next section) and on the following criteria:
we exclude from our sample those MSPs classified either as redbacks or black widows (e.g., J1959$+$2048, 
J2214$+$3000 and J2241$-$5236).
For these compact systems (with orbital periods of less than a day), the bulk of the X-ray emission is believed to arise from 
an intra-binary shock between the wind of relativistic particles ejected by the pulsar with matter in-flowing from the late-type 
companion (e.g. \citealt{romani16,an18,kandel19}), rather than from acceleration processes taking place in the 
pulsar magnetosphere.
Therefore, we retain only those MSPs that are either isolated (e.g., J2124$-$3358), or are harbored in binary systems 
with a white dwarf companion (J0437$-$4715, J0614$-$3329, J0751$+$1807 and J1231$-$1411; \citealt{guillot16} 
and references therein; \citealt{ransom11,lundgren95}). 
%

\subsection{Archival analysis and further sample selection}
\label{sec:selection2}

For each source of our sample, we retrieved all archival X-ray observations carried out over the past $\sim20$ yrs with the 
X-ray instruments on board the \xmm, \cxo\ and \nustar\ satellites, which provide the highest quality data in terms of energy resolution.
No evidence of variable X-ray emission has been reported on timescales of years in any of the pulsars of our sample. Therefore, 
for the sources that were observed multiple times with the same instrument, we focused only on the longest exposures available, i.e. 
those providing the largest photon counting statistics for an accurate spectral modelling. 
We note that the procedure of stacking spectra acquired at different epochs likely introduces additional 
systematic uncertainties in several cases, owing to the different instrumental setups adopted in distinct observations.

Following this analysis, for which procedural details are given below, 
we additionally excluded from our sample all those pulsars for which the photon counts were too few (typically 
$\lesssim$100) to allow for a tight constraint on the spectral shape of the pulsar emission. 
In particular, we removed  
pulsars where the power law photon index could not be measured with an accuracy of $\lesssim$50\% on the 
best-fitting value (typically those with an unabsorbed flux of $\lesssim2\times10^{-13}$ \flux\ over the 0.3--10~keV 
energy band), e.g. J1016$-$5857, J1023$-$5746, J1048$-$5832, J1112$-$6103, J1135$-$6055, J1420$-$6048, 
J1459$-$6053, J1732$-$3131, J1907$+$0602, J1958$+$2846.
For these pulsars the statistics is so low that either it cannot be assessed whether the spectral shape 
is thermal or non-thermal, or, even when it is non-thermal, the photon index is poorly constrained. 
The final sample used in our paper is shown in Table~\ref{tab:pf}, and consists of 33 PSRs and 8 MSPs.

There is another observational caveat to consider.
Ideally, if the pulsar were isolated and the background were negligible, an appropriate choice for the source extraction 
region in the data would yield a reliable estimate of the pulsar spectral shape and flux. 
However, for pulsars that are either embedded in, or surrounded by a bright pulsar wind nebula (PWN) or, more generally, for pulsars located 
in a sky region where the contamination from background emission is relatively high and difficult to be removed, extracting the pulsar emission
is not straightforward.
We checked the literature for the presence of PWNe around each  target (see, e.g., Tables~15 and 16 by \citealt{2fpc}).
In particular, we checked whether the PWN either embeds 
or surrounds the pulsar, or whether it is located sufficiently far from it so that the PWN emission can be disentangled 
properly from the pulsar emission with the available data. 
We note, however, that in some cases the PWN is so compact 
that it appears unresolved in either \xmm\ or \nustar\ data. 
Therefore, as a general strategy, we gave priority to \cxo\ data 
for all cases where a PWN has been reported in the literature (regardless of the size and flux of the PWN), owing to the 
sub-arcsecond angular resolution of the ACIS. 
For those cases where a PWN has been detected and only 
archival \xmm\ data exist (J0357$+$3205, J0614$-$3329, J1741$-$2054 and J2055$+$2539), we verified that the 
setup of the pn provided sufficient angular resolution to disentangle the PWN and the pulsar emission.

However, even in the \cxo\ data of some pulsars there might be still some residual contamination due to the emission 
from the PWN (especially for those cases where the PWN is very compact, extending only for a few arcseconds around 
the pulsar).
This could distort the measurement of the spectral shape of the pulsar emission,
and poses the question of whether it is more desirable to consider the overall pulsar emission including such residual contamination, 
quantifying it whenever possible, or just the pulsed pulsar emission. 
The latter choice would solve the problem of possible residual contamination, 
but certainly represents only a lower limit to the emission from the pulsar itself, as it does not account for additional 
unpulsed DC emission.
In the next section, we will describe this issue in more detail.

\section{Pulsed and unpulsed radiation}
\label{sec:approx}

Pulsar radiation is observed in the form of both pulsed (i.e. modulated at the spin period) and unpulsed 
(i.e. steady at all rotational phases) emissions over the X-ray and gamma-ray bands. 
%
%
The relative contribution of the pulsed and unpulsed components to the total observed 
emission is parametrized by the pulsed fraction (PF). 
%
It can be computed as
$PF = (max-min)/(max+min)$, where $max$ and $min$ are the count rates at the maximum and the minimum of the pulse profile, respectively. 
However, this method gives a reliable estimate for the pulsed fraction only for a sinusoidal profile.
Alternatively, the PF can be estimated via modelling of the pulse profile with one or more sinusoidal functions with periods fixed at the fundamental and 
the higher harmonic components, and dividing the semi-amplitude of the fundamental by the average count rate. 
Other definitions for the PF, such as the rms PF and the area PF, have been adopted in the literature (see e.g. \citealt{gonzalez10} 
and \citealt{tendulkar15}).
Values for the PF in the X-ray band have been estimated only for about half of the targets of our sample, 
using different definitions (as mentioned above). For these cases, the PFs span a very broad range, from $\sim7$\% to $\sim95$\% 
(see Table~\ref{tab:pf}). 
Based on these considerations, two different approaches (not devoid of issues) can be adopted to obtain an approximation of the intrinsic pulsar spectrum and flux.

\begin{table*}
\begin{center}
\caption{Compilation of all values for the PF for the non-thermal X-ray emission of our sample of X-ray and gamma-ray 
pulsars, selected according to the criteria outlined in Section~\ref{sec:selection}. The fifth column indicates whether a X-ray nebula has been detected around or close to the pulsar as reported in the literature.} 
\label{tab:pf}
\vspace{-.5cm}
\resizebox{2.1\columnwidth}{!}{
\begin{tabular}{llllcl} \hline \hline 
						& Pulsed fraction (\%)  & Energies (keV)  		& Notes   												& PWN	& References					\\						
{\bf PSRs}					&             		   	   & 			         	&													&		& \\													
\hline			
J0007$+$7303				& --				 & 		 		 	& No non-thermal pulsations detected						& N	& \citet{lin10}		\\ \vspace{0.08cm}			
J0030$+$0451				& --				 &					& No non-thermal pulsations detected						& N   & \citet{bogdanov09} \\ \vspace{0.08cm}
J0205$+$6449$\dagger$		& >21      			 & 0.08--10			& Strong contamination from the PWN 						& Y	& \citet{murray02}	\\ \vspace{0.08cm}			
J0357$+$3205				& --				 &		 		 	& Not possible to single out the non-thermal pulsations			& Y	& \citet{marelli13}   	\\ \vspace{0.08cm}			 
J0534$+$2200	 			& $\sim13$ ($\sim18$) & 20 (150)			& Increase of the PF with energy							& N	& \citet{eckert10} 	\\ \vspace{0.08cm}			
J0540$-$6919 				& $28.0\pm0.4$	& 0.5--7				& 													& Y	& \citet{kim19}			\\ \vspace{0.08cm}		
J0633$+$0632				& --				&					& No non-thermal pulsations detected 						& Y	& \citet{karpova17}  \\ \vspace{0.08cm}	
J0633$+$1746				& $\sim43$		& 3--20				& 													& Y	& \citet{Mori2014} \\ \vspace{0.1cm} 
J0659$+$1414				& $71^{+14}_{-13}$  & 3--20 				&													& N	& \citet{arumugasamy18}  \\	\vspace{0.1cm} 	
J0835$-$4510 				& $7.1\pm1.1$		& 0.1--10				&													& N	& \citet{helfand01}	\\	\vspace{0.08cm}	 
J1057$-$5226				& -- 				& --	 				& Not reported for the non-thermal emission only				& N	& \citet{mineo02}   \\	\vspace{0.1cm}	
J1124$-$5916				& --				&					& No non-thermal pulsations detected		 				& Y	& \citet{kuiper15}	\\	\vspace{0.08cm}		
J1357$-$6429				& --				&					& No non-thermal pulsations detected						& Y	& \citet{kuiper15} 	\\ \vspace{0.08cm}		
J1420$-$6048				& -- 				&					& No pulsations detected									& N	& \citet{kuiper15}	\\ \vspace{0.08cm}			
J1513$-$5908$\dagger$		& $\sim88$ 		& 0.5--79				&													& Y	& \citet{chen16}; 	 \\ \vspace{0.08cm}
						&				&					&													& N	& \citet{hu17} \\ 
J1709$-$4429				& -- 				& --  					& Not possible to single out the non-thermal pulsations			& Y	& \citet{gotthelf02} \\  \vspace{0.08cm}
						&				&					&													& N	& \citet{mcgowan04} \\ \vspace{0.08cm}
J1718$-$3825				& --				&					& No pulsations detected									& Y	& \citet{hinton07} 	 \\ \vspace{0.08cm}			
J1741$-$2054				& $\sim38$ 		& 0.3--10				&													& Y	& \citet{marelli14}	\\	\vspace{0.1cm}
J1801$-$2451				& --				&					& No pulsations detected									& Y	& \citet{kaspi01} 	\\	\vspace{0.08cm}	
J1809$-$2332				& --				&					& No pulsations detected									& Y	& \citet{vanetten12} 	\\	\vspace{0.08cm}	
J1813$-$1246$\dagger$		& $96\pm3$  		& 0.3--10 				& No variation of the PF with energy							& Y	& \citet{marelli14b}	\\	\vspace{0.08cm}			
J1833$-$1034				& $<4.2$*			& 10--79				& No pulsations detected									& Y	& \citet{nynka14} 	\\	\vspace{0.08cm}	
J1836$+$5925				& -- 				& --					& No non-thermal pulsations detected						& N	& \citet{lin14}		\\	\vspace{0.08cm}	
J1838$-$0537				& --				&					& No pulsations detected									& N	& \citet{kuiper15}	\\	\vspace{0.08cm}	
J1846$-$0258$\dagger$ 		& $\sim31$  ($>44$) [$\sim100$] & 0.5--10 (20--100) [150]	&										& Y	& \citet{kuiper09}	\\	\vspace{0.08cm}	
J1952$+$3252				& --				&					& No pulsations detected	 								& Y	& \citet{chang00}	\\	\vspace{0.08cm}	
J2021$+$4026				& --				& --					& No significant non-thermal pulsations above 2~keV  			& N	& \citet{lin13}		\\	\vspace{0.08cm}	
J2022$+$3842$\dagger$ 		& $84\pm3$  		& 0.5--12				&													& Y	& \citet{arumugasamy14} 	\\	\vspace{0.08cm}	
J2030$+$4415				& --				&					& No pulsations detected									& N	& \citet{marelli15} 	\\	\vspace{0.08cm}	
J2043$+$2740				& $<57$** 		& 0.3--10				& No pulsations detected									& N	& \citet{becker04}	\\	\vspace{0.08cm}	
J2055$+$2539				& $25\pm3$ 		& 0.3--10  				& No variation of the PF with energy							& Y	& \citet{marelli16}	\\		
J2229$+$6114$\dagger$		& >75 			& 0.8--10				& Pulsations detected up to $\sim22$~keV					& Y	& \citet{halpern01}; \\
						&				&					&													& 	& \citet{kuiper15} \\		
\hline
{\bf MSPs}	\\																	
\hline 
J0218$+$4232				& $69\pm6$		& 0.6--12				&													& N	& \citet{webb04b};	\\ \vspace{0.08cm}	
						&				&					&													&	& \citet{Gotthelf2017} \\  \vspace{0.08cm}
J0437$-$4715				& $24\pm6$  		& 2--20				& Likely an underestimate owing to drifts in the \nustar\ clock		& N	& \citet{guillot16} \\	\vspace{0.08cm}		
J0614$-$3329				& --				&					& No pulsations detected									& Y	& \citet{ransom11} \\	\vspace{0.1cm}																					
J0751$+$1807				& $52\pm8$  		& 0.6--7	  			& Possible increase of the PF with energy						& N	& \citet{webb04}	\\	\vspace{0.08cm}			
J1231$-$1411				& $\sim38$  		& 0.35--1.5 			& Likely thermal pulsations								& N	& \citet{ray19}		\\	\vspace{0.08cm}	
J1824$-$2452A			& $\sim100$		& 3--79				&													& N	& \citet{Gotthelf2017} \\	\vspace{0.08cm}
J1939$+$2134				& $\sim100$		& 3--79				&													& N	& \citet{Gotthelf2017}	\\ \vspace{0.08cm}		
J2124$-$3358				& --				& 					& No non-thermal pulsations detected						& N	& \cite{zavlin2006}		\\
\hline
\end{tabular}
}
\end{center}
{\bf Notes.}
$*$ = upper limit quoted at a confidence level of 3$\sigma$;
$**$ = upper limit quoted at a confidence level of 2$\sigma$;
$\dagger$ = pulsars with a pulsed SED in the hard X-ray catalog of \citet{kuiper15}.
\end{table*}

\subsection{Estimating the spectrum of the pulsed X-ray emission}
\label{sec:below}

Since both the unpulsed pulsar emission and the background emission are indistinguishable 
in the pulse phase distribution, in order to eliminate the contamination 
of the background one might consider only the pulsed pulsar emission. 
The most compelling case for such an approach is the Crab pulsar;  only the pulsar pulsed spectrum can be extracted to gauge the pulsar flux,
the PWN being "on top" of the pulsar (see \citealt{madsen15} and references therein). 
This method was adopted by \citet{kuiper15}  (see also \citealt{kuiper09}). 

The pulsed emission would correspond to the total pulsar emission only if the 
measured PF is 100\% or slightly smaller (i.e., if the unpulsed component gives only a negligible 
contribution to the total pulsar emission). 
However, the PF measured 
in the data is <40\% in several cases (see again Table~\ref{tab:pf}), implying that the pulsed emission represents only a 
lower limit to the total pulsar emission for a large number of sources of our sample, and that the underestimate in the flux can reach a factor $\gtrsim2.5$. 

Moreover, in almost all pulsars of our sample, the PF has been 
observed to change as a function of energy, with a trend that varies from case to case (\citealt{kuiper15} and references 
therein; see also Table~\ref{tab:pf}). 
Therefore, to have the most reliable approximation to the pulsar flux, one should 
compute the pulsed fluxes over restricted energy intervals, and scale them by the corresponding PFs estimated 
over the same energy ranges. 
However, also this procedure might not necessarily yield a realistic estimate of the total flux 
of the pulsar, as it implicitly assumes that no difference exists between the spectral shapes of the pulsed and pulsed plus 
unpulsed emission (which is not necessarily the case).

\subsection{Estimating the spectrum of the total X-ray emission}
\label{sec:above}

For all those cases where the spin signal is not detected in the X-ray band, or the photon counting statistics 
is too low to extract a meaningful pulsed spectrum, the closest approximation to the intrinsic pulsar spectrum 
and flux can only be derived by a spatial analysis of the total emission (pulsed plus unpulsed).
This approach leads to estimate an upper limit to the real flux, owing to the possible residual contamination from the background.
It is expected to give a very close approximation to the intrinsic pulsar emission especially in the \cxo\ data, where this contamination can be minimized.

\subsection{Similarities and differences in the pulsed and total X-ray SEDs: a few examples}

\begin{figure*}
\begin{center}
\includegraphics[width=0.48\textwidth]{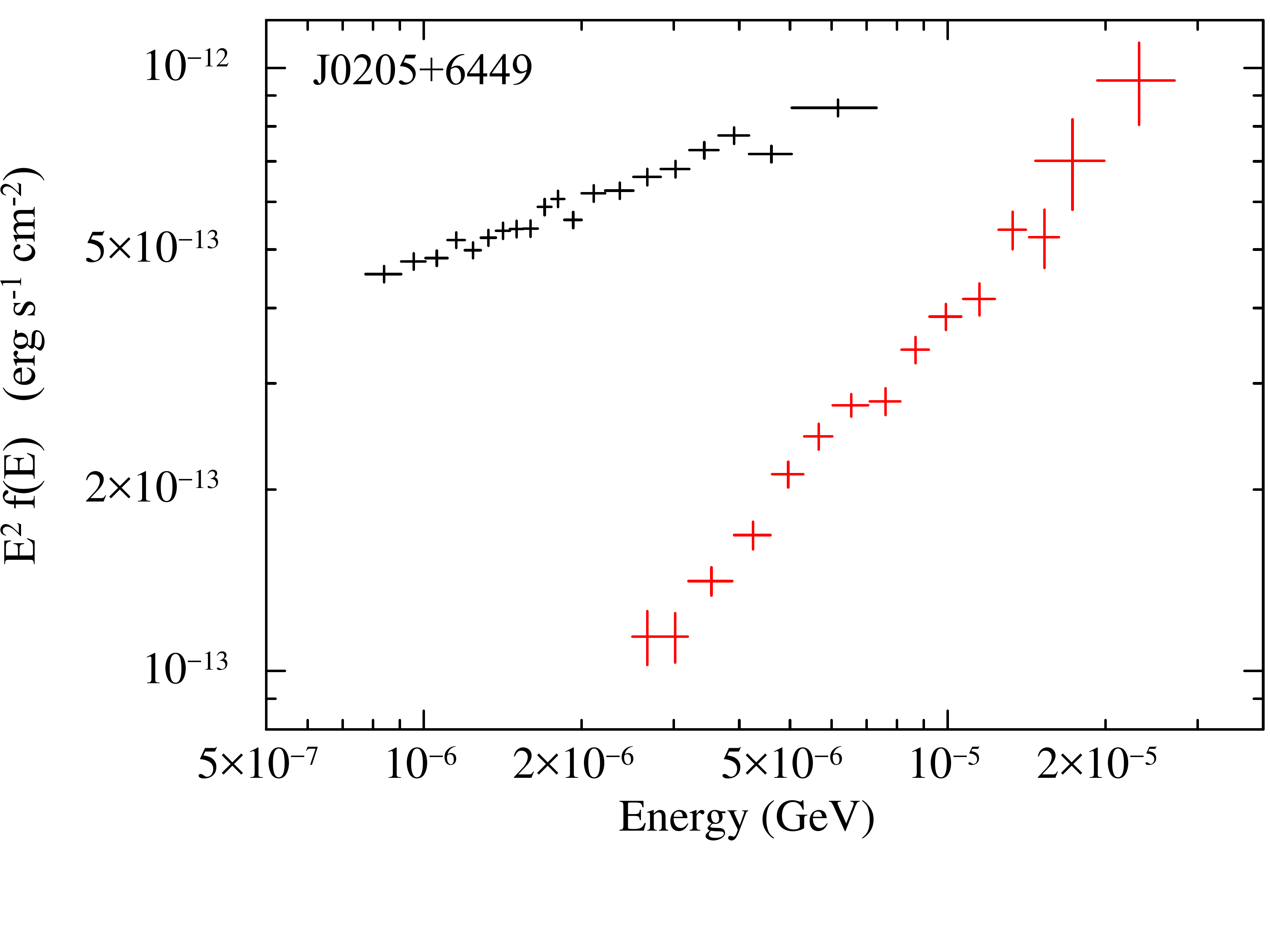}
\includegraphics[width=0.48\textwidth]{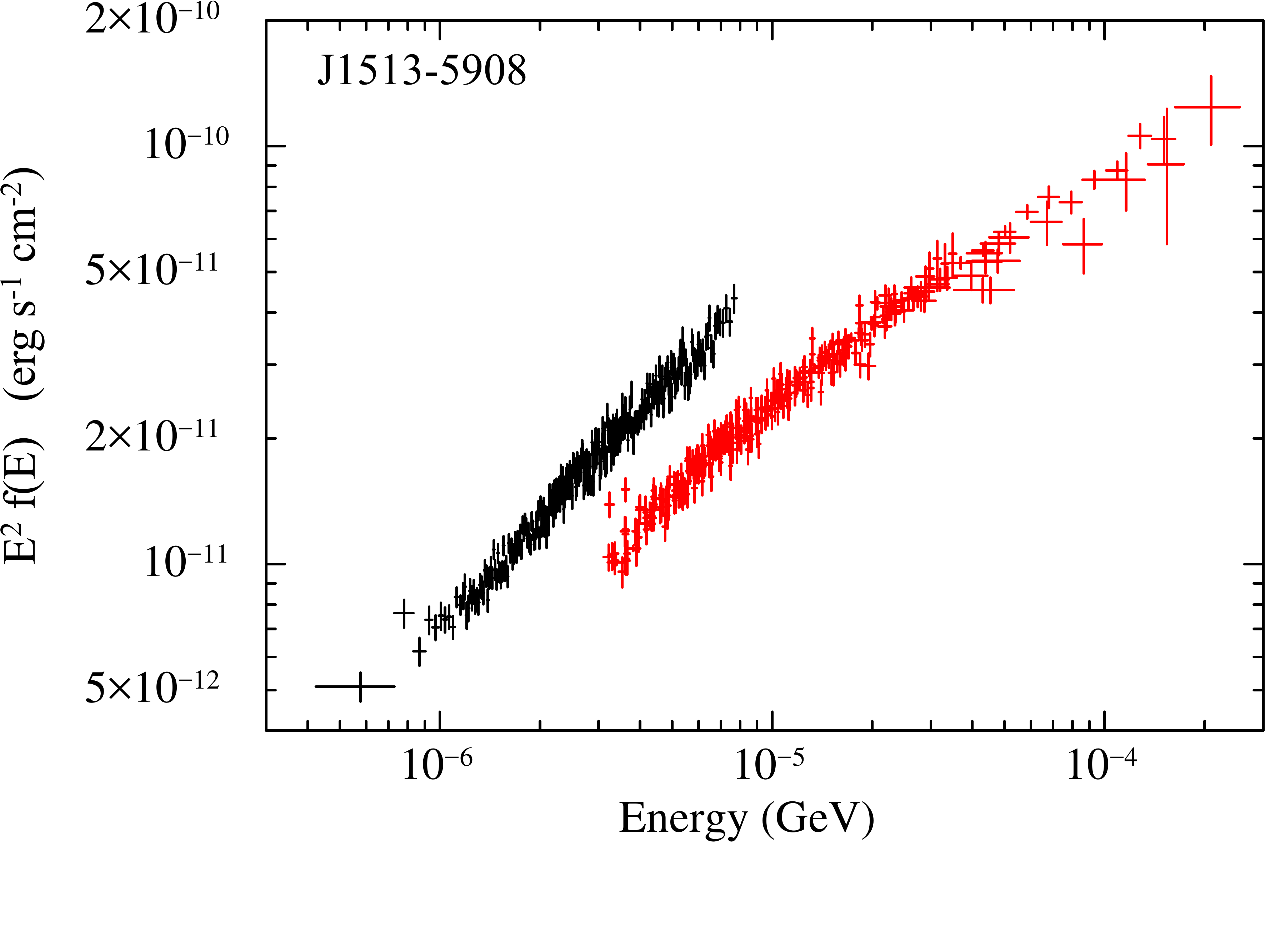}\\
\includegraphics[width=0.48\textwidth]{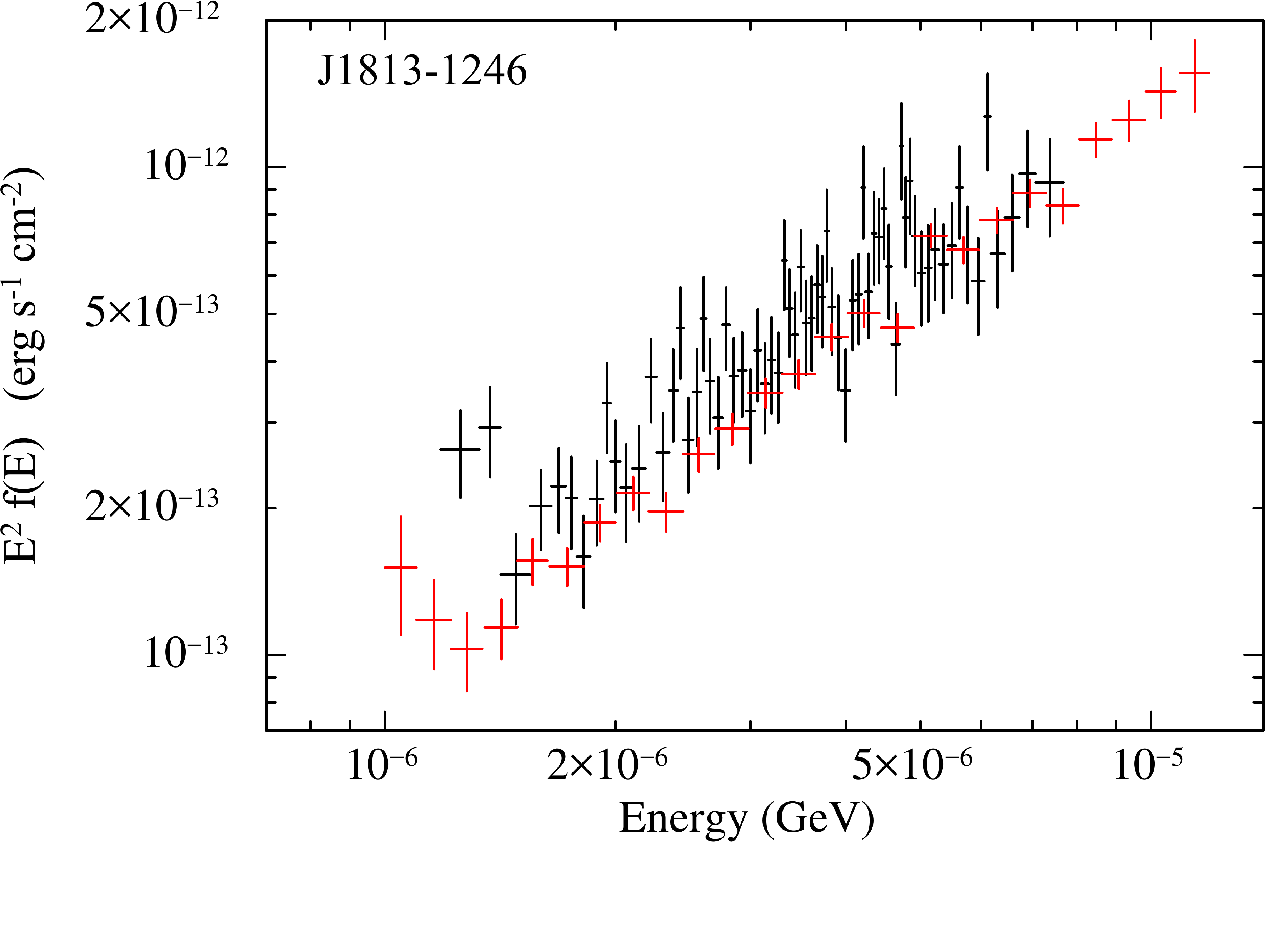}
\includegraphics[width=0.48\textwidth]{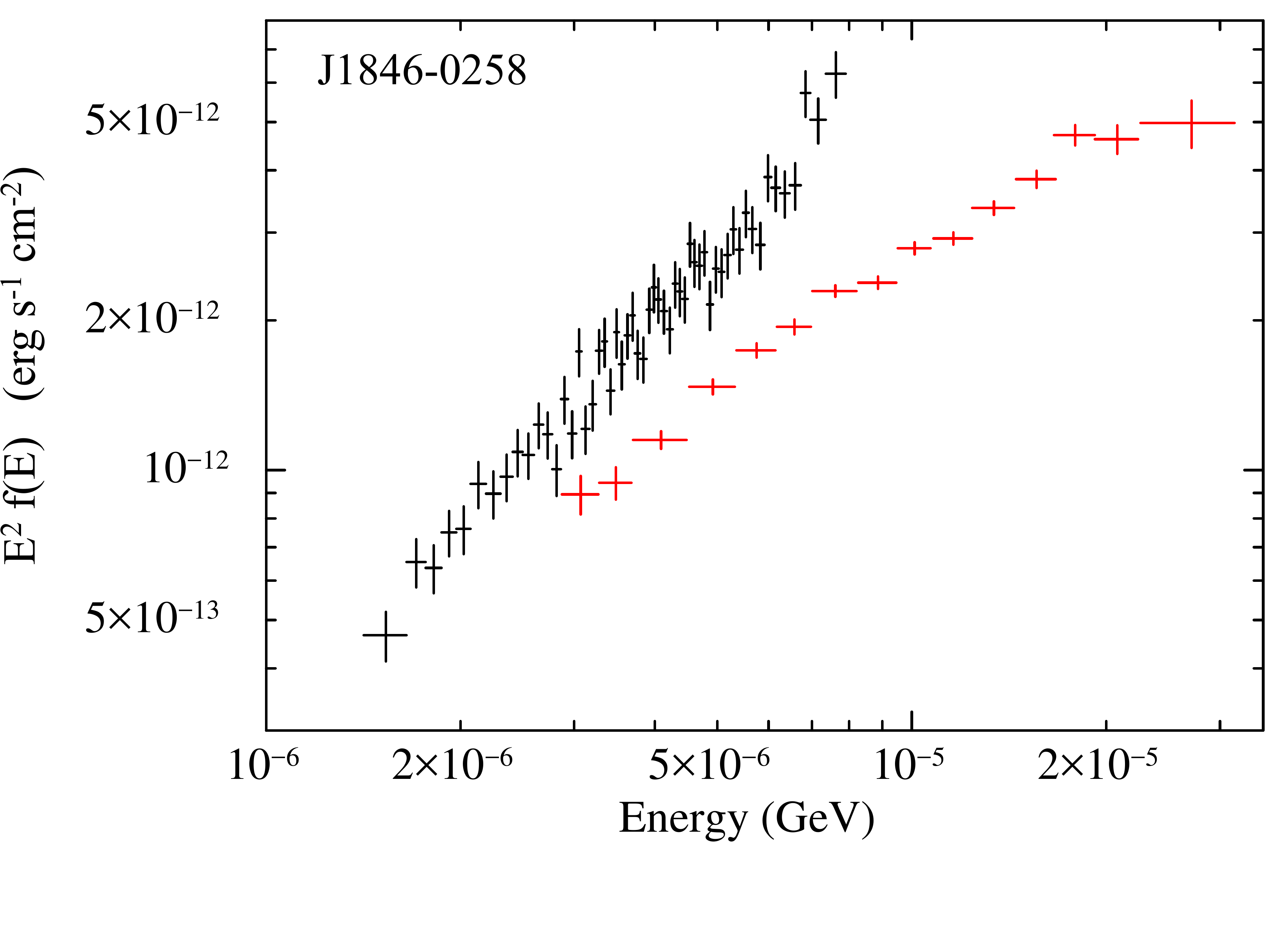}\\
\includegraphics[width=0.48\textwidth]{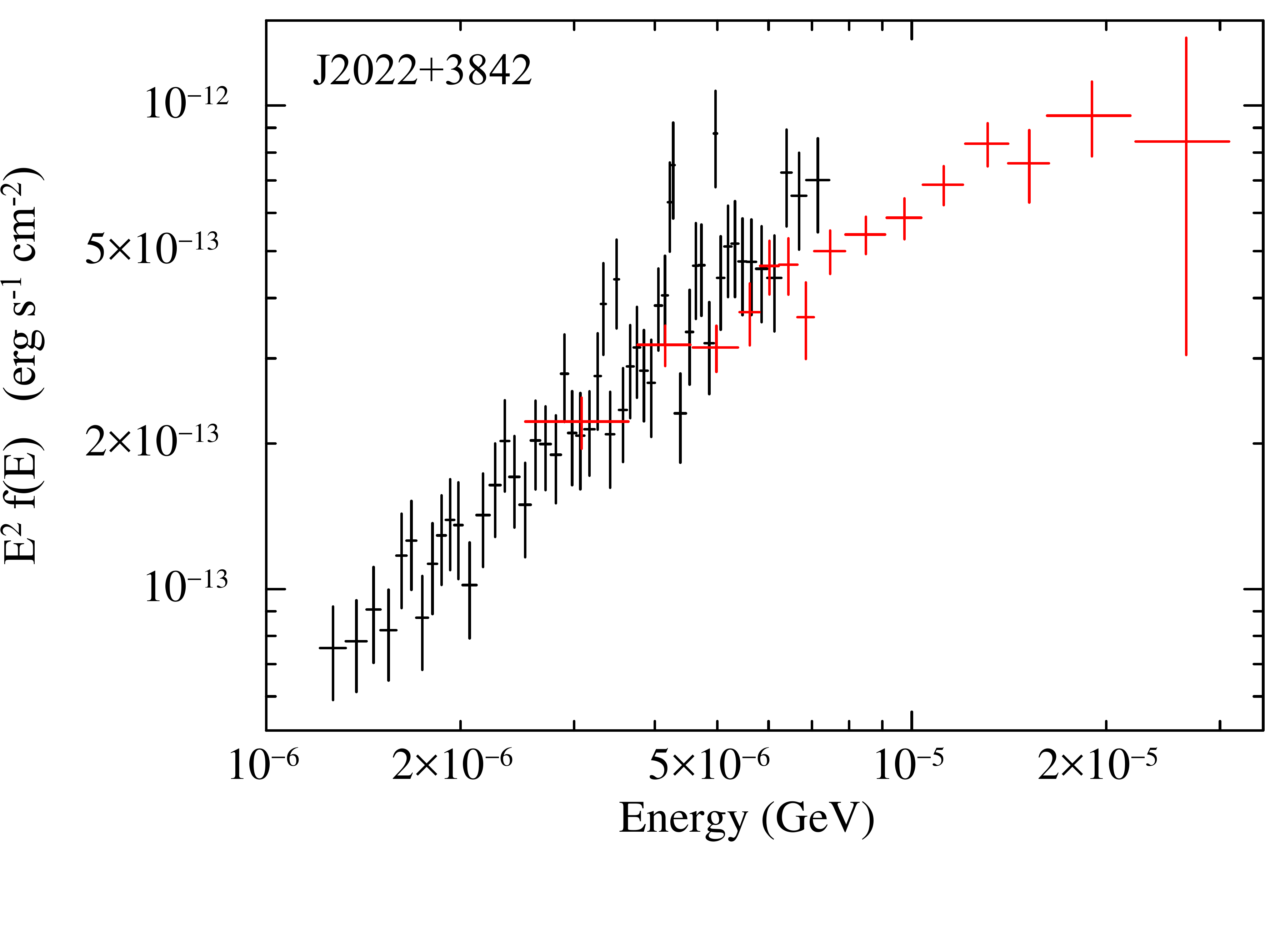}
\includegraphics[width=0.48\textwidth]{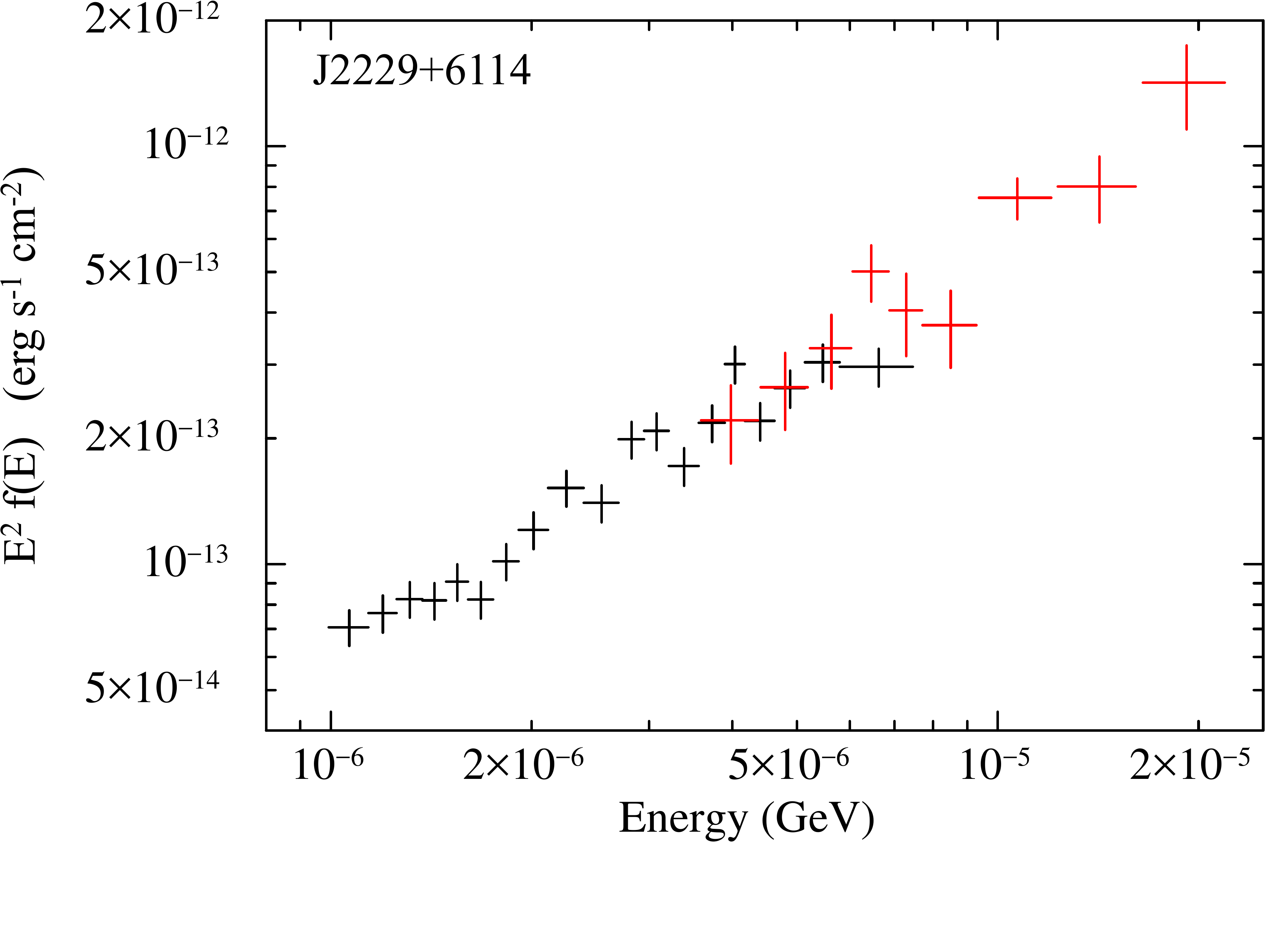}
\vspace{-0.5cm}
\caption{Total (pulsed plus unpulsed) and pulsed-only X-ray SEDs for a sample of pulsars. The former were derived in this work and are shown in black, the latter are taken from \citet{kuiper15} and are shown in red. 
}
\label{fig:comparison}
\end{center}
\end{figure*}

In Figure~\ref{fig:comparison} we show a comparison for a few X-ray SEDs extracted considering either the total (pulsed plus unpulsed) 
emission (in black; this work, with procedural details given below) or the pulsed-only emission (in red; from \citealt{kuiper15}). 
In most cases, these SEDs overlap in energy only partially, as the pulsed SEDs were typically extracted at higher energies with respect to the 
pulsed plus unpulsed SEDs that are extracted here over the soft X-ray energy range (below 10~keV).

Clearly, the total (pulsed plus unpulsed) 
spectrum of J0205$+$6449 and J1513$-$5908 cannot be reproduced just by scaling the pulsed spectrum for the (average) value 
of the PF ($\gtrsim21$\% and $\sim88$\%), indicating a strong contamination from diffuse emission in the pulsar spectral
data, which can hardly be removed  (this is clearly more evident for the case of J0205$+$6449, 
and it is consistent with the detection of the PWN also at GeV energies \citep{Li173C}.

On the other hand, the total and pulsed SEDs of J1813$-$1246, J1846$-$0258, J2022$+$3842 and J2229$+$6114 are consistent 
with each other once the pulsed SED is rescaled by the value for the corresponding PF, implying a comparatively small 
contamination from the diffuse and/or background emission. 
In particular, the pulsed SEDs of J1813$-$1246, J2022$+$3842 
and J2229$+$6114 are slightly below the corresponding total SEDs (note that in all these cases the PFs are 
$>75$\%; see Table~\ref{tab:pf}). 
Although the pulsed SED of J1846$-$0258 appears considerably below the SED for the total emission, the two SEDs turn out to be 
consistent with each other once the rather small PF is taken into account ($\sim31$\%; see again Table~\ref{tab:pf}).

\subsection{Recommended approach}

The examples above show that many caveats should be kept in mind in identifying the pulsar emission and in drawing firm 
conclusions based on the theoretical modelling of pulsars spectral data. 
The X-ray and gamma-ray spectral data sets should be extracted using, as far as possible, the same approach so as to guarantee a
homogeneous investigation of the pulsar emission mechanisms. 
Our recommended scheme is to estimate the pulsar emission from the pulsed component
for all cases where the X-ray pulsed spectra can be established well, e.g., J0205$+$6449, the 
Crab pulsar (J0534$+$2200), the Crab twin (J0540$-$6919), the Vela pulsar (J0835$-$4510), J1513$-$5908, 
J1813$-$1246, J1846$-$0258, J2022$+$3842 and J2229$+$6114 (see \citealt{kuiper15,madsen15}). 
We shall use data from \cite{kuiper15} in these cases.
The influence of the nebula can be large in some of these cases (e.g., for the Crab pulsar, or  J0205$+$6449 (3C58), or J1513$-$5908 (MSH 15-52),  
whereas for others (e.g., see Figure \ref{fig:comparison}), given the high PF, the difference between the two approaches is minimal. 
As we mentioned before, the contamination of the PWNe for J1813$-$1246, J1846$-$0258, J2022$+$3842 and J2229$+$6114 seems small.
For all the other cases, we shall characterize the pulsar emission with the method outlined in Section~\ref{sec:above} (using the X-ray analysis herein presented).
In the case of Geminga (J0633$+$1746), we use the data from \cite{Mori2014}; the results of our analysis are fully compatible with theirs.

\begin{table*}
\small
\begin{center}
\caption{Journal of the X-ray observations reanalysed here. } 
\label{tab:log}
\begin{tabular}{llccccl} \hline \hline 
			    				& Satellite/Instr. 	& Obs. ID		& Mode		& Start  time  (UTC)	& Net exposure (ks)  & Net count rate (counts s$^{-1}$) 	\\
 {\bf PSR}        					&       			&          		&			& 	     			&   				& 	\\
\hline 			
J0007$+$7303				& CXO/ACIS-S 		& 3835 		& TE VF 		& 2003-04-13 10:09:53 	& 49.5		& $0.0039\pm0.0003$	\\ 
J0357$+$3205				& XMM/EPN		& 0674440101 	& LW 		& 2011-09-15 02:39:47 	& 99.8 		& $0.0187\pm0.0005$	\\ 
J0633$+$0632 				& XMM/EPN		& 0764020101	& SW 		& 2016-03-31 05:55:20 	& 64.1 		& $0.038\pm0.001$		\\ 
J0659$+$1414				& XMM/EPN		& 0762890101 	& SW 		& 2015-09-19 20:16:09 	& 87.2 		& $3.512\pm0.006$		\\ 
J1057$-$5226				& XMM/EM 		& 0113050201  & FF        		& 2000-12-15 17:08:15 	& 53.4  		& $0.141\pm0.002$		\\ 
J1124$-$5916  		 	        & CXO/ACIS-I     	& 6677      	& TE VF		& 2006-10-16 17:52:21 	& 159.1     	& $0.0584\pm0.0006$	\\ 
J1357$-$6429				& CXO/ACIS-I      	& 10880      	& TE VF    	& 2009-10-08 06:56:33 	& 59.2    		& $0.0063\pm0.0003$	\\
J1420$-$6048				& XMM/EPN     		& 0606380101  & FF      		& 2009-08-04 11:58:18 	& 20.3     		& $0.015\pm0.001$		\\
J1709$-$4429				& CXO/ACIS-I     	& 4608  		& TE VF      	& 2004-02-01 02:05:30 	& 98.9     		& $0.0276\pm0.0005$	\\
J1718$-$3825				& CXO/ACIS-S      	& 12547		& TE VF   		& 2011-07-17 07:44:25 	& 39.6    		& $0.0072\pm0.0004$	\\
J1741$-$2054				& XMM/EPN 		& 0693870101 	& SW		& 2013-02-28 19:50:39	& 44.7 		& $0.195\pm0.002$		\\ 
J1801$-$2451				& CXO/ACIS-S 		& 753 		& TE F 		& 2000-04-12 12:41:23  	& 19.7 		& $0.020\pm0.001$		\\ 
J1809$-$2332				& CXO/ACIS-I		& 12546 		& TE VF 		& 2011-07-28 20:45:43	& 29.7 		& $0.0078\pm0.0005$	\\ 
J1833$-$1034				& CXO/ACIS-S  	& 1433 		& TE F 		& 1999-11-15 22:32:21	& 15.0 		& $0.213\pm0.004$		\\ 
J1836$+$5925				& XMM/EPN 		& 0693090101 	& SW  		& 2013-02-14 10:53:51 	& 30.2 		& $0.0140\pm0.0008$	\\ 
J1838$-$0537				& XMM/EPN		& 0720750201 	& FF 		& 2013-10-14 20:14:52 	& 25.8 		& $0.0029\pm0.0005$	\\ 
J1952$+$3252				& CXO/ACIS-S  	& 1984  		& TE VF   		& 2001-07-12 06:24:50 	& 73.8      		& $0.270\pm0.002$		\\ 
J2021$+$4026				& XMM/EPN     		& 0763850101  & SW      		& 2015-12-20 10:39:21 	& 94.0      		& $0.0089\pm0.0004$	\\ 
J2030$+$4415				& CXO/ACIS-S      	& 14827      	& TE VF 	 	& 2014-04-15 01:48:00 	& 24.7      		& $0.0024\pm0.0003$	\\ 
J2043$+$2740				& XMM/EPN      	& 0037990101  & SW      		& 2002-11-21 23:25:45 	& 11.6     		& $0.011\pm0.001$		\\
J2055$+$2539			        & XMM/EPN    		& 0724090101  & LW       		& 2013-05-15 03:29:36 	& 10.4    		& $0.0065\pm0.0003$	\\
\hline
{\bf MSP} \\
\hline
J0437$-$4715		& NuS/FPMs		& 30001061006 & S  		& 2015-01-02 17:16:07  	& 63.1		& $0.0429\pm0.0008$	\\ 
J0614$-$3329		& XMM/EPN		& 0653190101	& FF 		& 2010-10-04 00:25:52 	& 11.3 		& $0.040\pm0.002$		\\ 
J0751$+$1807		& XMM/EM		& 0111100301 	& FF 		& 2000-10-01 03:57:16 	& 30.0 		& $0.0044\pm0.0004$	\\ 
J1231$-$1411		& XMM/EPN      	& 0605470201  & FF      		& 2009-07-15 09:05:54 	& 18.6      		& $0.082\pm0.002$		\\ 
J2124$-$3358		& CXO/ACIS-S   	& 17900  		& TE VF   		& 2016-07-07 11:36:49 	& 91.6     		& $0.0094\pm0.0003$	\\
\hline
\end{tabular}
\end{center}
{\bf Notes.}
TE stands for `timed exposure' with either faint (F) or very faint (VF) telemetry format, CC for `continuous clocking', SW for `small window', 
LW for `large window', S for `science mode', FF for `full frame' . The average net count rates refer to the 0.3--10~keV energy band for \swift\ 
and \xmm\ data, to the 0.3--8~keV energy band for \cxo\ data, and to the 3--79~keV energy range for \nustar\ data.
\end{table*}

\subsection{Origin of the phase-averaged gamma-ray spectra in our compilation}

Gamma-ray spectral data such as those reported by \cite{2fpc} are extracted using a sort of conceptual mixture of the approaches described in Sections~\ref{sec:below} and \ref{sec:above},
owing to the significantly broader point spread function of \fermi\ LAT. 
Depending on the viewing geometry and the emission mechanisms, 
a pulsar can have up to a 100\% duty cycle in gamma-rays, and significant magnetospheric emission can also exist away from the peaks of the light curves.
The commonly adopted procedure for gamma-ray data is to fit any putative PWN contribution simultaneously with the PSR and the diffuse background 
contribution. 
The magnetospheric emission is known to produce SEDs that are well fit by an exponential cutoff at GeV energies \cite{2fpc}. 
The spectra of PWNe and of diffuse emission are in turn described using a power-law with a hard index up to tens of GeV (for the case of PWNe)
or present only at lower energies with a very soft index (for the case of diffuse emission).
The former can be used to distinguish the PWN component, as was the case in, e.g., 3C58 \citep{Li173C}.
The latter can be used to distinguish defects in the diffuse model \citep{acero13,Acero2013b}.
PWNe are also often (but not always) extended and resolved at GeV energies.

In most cases, then, the PWN contribution is obtained from the region of the light curve off the peaks, 
after the photons there are modelled to see whether their spectrum does not cut off (as the pulsar emission probably would).
The PWN contribution so evaluated is then subtracted away.  
While one can be confident that what is left must come from the pulsar's magnetosphere, there is always the possibility that relevant photons may be removed (particularly, if we are misunderstanding the observed pulsar cutoffs), see \cite{2fpc} for further discussion.
Only a few PWNe have been detected in gamma-rays, with fluxes that are typically much smaller than those of the associated pulsars (e.g., see \citealt{acero13,li16,Li173C}).

In our compilation, we mostly used the gamma-ray data set available from the 2PC catalogue \citep{2fpc}, except for a few targets where additional dedicated studies
have been performed. This is the case of J0007$+$7303, the pulsar in CTA 1 
\citep{li16}; J0205+6449, the pulsar in 3C 58 \citep{Li173C}; Geminga -- J0633+1746 \citep{Abdo10a}; and Vela -- J0835-4510 \citep{Abdo10b}, where 
uncertainties on the spectral parameters are smaller than 1\%. 
For MSPs we use data from \cite{2fpc} and \cite{Xing2016}, which are compatible with each other within the uncertainties.
Note that for J0205$+$6449, the highest energy data have been associated to the PWN 3C58, not the pulsar \citep{Li173C}.
We also use the data from \cite{Lemoine-Goumard2011} for J1357$-$6429.
For the Crab twin in the Large Magellanic Cloud we use the gamma-ray data from \cite{Ackermann2015}. 
A few pulsars are dimmer in the high-energy gamma-ray band, with only a few data points available in their SEDs.
This is the case of J1513$-$5908 \citep{kuiper18}, J1846$-$0258 \citep{kuiper16a,kuiper18} and J2022$+$3842 \citep{pilia10,ohuchi15,smith16}.
%

\section{X-ray data analysis}
\label{sec:analysis}

This section describes the processing and analysis of pulsar data acquired with the X-ray instruments on board the \xmm, \cxo\ 
and \nustar\ satellites that were reanalysed here.  
Details of the observations used in our analysis are reported in Table~\ref{tab:log}.

\subsection{\xmm\ data}

We considered only the data acquired with the European Photon Imaging Cameras (EPIC) set in imaging modes. 
In particular, 
we considered data taken with the pn instrument (the one with the largest collecting area; \citealt{struder01}) in all cases except 
for the MSPs J0030$+$0451, J0751$+$1807 and J1057$-$5226. 
For these two cases, the pn camera was configured in the timing mode to 
search for X-ray pulsations, hence we considered the data acquired with the two MOSs cameras \citep{turner01} in imaging 
mode to better estimate the background level. 

The pn camera was configured either in the full frame mode (FF; time resolution of 73.4~ms), in large window mode (LW; 
time resolution of 47.7~ms), or in small window mode (SW; time resolution of 5.7~ms) in the different observations. 
MOSs 
data were operating in the full frame mode (time resolution of 2.6~s) during the observations of J0030$+$0451, J0751$+$1807 
and J1057$-$5226.

We retrieved the raw observation data files from the \xmm\ Science Archive, and produced calibrated, concatenated photon 
event lists using the \textsc{epproc} and \textsc{emproc} tools of the \xmm\ Science Analysis System (\textsc{sas} version 
17.0.0) and the most up-to-date calibration files available (XMM-CCF-REL-358). 
For each observation, we built a light curve 
of single pixel events (\textsc{pattern} = 0) for the entire field of view, and removed episodes (if any) of strong background 
flaring activity. 
We extracted the source photons from a circular region centred on the most accurate source position as 
reported in the literature, and with a typical radius ranging between 15 and 40 arcsec. 
The best value for the extraction radius 
was determined using the \textsc{eregionanalyse} tool, accounting for the source flux, the presence of a compact PWN, the 
flux of the PWN itself, and the distance of the target from the edge of the CCD. 
The background was extracted from a circle 
located on the same CCD, sufficiently far from the pulsar and, when present, the diffuse emission associated to the PWN.

For the pn data we retained only single and double pixel events  (\textsc{pattern} $\leq4$), whereas for MOS data we selected 
single-to-quadruple pixel events (\textsc{pattern} $\leq12$). 
Only events optimally calibrated (\textsc{flag} = 0) were considered 
for the following spectral analysis. 
We generated the redistribution matrices and effective area files with \textsc{rmfgen} and 
\textsc{arfgen}, respectively. 
The spectra were then rebinned so as to contain a minimum of 20 photon counts in each spectral bin.

\subsection{\cxo\ data}

The Advanced CCD Imaging Spectrometer (ACIS; \citealt{garmire03}) on board the {\em Chandra X-Ray Observatory} 
consists of an imaging (ACIS-I) and a spectroscopic (ACIS-S) CCD arrays. 
It can operate either in the timed exposure 
(TE) mode (time resolution of 3.24~s or a sub-multiple, if only a sub-array of a chip is being read-out) or the continuous 
clocking (CC) mode (time resolution of 2.85~ms, no imaging capabilities).
 
We analysed the data following the standard analysis threads for a point-like source with the \cxo\ Interactive Analysis of 
Observations software (\textsc{ciao}, v. 4.11; \citealt{fruscione06}) and the calibration files stored in the \cxo\ \textsc{caldb} 
(v. 4.8.2). 
We generate new `level 2' events files using the \textsc{chandra$_{-}$repro} task. 
For TE-mode data and on-axis 
targets, we collected the source photons from a circular region around the source position with a radius of 1 or 2 arcsec, 
depending on the presence/absence of a compact PWN around the pulsar. 
We extracted the background using an annulus 
centred on the source location with inner and outer radius of 4 and 8 arcsec for the case of J2030$+$4415 (where no extended 
emission has been reported), and a circle of radius 6 arcsec far from the PWNe in all other cases. 
The observation of 
J1513$-$5908 was the only one carried out with the ACIS set in CC mode. 
However, the roll angle of the telescope was 
chosen so as to exclude the bright features of the PWN from the strip, hence guaranteeing a comparatively small contamination 
from the PWN emission in the extraction of the pulsar spectral data \citep{hu17}. 
In this case, the source events were collected 
through a rectangular region with a length of 3 arcsec along the readout direction of the CCD. 
The background was extracted 
using two boxes of the same size oriented along the image strip, symmetrically placed with respect to the target and sufficiently 
far from the position of the source so as not to include emission from the PWN. 
We created the source and background spectra, 
the redistribution  matrices and the ancillary response files using \textsc{specextract}.

\subsection{\nustar\ data}

We processed the \nustar\ \citep{harrison13} data of J0437$-$4715 using the script \textsc{nupipeline} (v.~0.4.6) under the 
\nustar\ Data Analysis Software (\textsc{nustardas} v.~1.9.3) and the most recent instrumental calibration files (v20181030). 
We filtered the data excluding the time intervals related to passages of the satellite through the South Atlantic Anomaly (SAA). 
For 
each focal plane module (FPM), we collected the source counts within a circular region of radius 40 arcsec and background 
photons from a circle of radius 80 arcsec located on the same detector chip and far from the source. 
Source and background 
spectra, response matrices and ancillary files were created separately for each FPM using \textsc{nuproducts} (v.~0.3.0). 
The spectra were then grouped so as to have at least 20 counts in each spectral channel.

\begin{table*}
\small
\begin{center}
\caption{Results of X-ray spectral analysis for all the gamma-ray pulsars in the 2PC catalog for which a significant detection of a non-thermal component was found at X-ray energies, considered in this work or reported earlier (see references).} 
\label{tab:results}
\resizebox{2.1\columnwidth}{!}{
\begin{tabular}{lllllllcl} \hline \hline 
{\bf PSR}	    		& Best-fitting   		& $\chi^2_{{\rm red}}$ (dof)	& $N_{\rm H}$			& PL index		& Non-th. unabs. flux				& Energies 		& Non-th. flux fraction  	& Reference			\\
         			& model			&	     					& (10$^{21}$ cm$^{-2})$	&				& (10$^{-13}$ \flux)			 	& (keV)			& (\%) 				  			\\
\hline 			
J0007$+$7303		& PL 			& 0.92 (7)					& $<2.0$				& $1.7\pm0.3$		& $0.62\pm0.08$ 				& 0.3--10			& 100 				& This work			\\ 
J0205$+$6449		& PL				& ...						& 3.4 (fixed)			& $1.03\pm0.02$	& $55\pm3$ 					& 0.56--267.5		& 100 				& \cite{kuiper10}, \cite{kuiper15}			\\	
J0357$+$3205		& BB+PL 	 		& 1.38 (27)				& $1.8\pm0.4$			& $2.4\pm0.2$		& $0.59\pm0.06$ 		 		& 0.3--10			& 50					& This work			\\ 
J0534$+$2200		& curved  		 	& ... 						& 3.61 (fixed)			& ...				& $19480\pm50$ , $27660\pm80$  	& 2--10, 20--100	& 100				& \cite{kuiper15}  \\			
J0540$-$6919		& curved	 		& ...						& 3.7 (fixed) 			& ...				& $67^{+4}_{-7}$, $61^{+6}_{-21}$ 	& 2--10, 20--100	& 100 				& \cite{campana08}, \cite{kuiper15}  \\	
J0633$+$1746		& 2BB+PL			& 0.99 (437)				& $0.15\pm0.03$		& $1.70\pm0.04$	& $\sim8$						& 0.2--20			& 22					& \cite{Mori2014} \\
J0659$+$1414		& 2BB+PL	 		& 1.24 (262)				& $0.37\pm0.04$		& $2.0\pm0.1$		& $2.3\pm0.1$  				& 0.3--10			& 2 					& This work				 \\ 
J0835$-$4510		& PL	 			& ...						& 0.33 (fixed)			& $1.06\pm0.05$	& $8\pm1$ , $110\pm40$			& 2--10, 20--100	& 100 				& \cite{kuiper15} \\			
J1057$-$5226		& 2BB+PL 	  	& 0.87 (27)      				& $<0.26$      	      		& $1.8\pm0.4$		& $1.1^{+0.3}_{-0.2}$ 			& 0.3--10			& 6					& This work			\\ 
J1124$-$5916		& PL            	  	& 1.07 (77)      				& $6.0\pm0.3$      	      	& $1.88\pm0.04$	& $13.2\pm0.3$				& 0.3--10			& 100 				& This work			\\ 
J1357$-$6429		& BB+PL           	& 1.09 (13)     				& $7\pm6$     	     		& $1.8\pm0.5$		& $1.2\pm0.3$	 				& 0.3--10			& 24 					& This work				 \\
J1420$-$6048		& PL            	  	& 0.94 (17)     				& $37\pm18$    	   	& $0.7\pm0.4$		& $2.8\pm0.4$					& 0.3--10			& 100 				& This work			\\
J1513$-$5908		& curved  	 		& ...						& 9.5 (fixed)			& $1.233\pm0.005$	& $252\pm5$ , $959\pm13$  		& 2--10, 20--100	& 100				& \cite{kuiper15}			\\
J1709$-$4429		& BB+PL         	  	& 1.11 (20)     				& $6\pm2$  	  		& $1.4\pm0.2$		& $3.1\pm0.3$ 					& 0.3--10			& 44					& This work			\\
J1718$-$3825		& PL           	  	& 0.82 (11) 			    	& $4\pm2$     	     		& $1.4\pm0.2$		& $1.3\pm0.1$					& 0.3--10			& 100 				& This work			\\
J1741$-$2054		& BB+PL 	 		& 1.16 (57) 				& $1.7\pm0.4$ 		 	& $2.74\pm0.09$	& $6.8\pm0.6$ 					& 0.3--10			& 32 					& This work				 \\ 
J1747$-$2958		& PL				& 1.16 (56)				& $26\pm2$			& $1.3\pm0.1$		& $24\pm1$					& 0.2--10			& 100 				& \cite{li18} 			\\
J1801$-$2451		& PL	 			& 1.75 (16) 				& $56\pm13$ 		 	& $1.6\pm0.4$		& $11\pm4$					& 0.3--10			& 100 				& This work			\\ 
J1809$-$2332		& PL	 			& 0.96 (8) 					& $<2.5$ 		 		& $1.8\pm0.3$		& $1.7\pm0.3$					& 0.3--10			& 100 				& This work			\\ 
J1813$-$1246		& PL	 			& ...						& ...					& $0.85\pm0.03$	& $9.6\pm0.2$ 					& 2--10			& 100 				& \cite{kuiper15}			 \\
J1826$-$1256		& PL				& 1.08 (16)				& $23\pm5$			& $1.3\pm0.2$		& $3.4^{+0.5}_{-0.3}$			& 0.2--10			& 100 				& \cite{li18}			\\
J1833$-$1034		& PL	 			& 1.11 (62) 				& $32\pm2$ 		 	& $1.37\pm0.09$	& $86\pm3$					& 0.3--10			& 100 				& This work			\\ 
J1836$+$5925		& BB+PL 	 		& 0.54 (14) 				& $1.3\pm0.9$ 			& $2.0\pm0.3$		& $0.37\pm0.05$ 				& 0.3--10			& 40					& This work			\\ 
J1838$-$0537		& PL	 			& 0.56 (5) 					& $70^{+77}_{-41}$ 		& $1.2\pm1.0$		& $0.7^{+1.1}_{-0.2}$			& 0.3--10			& 100 				& This work			 \\ 
J1846$-$0258		& PL  	 		& ...						& 39.6 (fixed)			& $1.20\pm0.01$	& $23.8\pm0.2$, $152\pm1$		& 2--10, 20--100	& 100 				& \cite{kuiper09}			\\
J1952$+$3252		& PL	 	 		& 1.36 (16)  	    			& $3.9\pm0.2$         	 	& $1.59\pm0.03$	& $42\pm0.4$					& 0.3--10			& 100				& This work			 \\ 
J2021$+$3651		& BB+PL			& 1.07 (36)				& 6.9 (fixed)			& $1.7\pm0.3$		& $0.8\pm0.1$					& 0.2--10			& 17 					& \cite{li18}				\\
J2021$+$4026		& BB+PL          	  	& 1.13 (42)     	 			& $14\pm3$      	      	& $1.4\pm0.5$		& $0.33\pm0.09$ 				& 0.3--10			& 13 					& This work				\\ 
J2022$+$3842		& PL  	 		& ...						& 17 (fixed)			& $1.20\pm0.02$	& $5.5\pm0.2$					& 2.5--32			& 100 				& \cite{kuiper15}			\\
J2030$+$4415		& PL            	  	& 1.23 (8)      				& $<2.2$      	      		& $2.4\pm0.5$		& $0.3\pm0.1$					& 0.3--10			& 100 				& This work			\\ 
J2043$+$2740		& PL            	  	& 1.31 (5)     				& $<0.5$     	     		& $2.8\pm0.3$		& $0.21\pm0.03$				& 0.3--10			& 100 				& This work			\\
J2055$+$2539		& PL            	  	& 1.04 (20)     				& $2.4\pm0.5$     	     	& $2.2\pm0.2$		& $0.34\pm0.03$				& 0.3--10			& 100 				& This work			\\
J2229$+$6114		& PL	 			& ...						& 6.3 (fixed)			& $1.11\pm0.03	$	& $5.2\pm0.3$					& 2.5--32			& 100 				& \cite{kuiper15}			\\
\hline
{\bf MSP} \\
\hline
J0218$+$4232		& PL				& 0.88 (63) 				& $0.5\pm0.3$ 			& $1.1\pm0.1$		& $20\pm2$					& 3--79			& 100 				& \cite{Gotthelf2017}	\\
J0437$-$4715		& PL 	 		& 1.40 (24)				& $<19$				& $1.81\pm0.04$	& $98\pm4$					& 3--79			& 100 				& This work	 \\ 
J0614$-$3329		& PL  			& 0.54 (17)				& $1.0\pm0.4$			& $3.4\pm0.3$		& $1.5\pm0.4$ 					& 0.3--10			& 100 				& This work	\\ 
J0751$+$1807		& PL 			& 1.52 (5) 					& $<1.4$ 		 		& $2.7\pm0.6$		& $0.5\pm0.2$	 				& 0.3--10			& 100 				& This work	\\ 
J1231$-$1411		& PL            	 	& 0.81 (18)      				& $1.7\pm0.2$      	      	& $4.0\pm0.2$		& $6\pm1$ 					& 0.3--10			& 100 				& This work	\\ 
J1824$-$2452A	& PL				& 0.91 (226)				& $4.01\pm0.04$		& $1.28\pm0.05$	& $22\pm2$					& 3--79			& 100 				& \cite{Gotthelf2017}	\\	
J1939$+$2134		& PL				& 0.91 (83)				& $18\pm3$ 			& $1.2\pm0.1$		& $18\pm3$					& 3--79			& 100 				& \cite{Gotthelf2017}	\\	
\hline
\end{tabular}
}
\end{center}
\end{table*}

\section{Results}
\label{sec:xspec}
\subsection{X-ray spectral modelling}

All spectra were modelled within the \textsc{xspec} spectral fitting package (v. 12.10.1; \citealt{arnaud96}). 
We started by 
fitting each spectrum with either an absorbed power law (PL) or an absorbed blackbody (BB) model (\textsc{bbodyrad} 
in \textsc{xspec}). 
We described the absorption by the interstellar medium along the line of sight via the Tuebingen-Boulder 
model (\textsc{tbabs} in \textsc{xspec}), adopting the cross-sections by \citet{verner96} and the chemical abundances 
by \citet{wilms00}. 
Although in most cases such a model provided a satisfactory description of the data, in some cases 
structured residuals were observed over restricted energy intervals. 
For these cases, we included additional spectral 
components in the form of one or more blackbody (BB) components. 
We assessed the number of required components via the $F$-test 
setting a minimum threshold of 3$\sigma$ (99.7\%) for the statistical significance of the fit improvement. 

\begin{figure}
\begin{center}
\includegraphics[width=0.48\textwidth]{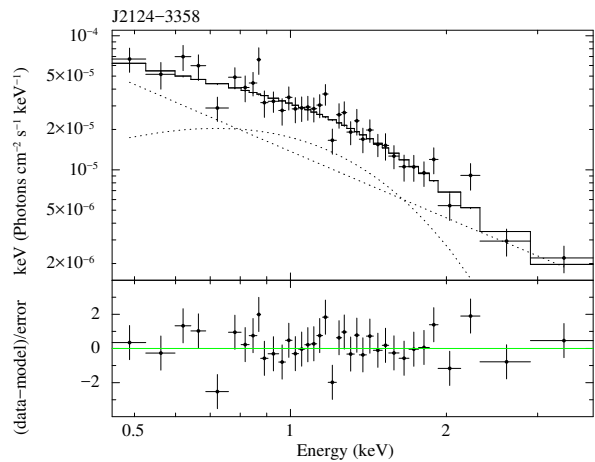}
\vspace{-0.5cm}
\caption{X-ray spectrum of J2124$-$3358 extracted using \cxo\ data. The best-fitting model is represented by an absorbed 
BB+PL with effective temperature of $k_{\rm B}T=0.25\pm0.03$~keV and PL photon index of $\Gamma=2.7^{+0.9}_{-0.3}$. }
\label{fig:j2124}
\end{center}
\end{figure}

In the case of J0030$+$0451 and J0633$+$0632, we found acceptable results using both a double BB model and a BB+PL
model. Hence, we cannot be confident that there exists a significant non-thermal component in these pulsars, and this is 
the reason why they are not included in Table \ref{tab:results}.
Another peculiar case is J2124$-$3358, where we could not find an energy range over which the non-thermal PL 
component clearly dominates over the thermal BB component (see Figure~\ref{fig:j2124}).

\begin{figure*}
\begin{center}
\includegraphics[width=0.49\textwidth]{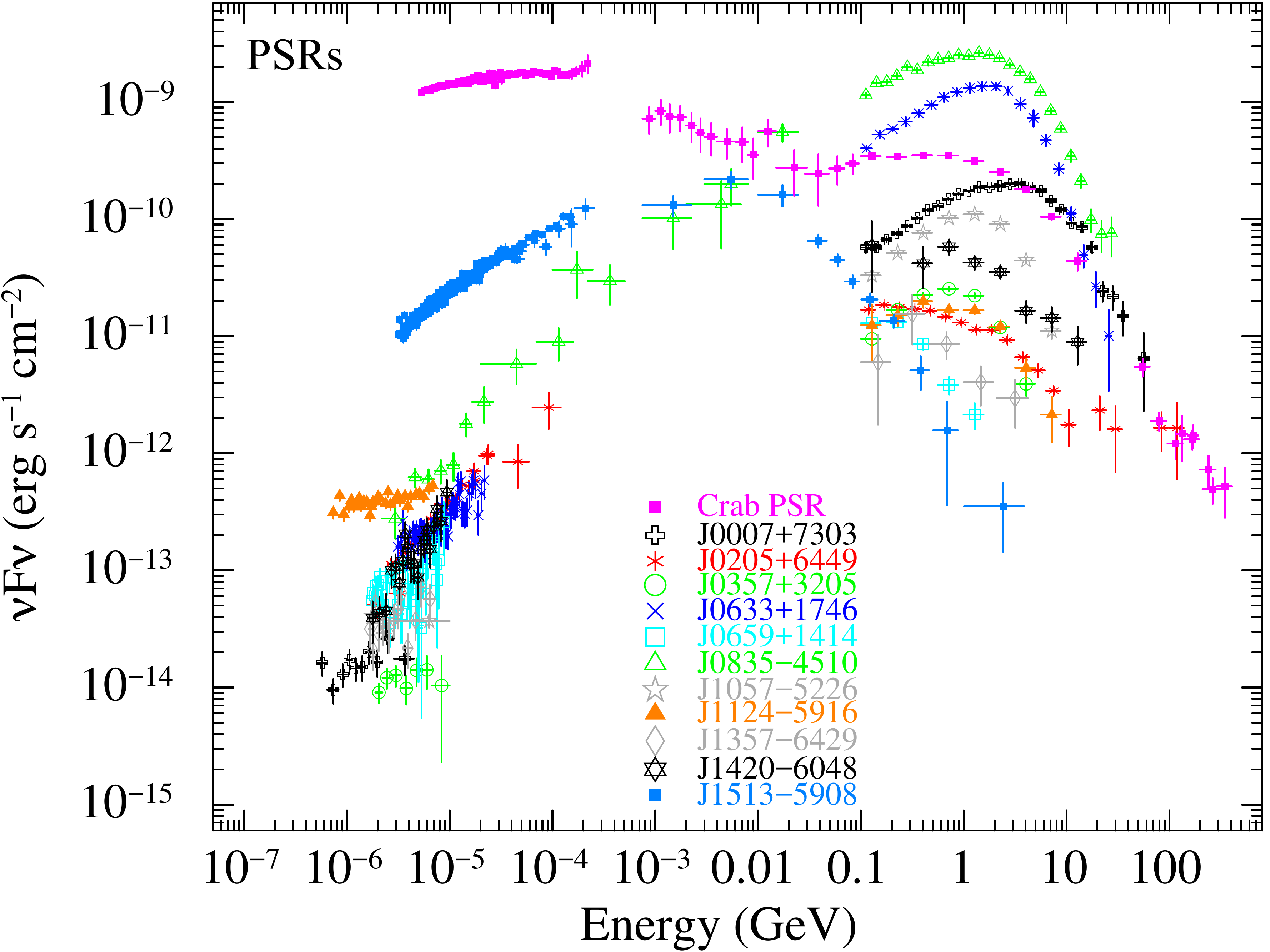}
\hspace{0.2cm}
\includegraphics[width=0.49\textwidth]{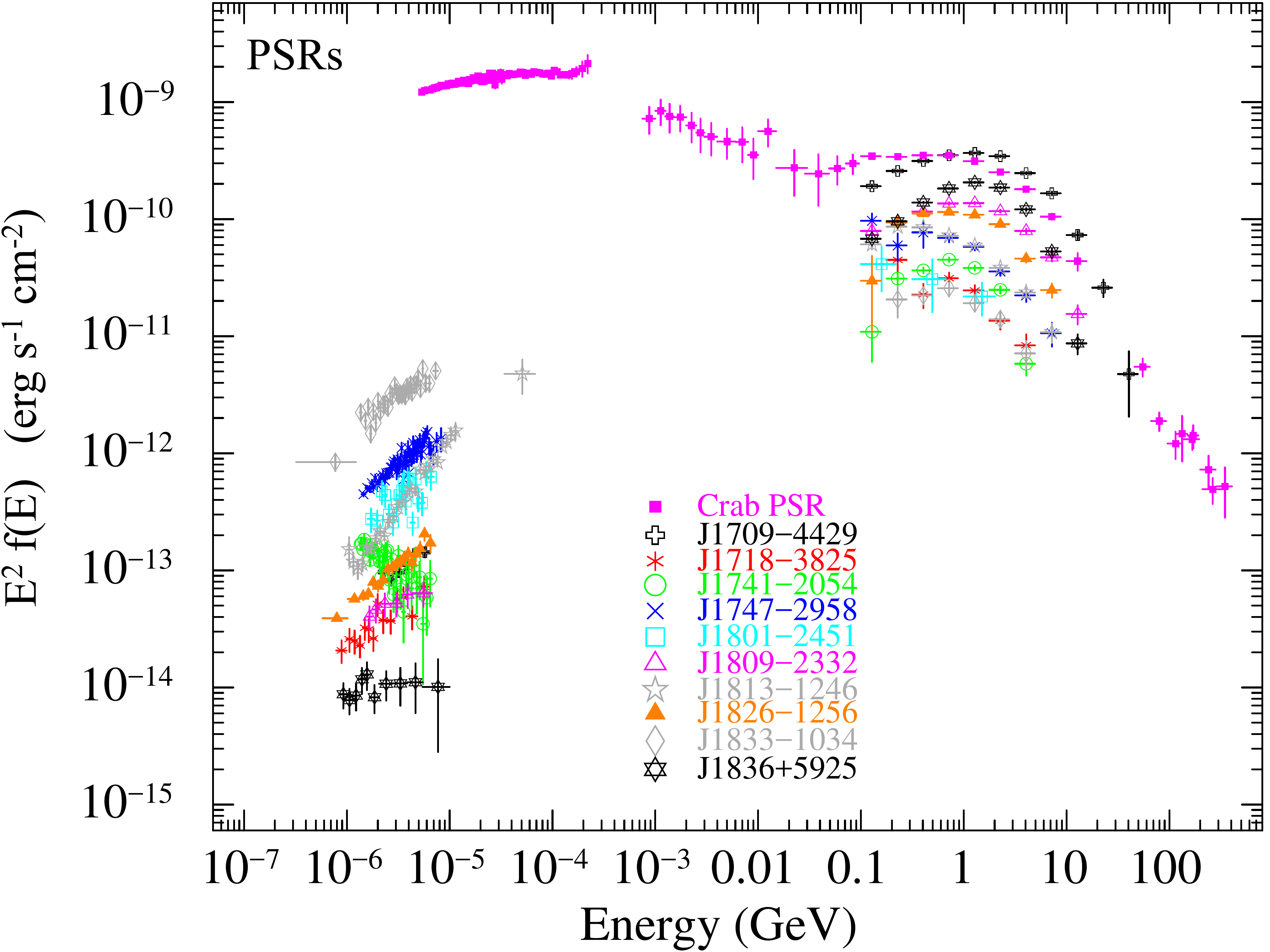}\\
\vspace{0.4cm}
\includegraphics[width=0.49\textwidth]{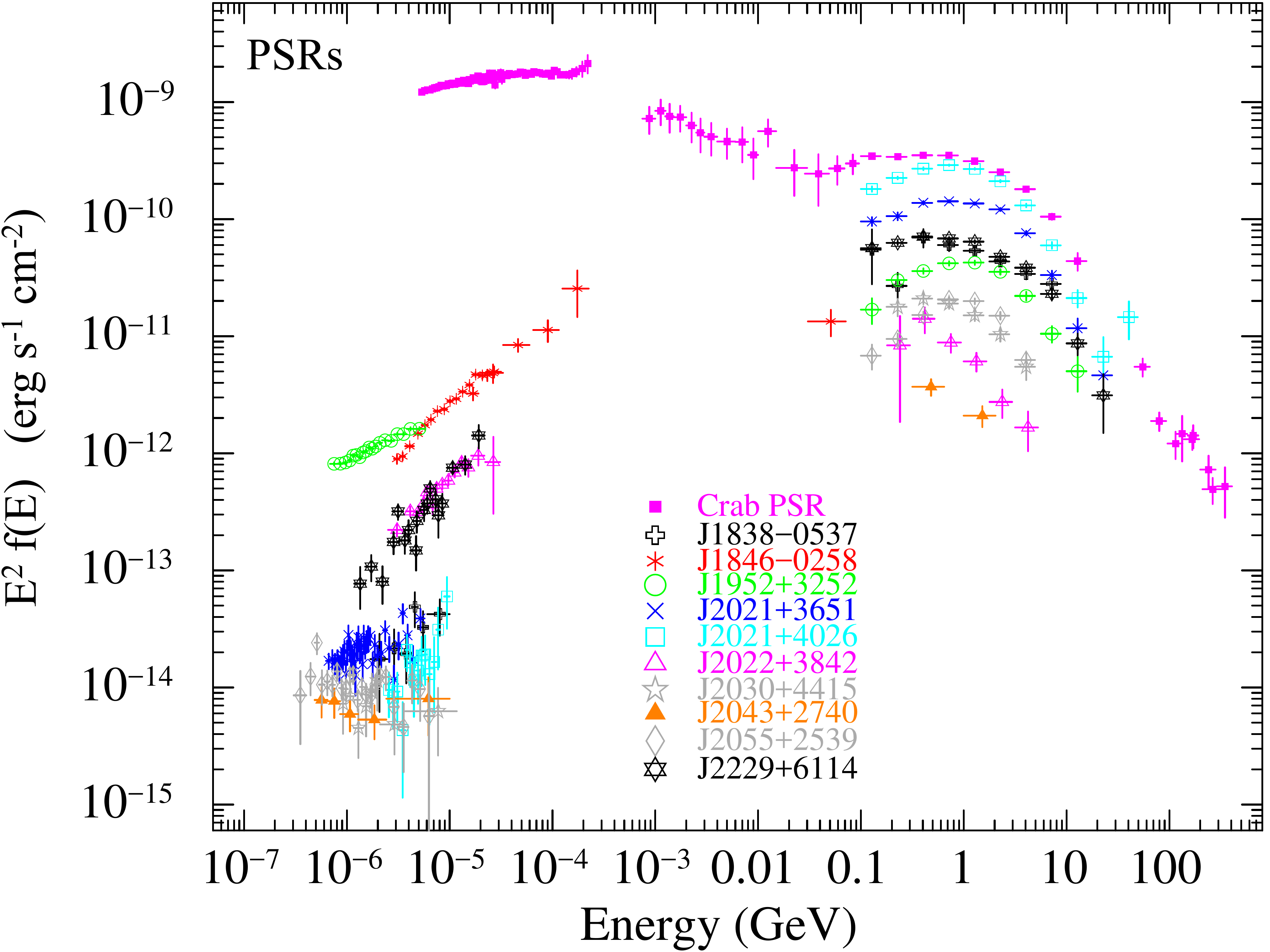}
\hspace{0.2cm}
\includegraphics[width=0.49\textwidth]{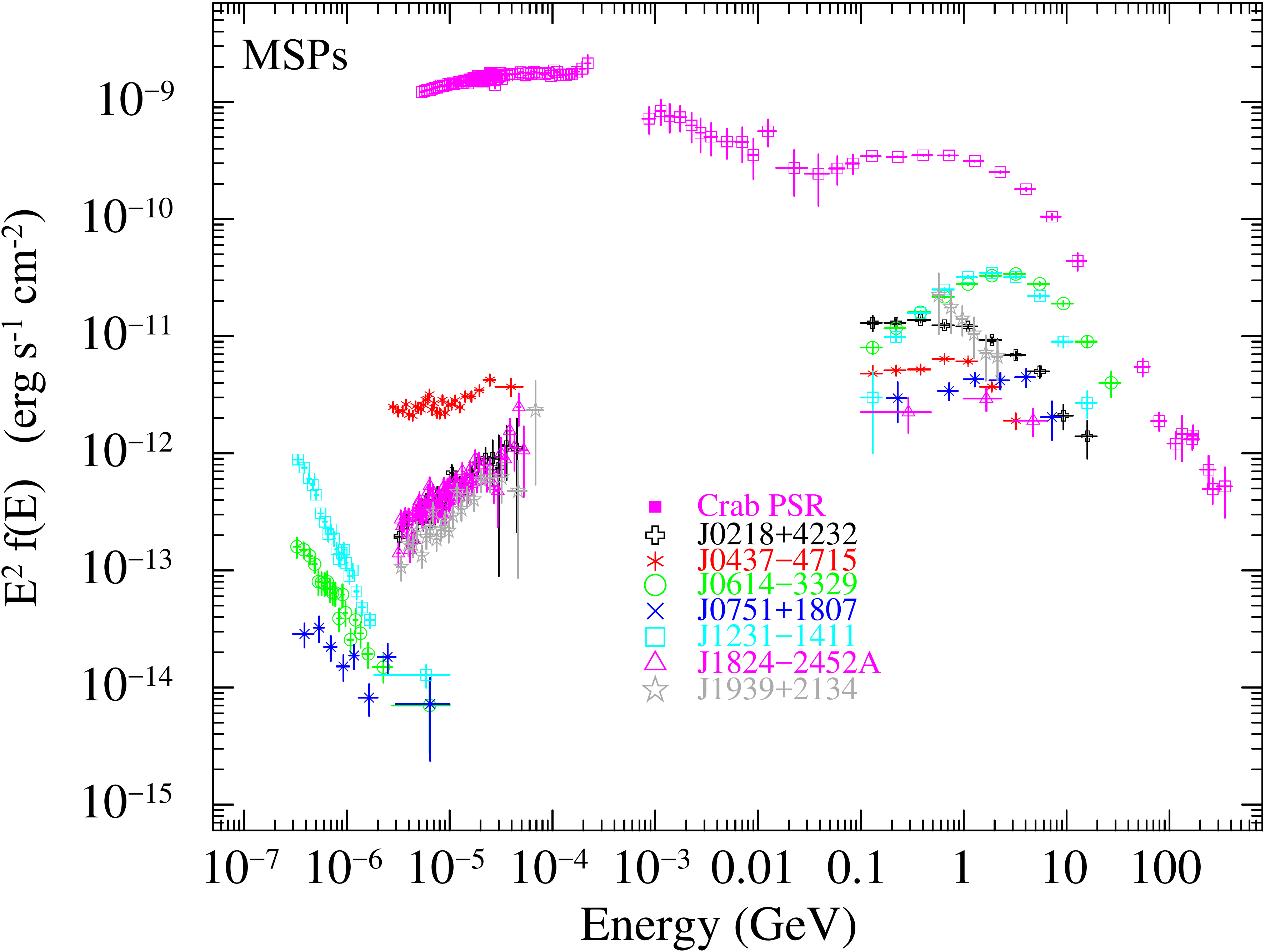}\\
\caption{Phase-averaged non-thermal X-ray and gamma-ray SEDs of pulsars. The SEDs are shown in four different panels for better 
visualization, with the MSPs shown in the bottom-right panel. This work contributed to the energy band below 10~keV (see Table~\ref{tab:results} 
for additional references). We have also included the SEDs extracted by \protect\cite{li18}, as well as those corresponding to the pulsed X-ray 
emission as reported by \protect\cite{kuiper15} (see their Figure~28) and by \protect\cite{Gotthelf2017}. The \fermi\ LAT high-energy gamma-ray 
spectra are taken from \protect\cite{Abdo10b,Abdo10a,pilia10,Lemoine-Goumard2011,2fpc,Ackermann2015,ohuchi15,kuiper16a,li16,smith16,Xing2016,Li173C,kuiper18}.
The broad-band SED of the Crab pulsar is shown in all panels for comparison.}
\label{fig:seds}
\end{center}
\end{figure*}

%
These findings indicate that, for 
these cases, the non-thermal nature of the pulsar emission is uncertain
(see also \citealt{zavlin2006}).
However, this does not necessarily rule out the presence of a significant non-thermal 
emission component at higher energies. 
For example, the total (pulsed plus unpulsed) X-ray spectrum of the Vela pulsar 
(J0835$-$4510) extracted using \xmm\ data is well described by a double BB model at energies below $\sim2$~keV (above 
this energy, the pulsar emission is overwhelmed by the PWN; see e.g. \citealt{manzali07}). 
However, the pulsed spectrum 
as detected with the Compton Gamma-Ray Observatory, the {\it RXTE} PCA and the INTEGRAL ISGRI extends up to the 
hard X-ray band, calling for a PL component at high (MeV) energies (e.g. \citealt{kuiper15}; see also Figure~\ref{fig:seds}). 

The results of the spectral fits for the remaining pulsars are reported in Table~\ref{tab:results}. 
The best-fit parameters and non-thermal X-ray fluxes obtained in this work are consistent 
within the uncertainties (at a confidence level $\leq2\sigma$ in all cases) with those reported in 
the literature using the same data sets (see \citealt{marelli11,marelli13,marelli14,marelli15,marelli16,ransom11,vanetten12,lin13,lin14,arumugasamy14,guillot16}).

The spectra we derived here, together with 
the best-fitting models are shown in the left-hand panels of Figure~\ref{fig:sedx1} in the Appendix~\ref{sec:appendixb}.

\subsection{Extraction of the non-thermal X-ray SEDs}
\label{sec:nthseds}

To extract the non-thermal X-ray SEDs of the pulsars listed in Table~\ref{tab:results}, we proceeded as follows. 
For those spectra 
adequately described by an absorbed power law (PL) model, we extracted 
the unfolded SEDs from the best-fitting model, plotting them in terms of $E^{2}f(E)$ (i.e. in units of \flux). Here $E$ represents the geometric mean 
of the lower and upper energies of the plotted energy bin, and $f(E)$ is the photon flux. 
In particular, for each case we extracted two unfolded SEDs: one obtained from the best-fitting absorbed model, and one derived after removing 
the absorption component from the model, so as to reproduce the intrinsic, de-absorbed SED of the pulsar.
We then scaled the fluxes estimated over each energy bin in the unfolded observed SED
by the ratio between the best-fitting de-absorbed and observed models over the corresponding energy bin, so as to extract the 
intrinsic X-ray SED over a wide energy range including the softest X-ray energies.
Instead, for those spectra that were well-described by an absorbed blackbody plus power law 
(BB+PL) model (or by two blackbodies plus power law, 2BB+PL as in the 
case of J0659$+$1414), we considered only the portion of the spectrum at higher energies where the fractional contribution of the 
PL component is dominant and the BB component gives only a negligible contribution to the emission. 
We defined 
the lower energy threshold as the energy above which the energy-integrated fractional contribution of the thermal component is 
$\lesssim3$\%. 
This value was evaluated case by case. 
We verified that slightly different choices for the threshold energy yielded 
similar results in the following analysis. 
We then extracted the de-absorbed SEDs in the same way as described above (note that 
the observed and de-absorbed unfolded spectra of the PL component are very similar with each other 
in these cases, as the spectral shape is not really affected by absorption effects at high energies).

All the X-ray SEDs of the non-thermal components extracted in the present work are reported in the right-hand panels of Figure~\ref{fig:sedx1} in 
Appendix~\ref{sec:appendixb}. The represented SEDs corresponding to PL photon indexes $\Gamma>2$ decrease 
with increasing energy ($E^{2}f[E] \propto E^{2-\Gamma}$).  
This is the case for three out of the four MSPs pulsars analyzed here. 
%
The three MSPs detected by \cite{Gotthelf2017} show SEDs in the X-ray band that are increasing with energy.
Hence, the non-thermal spectra over the same X-ray band, even at the same flux level, can differ significantly in slope.
For those pulsars whose SEDs are decreasing with energy, further components could show up at higher energies. If so, they will be useful for calibrating 
theoretical scenarios. Such components are actually predicted to appear according to synchro-curvature models 
(see Figure 10 by \citealt{Torres2019}), although they will be likely still too dim up to the MeV range to allow detection with current facilities.

The X-ray SEDs are also shown together with the gamma-ray SEDs in Figure~\ref{fig:seds}, using the same representation. 
This figure represents the state-of-the-art for the SEDs of pulsars showing a non-thermal component in their X-ray emission.

\begin{figure*}
\begin{center}
\includegraphics[width=1.0\textwidth]{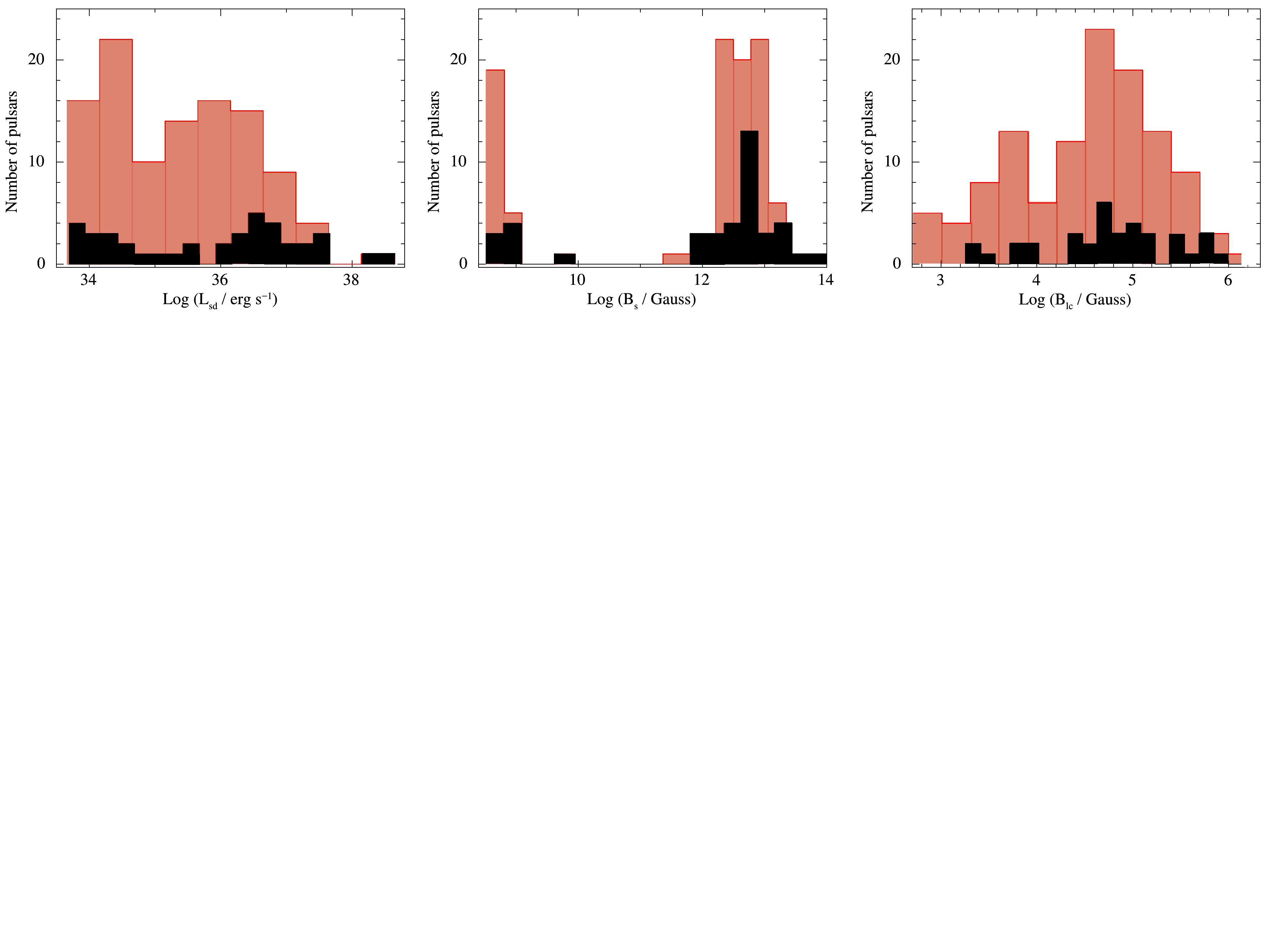}
\vspace{-9.2cm}
\caption{Distributions of all the gamma-ray pulsars listed in the 2PC catalogue (in red) and of the gamma-ray pulsars with a non-thermal X-ray emission component 
considered in this work (in black) as a function of the spin-down luminosity (left-hand panel), the dipolar component of the magnetic field at the surface (middle panel)
and the magnetic field at the light cylinder radius (right-hand panel).}
\label{fig:distrib_psrs}
\end{center}
\end{figure*}

\section{Discussion}
\label{sec:discuss}

In this paper we have compiled a sample of 40 pulsars detected in gamma-rays with a clear non-thermal 
X-ray emission component,  providing the full SEDs for all of them. 
These SEDs extend over more than eight and five decades in energy 
and flux, respectively. A large variety of spectral slopes is observed (Figure~\ref{fig:seds}).
This dataset is essential for model comparison, and is made available online by means of tables in ASCII format.
The columns in these tables list the minimum, maximum and central energy of the X-ray spectral bin,
as well as the flux and its corresponding 1-$\sigma$ uncertainty in cgs units.
We have also outlined in detail the caveats that need to be addressed to estimate the broad-band spectral shape and flux of 
pulsars in the most consistent way as possible.

Our sample includes several sources having a non-thermal emission that is relatively bright in the soft X-ray band,
but is too dim at higher energies to allow an adequate characterization using data from high-energy missions such as \integral, \suzaku\ 
and \rxte. 
These cases were not included in the soft gamma-ray catalog compiled by  \cite{kuiper15}, which only lists all pulsars for 
which non-thermal pulsed emission had been detected above 20\,keV.
However, in order to get a global picture of the non-thermal pulsar properties, we need to consider and eventually model the emission from 
as many pulsars as possible (including e.g. those not showing pulsations and those that are relatively faint in the X-ray band).
We attempt such a study in our accompanying paper \citep{Torres2019}.

Figure~\ref{fig:distrib_psrs} shows the distributions of the spin-down power, the dipolar magnetic field strength at the star surface $B_{\rm S}(t) = 6.4 \times 10^{19} (P \dot P)^{1/2}\; {\rm G}$
and the magnetic field strength at the light cylinder radius $B_{\rm lc}(t) = 5.9 \times 10^{8} P^{-5/2} \dot P^{1/2}\; {\rm G}$ 
for our sample (in black) and the overall population in the 2PC catalogue (in red; $P$ and $\dot{P}$ are the spin 
period and the time derivative of the spin period, respectively). 
Our sample of pulsars selected in the 2PC catalogue mimics the trends observed for the parent population of gamma-ray pulsars, that is,
there appears to be no obvious difference between our sub-sample and the rest of gamma-ray pulsars.
This is somewhat different from the results of the soft gamma-ray catalog by \cite{kuiper15}, 
where it appeared that the detected pulsating hard X-ray sources were younger and more luminous than the average of the gamma-ray emitting pulsars. 
However, their sample includes also a significant number of X-ray pulsars that were undetected in the gamma-ray band,
indicating that their SEDs reach maximal luminosities at MeV energies (e.g. like J1513-5908, J1846-0258 visible in Figure 3).
The different properties regarding spin-down luminosity could be  actually promoted by these pulsars.
Here, our sample includes pulsars with spin-down luminosity below 10$^{36}$ \lum\ and pulsars with dipolar magnetic field at the surface 
below 10$^{10}$ G (the MSPs). 
 Hence, in this work we are sampling pulsars over a broader range in spin-down luminosity and magnetic field compared to \cite{kuiper15}.
 All of them are still non-thermal.

A significant sample of gamma-ray pulsars with an adequate characterizazion of non-thermal emission in the X-ray band is building up, which 
can now be used for understanding the applicability of theoretical models in a variety of circumstances.
 Clearly, this sample is destined to increase in the upcoming years. 
On the one hand, the continuous survey of the gamma-ray sky with the \fermi\ LAT will lead to the discovery of more, fainter gamma-ray pulsars. 
A total of 234 gamma-ray pulsars have been discovered up to October 2018\footnote{See 
\texttt{https://confluence.slac.stanford.edu/display/GLAMCOG/}\\\texttt{Public+List+of+LAT-Detected+Gamma-Ray+Pulsars}}. 
This is already a factor of 2 larger than the number of sources listed in the 2PC.
On the other hand, follow-up observations with current and future missions (such as \emph{Athena}) will enable to
constrain the spectral shape and flux of these sources in the soft X-ray band. 
Moreover, missions such as \emph{Athena} itself, will be also key in order to characterise with unprecedented detail
the (so far poorly constrained) X-ray emission of the faintest pulsars (such as the ten pulsars mentioned in Section \ref{sec:selection2}).

\section*{Acknowledgements}

FCZ and DFT acknowledge support from the Spanish Ministry of Economy, Industry and Competitiveness grants PGC2018-095512-B-I00, SGR2017-1383, and AYA2017- 92402-EXP. 
FCZ is supported by a Juan de la Cierva fellowship.DV acknowledges support from the Spanish Ministry of Economy, Industry and Competitiveness grants AYA2016-80289-P and 
AYA2017-82089-ERC (AEI/FEDER, UE).


\appendix

\section{X-ray spectra and SEDs}
\label{sec:appendixb}

Figure~\ref{fig:sedx1} shows the X-ray spectra with the best-fitting models superimposed and the X-ray 
SEDs for pulsars obtained in this work.

\begin{figure*}
\begin{center}
\includegraphics[width=0.508\textwidth]{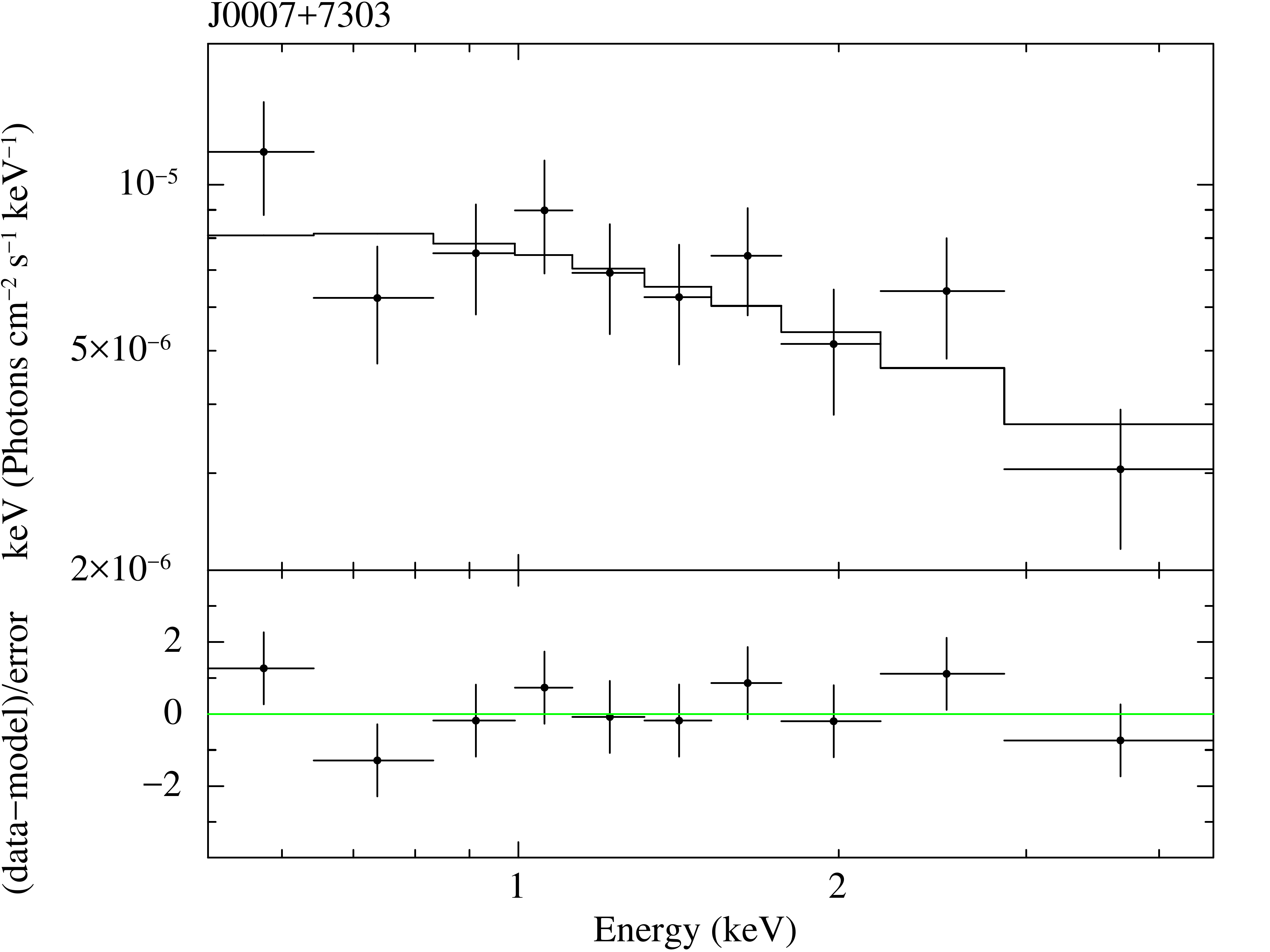}
\hspace{-0.44cm}
\includegraphics[width=0.508\textwidth]{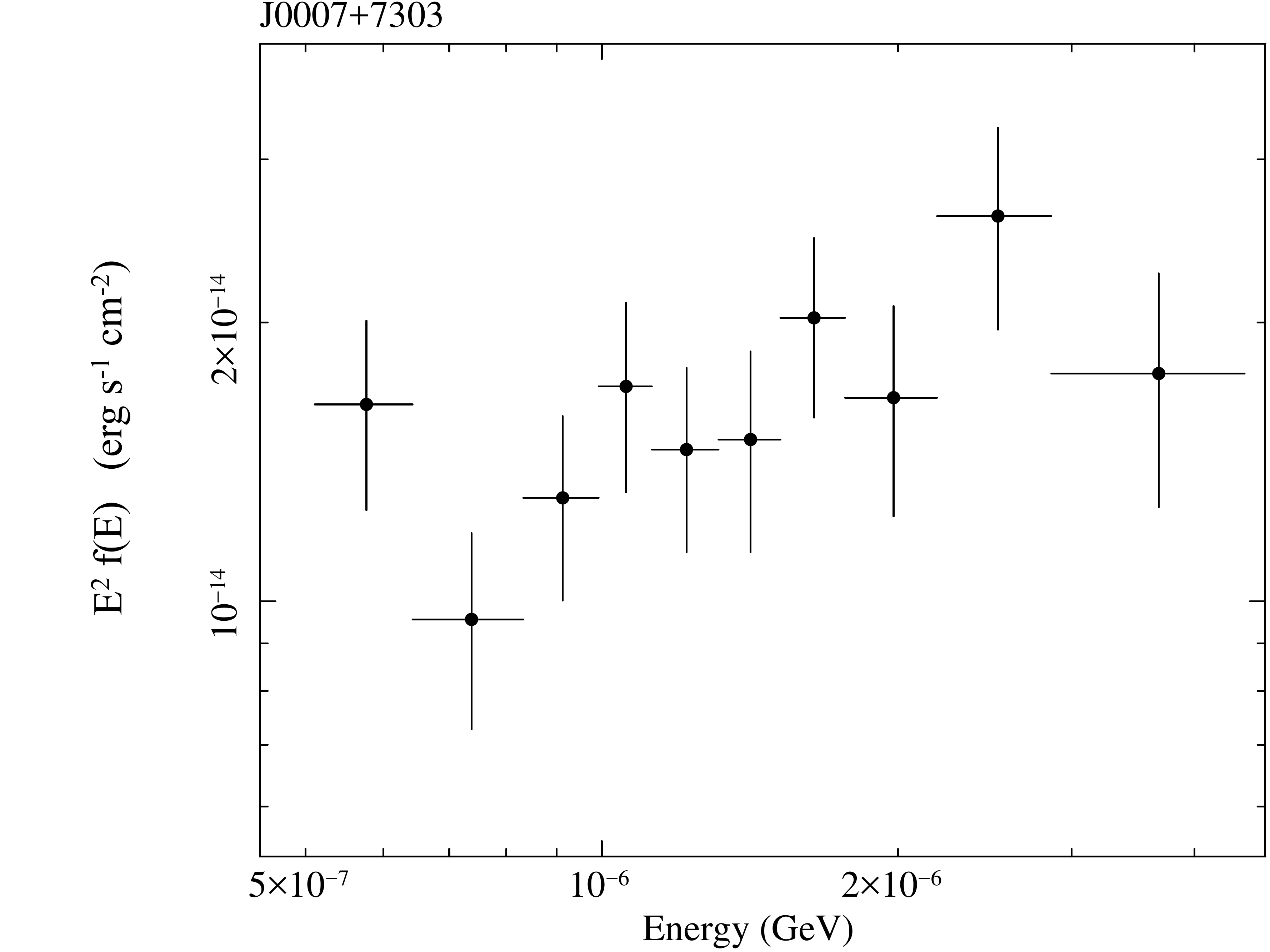}\\
\vspace{0.5cm}
\includegraphics[width=0.508\textwidth]{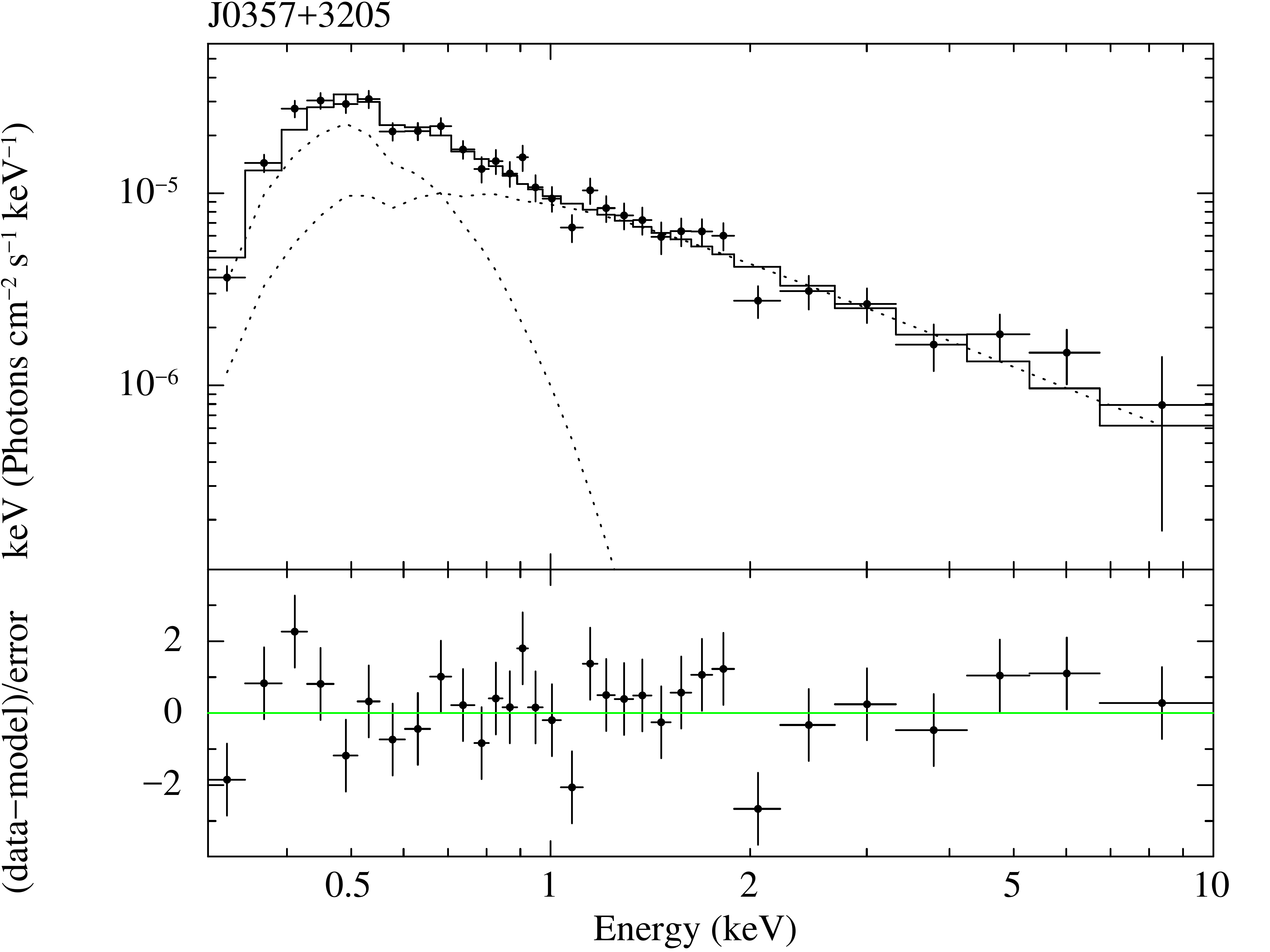}
\hspace{-0.44cm}
\includegraphics[width=0.508\textwidth]{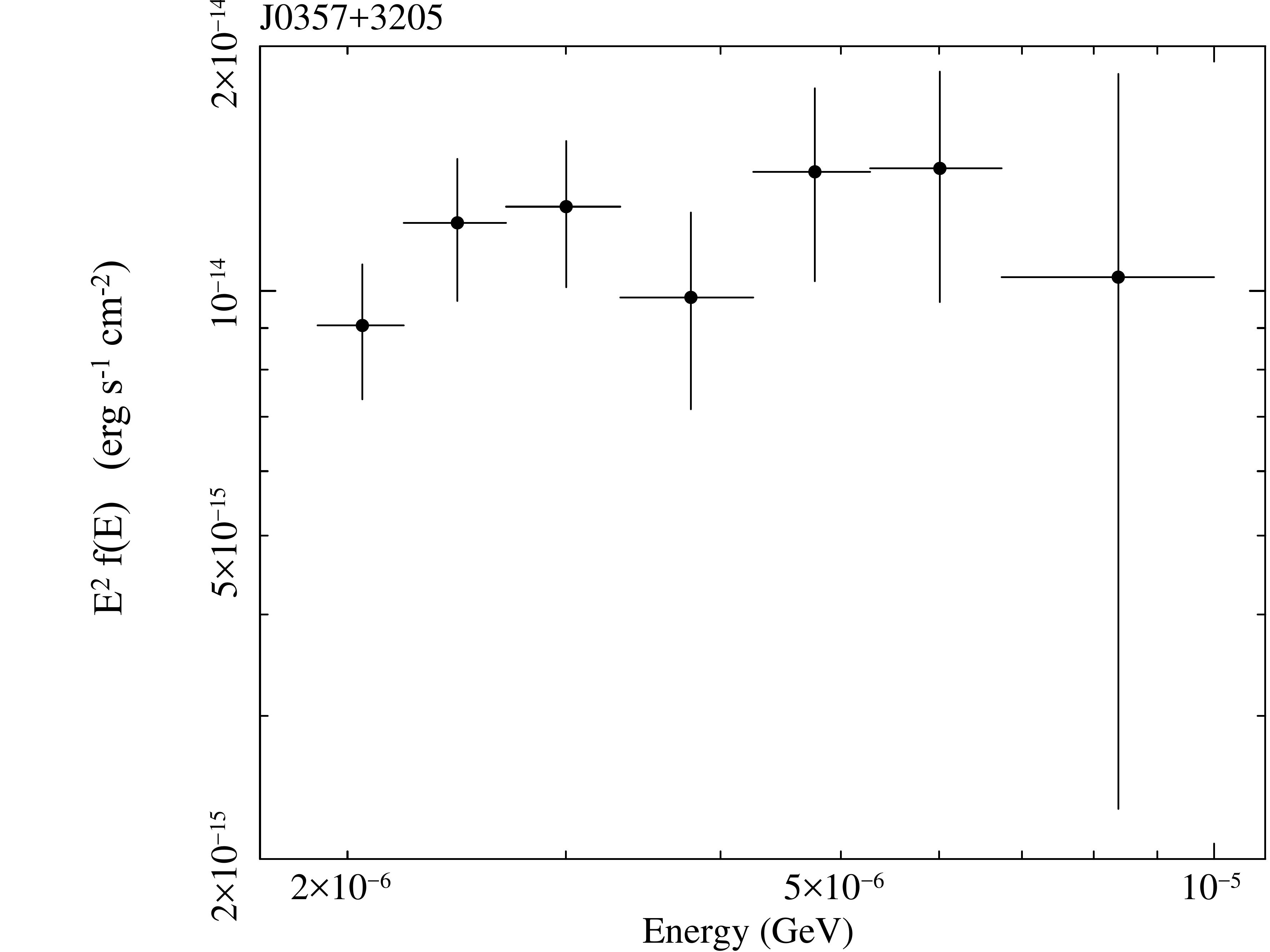}\\
\vspace{0.5cm}
\includegraphics[width=0.508\textwidth]{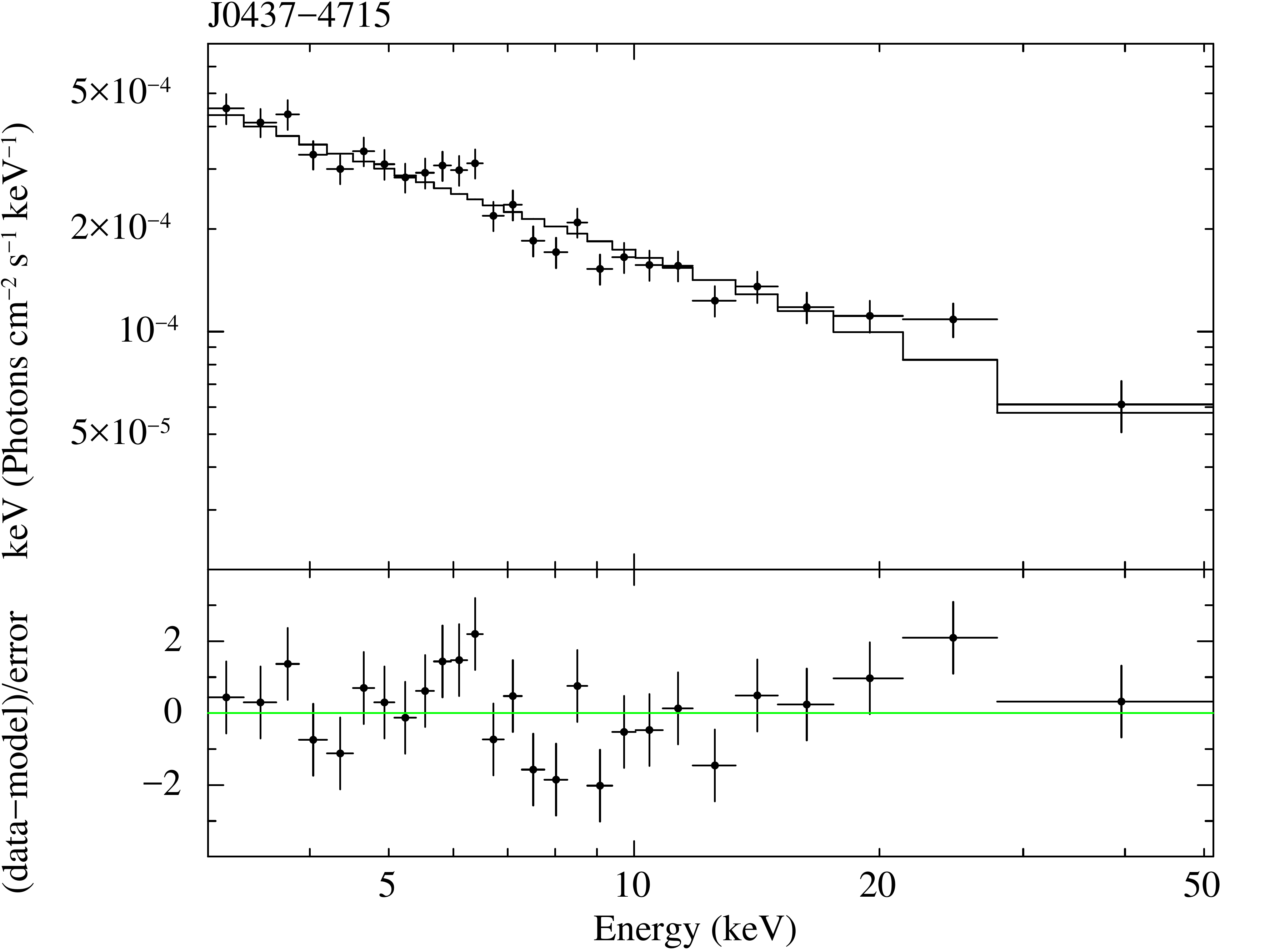}
\hspace{-0.44cm}
\includegraphics[width=0.508\textwidth]{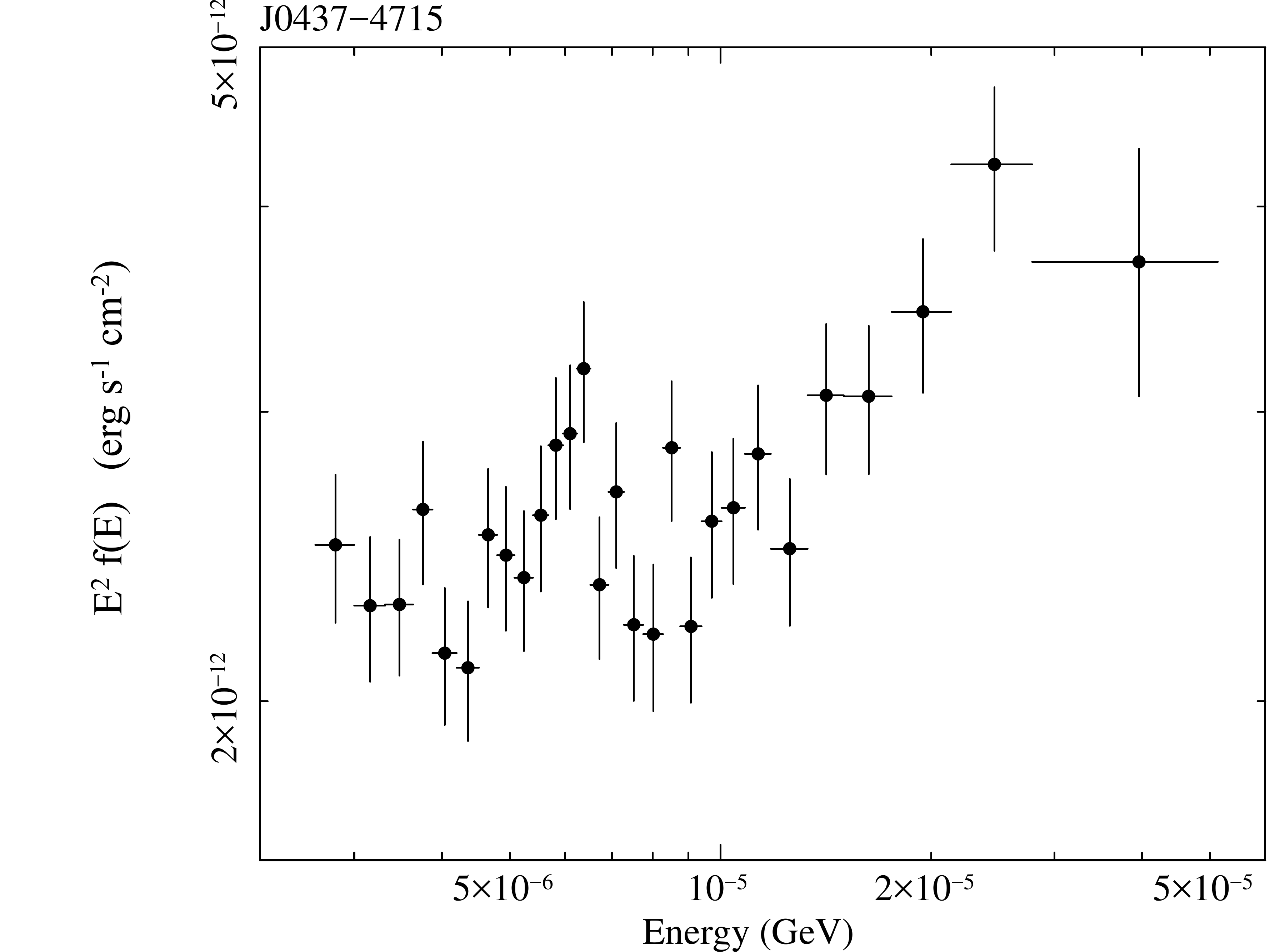}\\
\vspace{0.2cm}
\caption{Left-hand panels: background-subtracted unfolded spectra in the $E^2f(E)$ representation together with the best-fitting model 
(top; see Table \ref{tab:results}) and post-fit residuals (bottom). The contributions of the various additive components to the overall 
best-fitting model also are plotted in the cases where multiple components are needed to fit adequately the spectra. Right-hand panels: 
de-absorbed X-ray SEDs in the same units as shown in Figure~\ref{fig:seds}.}
\label{fig:sedx1}
\end{center}
\end{figure*}

\setcounter{figure}{1}
\begin{figure*}
\begin{center}
\includegraphics[width=0.508\textwidth]{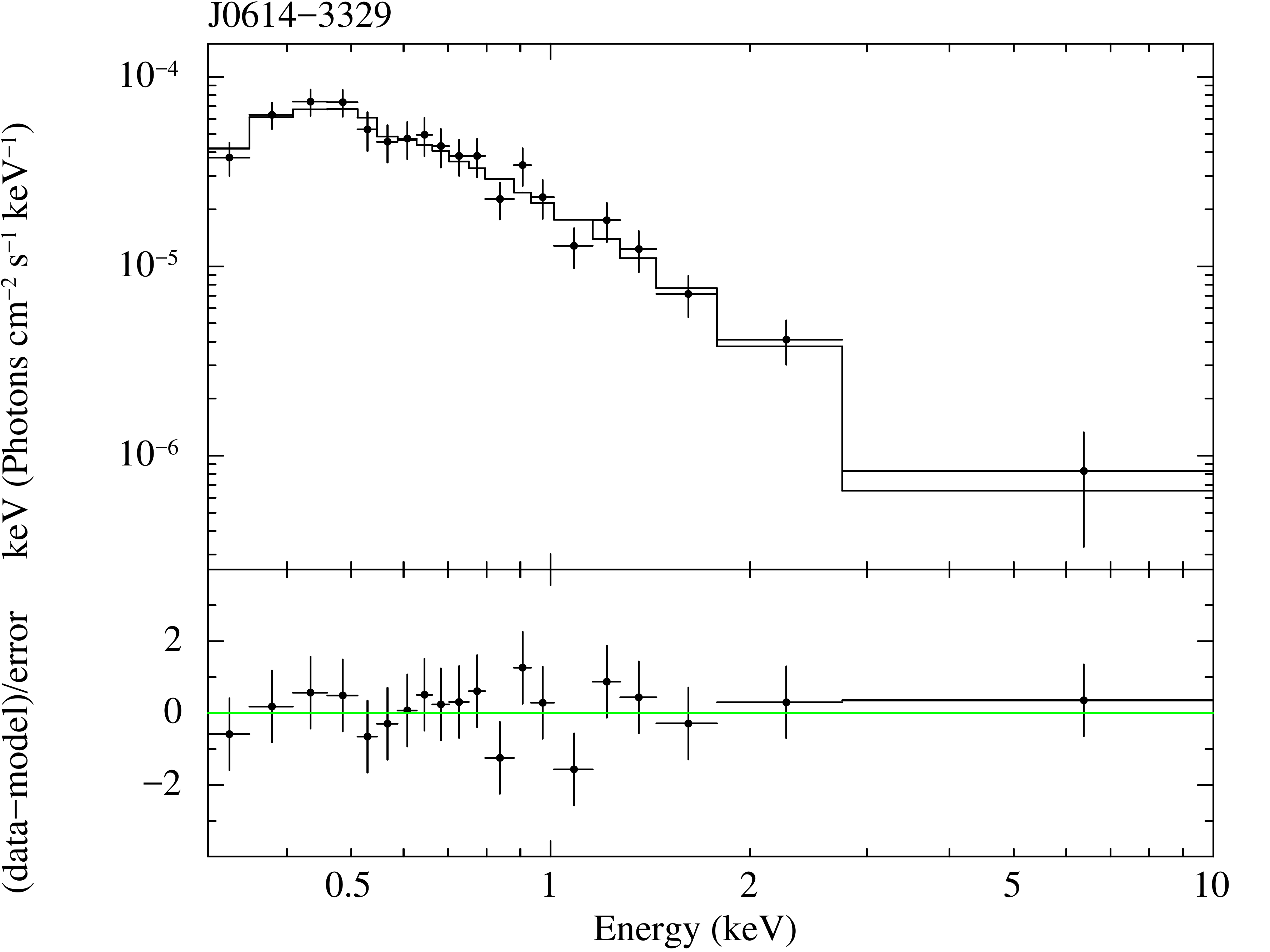}
\hspace{-0.44cm}
\includegraphics[width=0.508\textwidth]{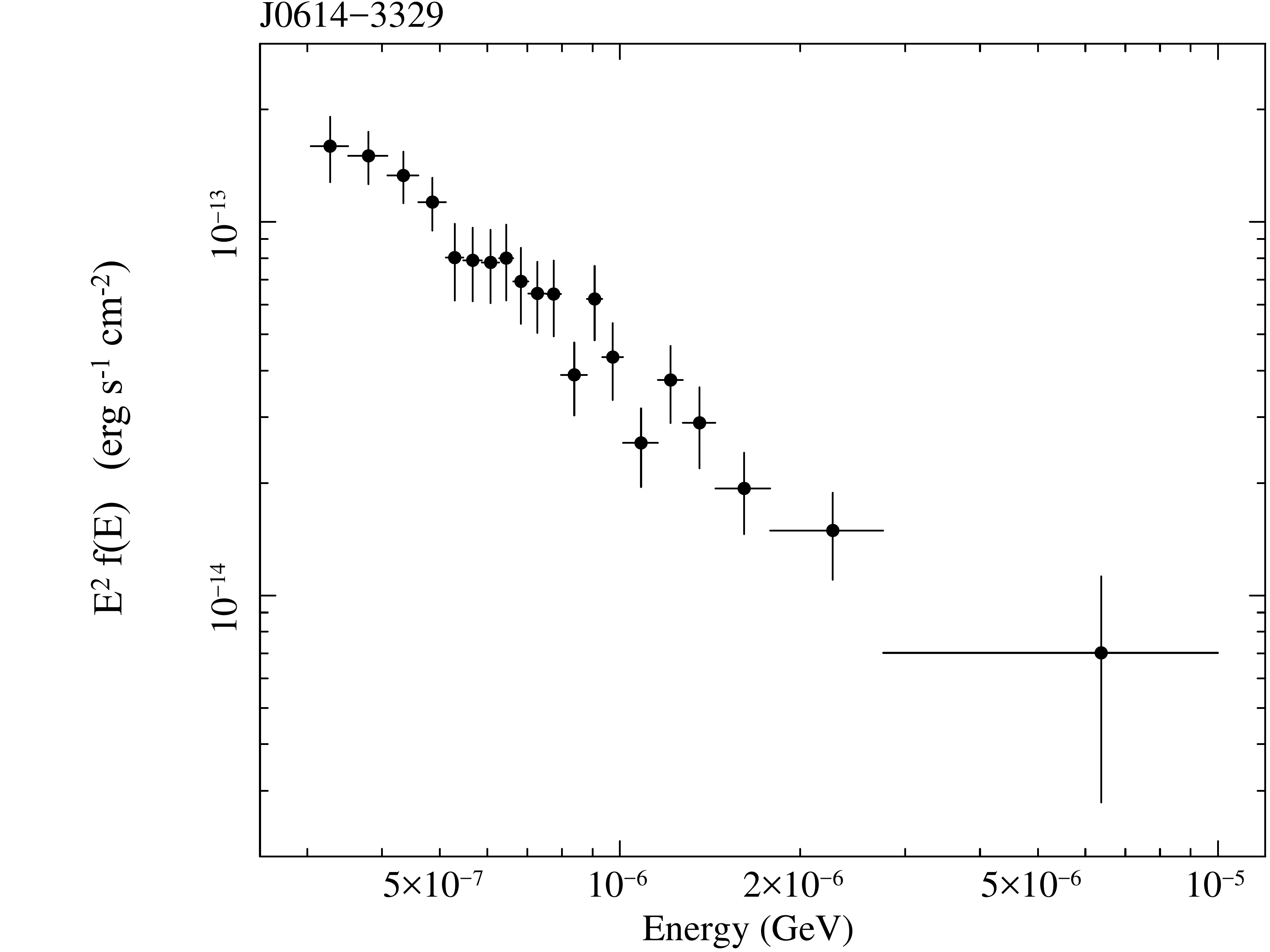}\\
\vspace{0.5cm}
\includegraphics[width=0.508\textwidth]{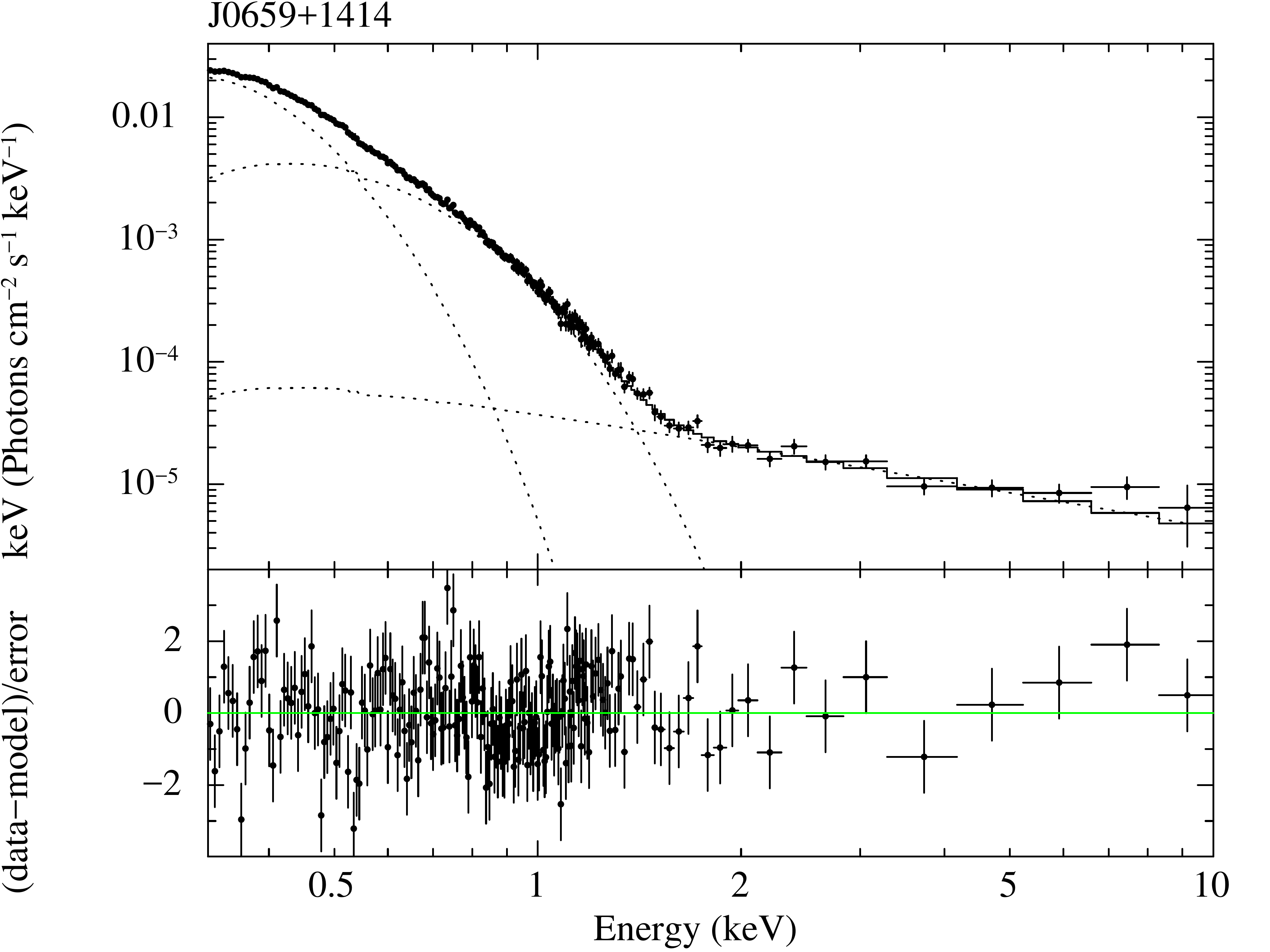}
\hspace{-0.39cm}
\includegraphics[width=0.505\textwidth]{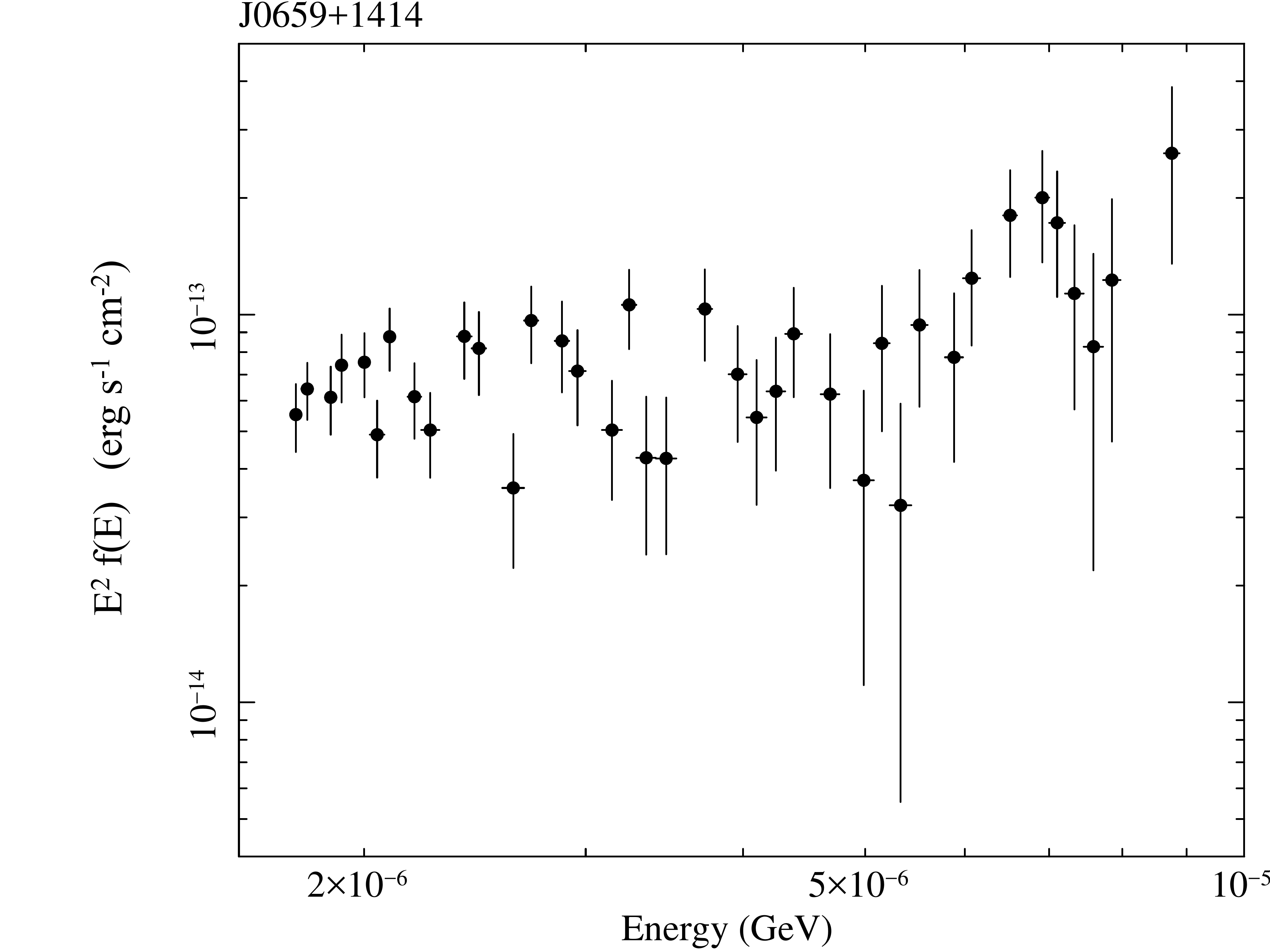}\\
\vspace{0.5cm}
\includegraphics[width=0.508\textwidth]{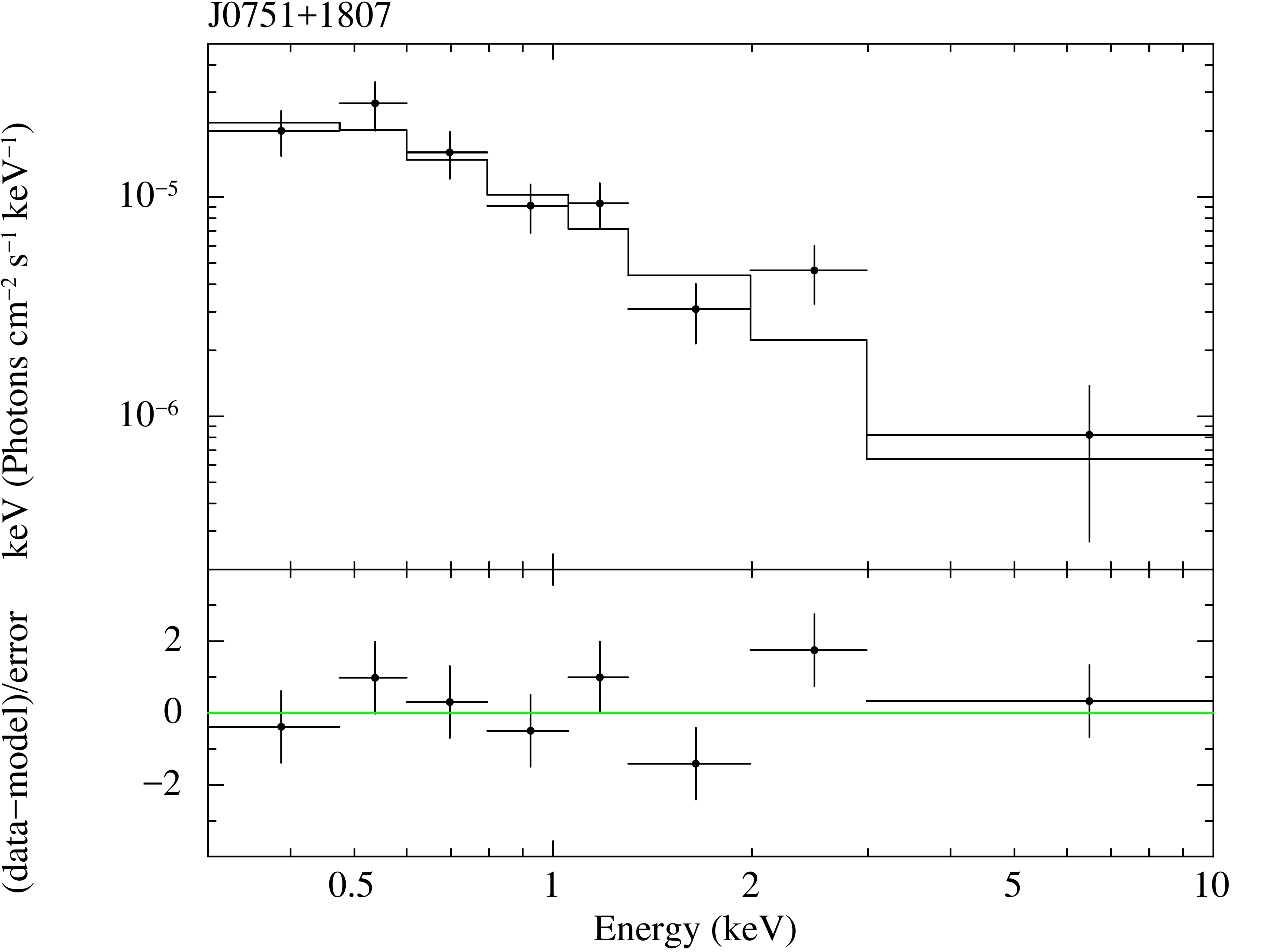}
\hspace{-0.44cm}
\includegraphics[width=0.508\textwidth]{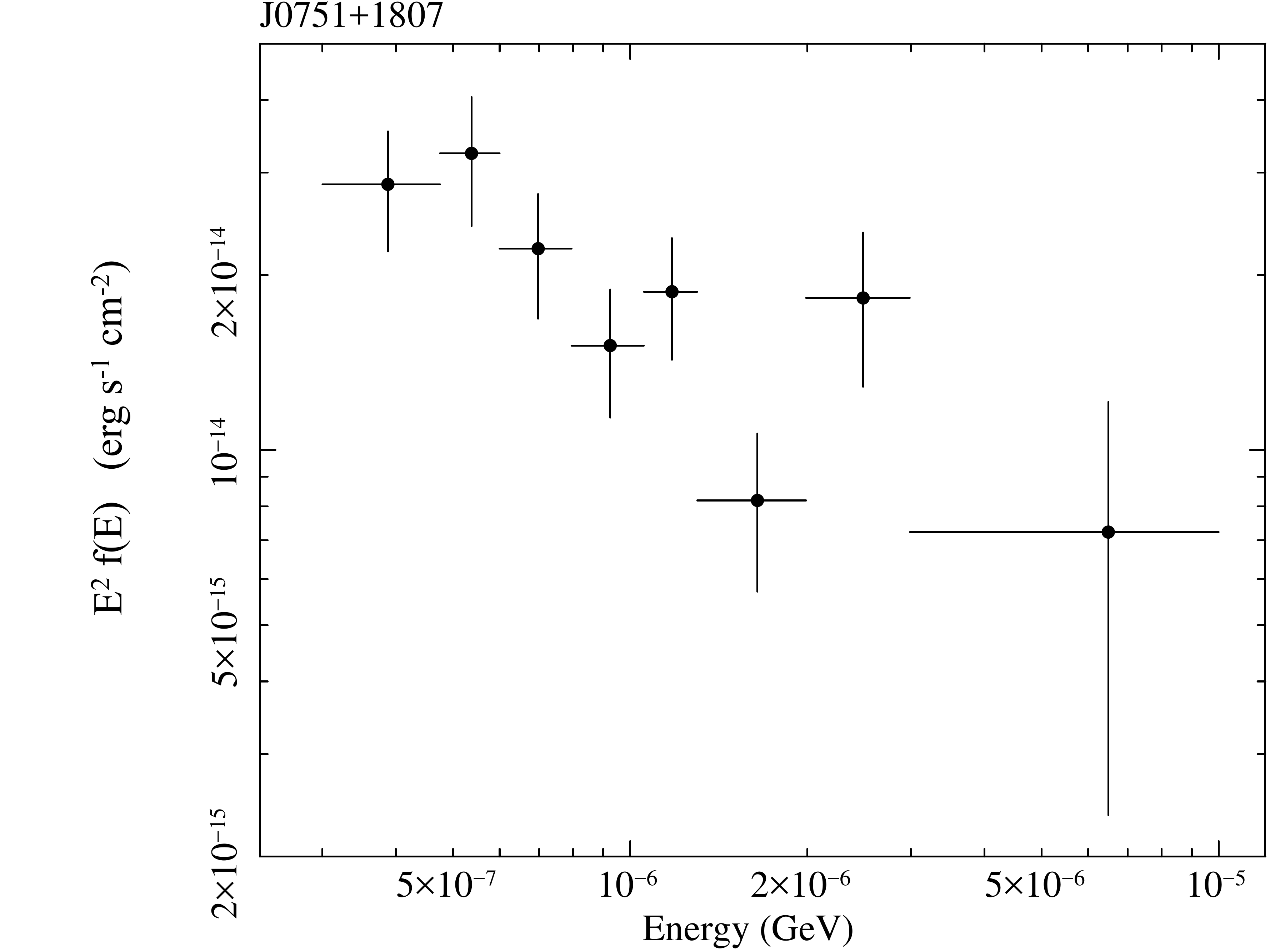}\\
\vspace{0.2cm}
  \contcaption{ }
\label{fig:sedx2}
\end{center}
\end{figure*}

\setcounter{figure}{1}
\begin{figure*}
\begin{center}
\includegraphics[width=0.508\textwidth]{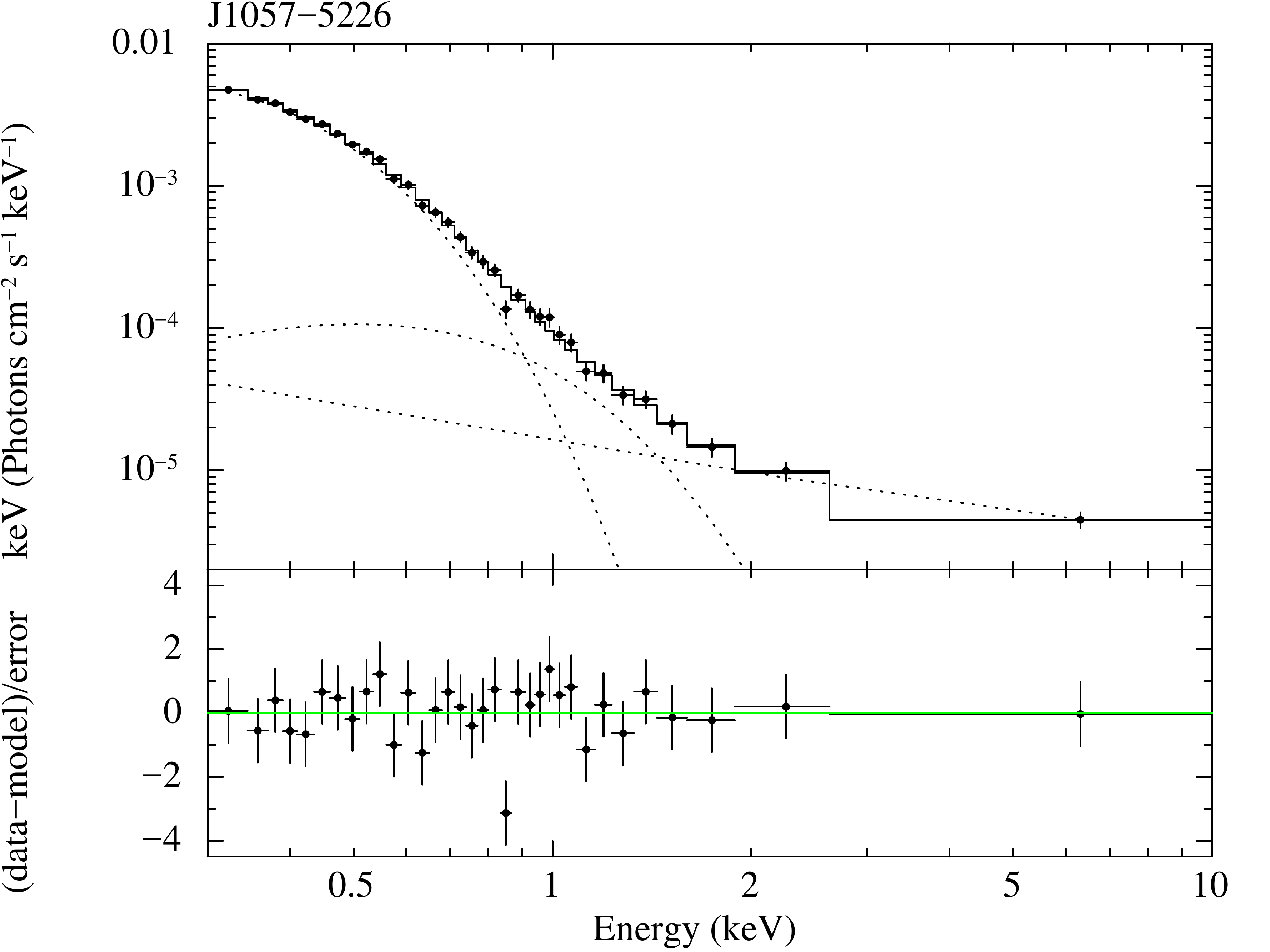}
\hspace{-0.44cm}
\includegraphics[width=0.508\textwidth]{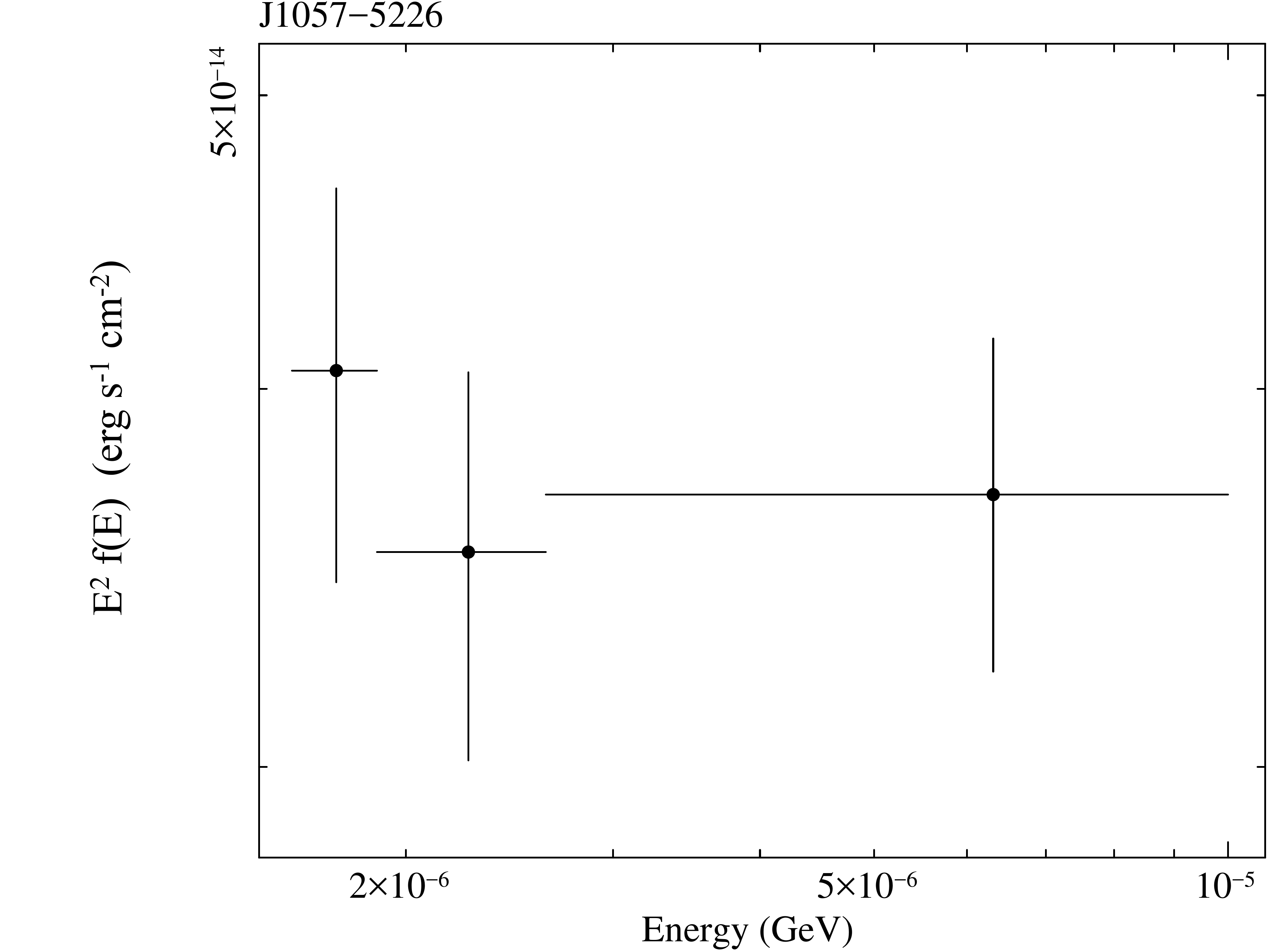}\\
\vspace{0.5cm}
\includegraphics[width=0.508\textwidth]{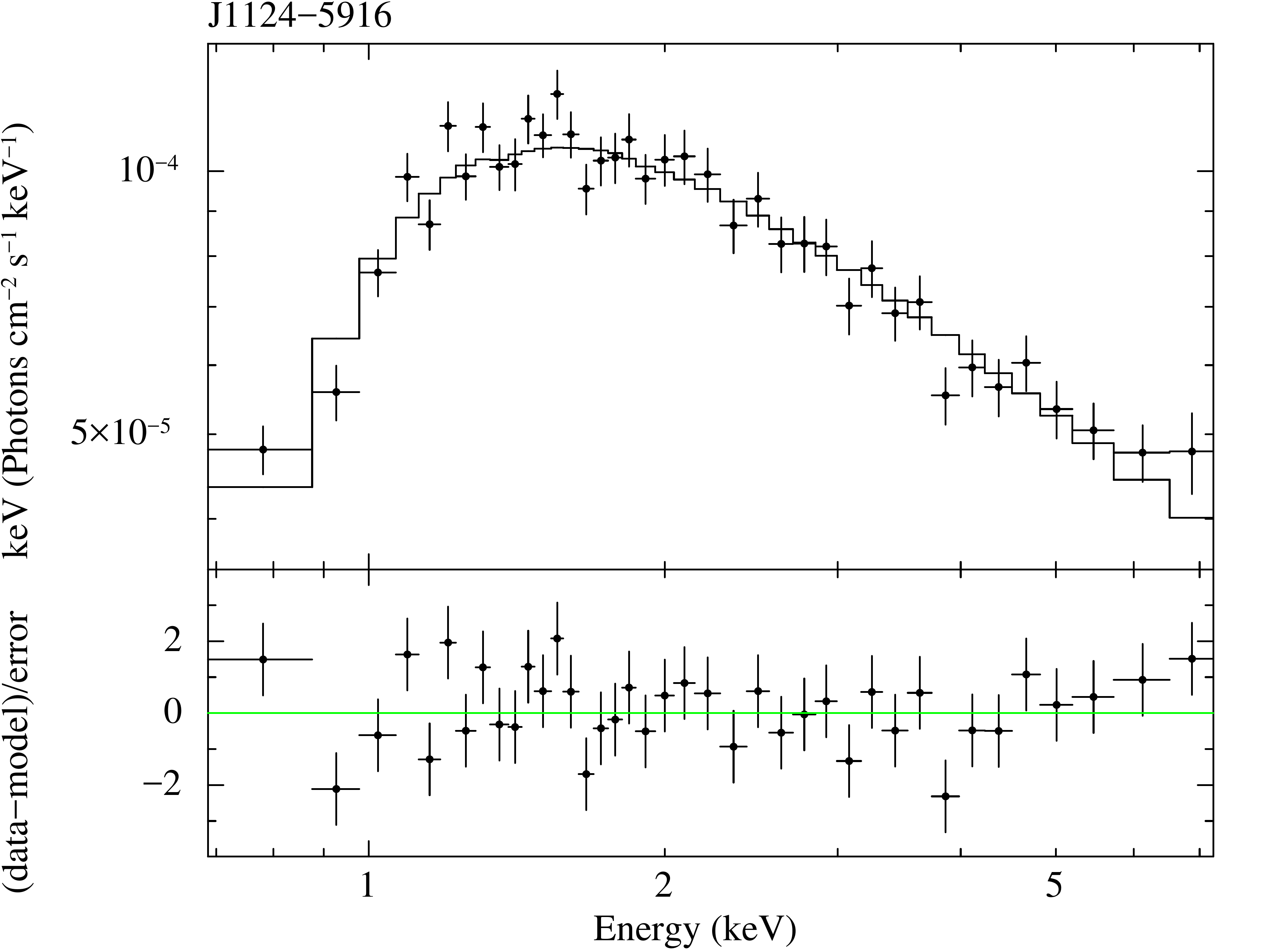}
\hspace{-0.44cm}
\includegraphics[width=0.508\textwidth]{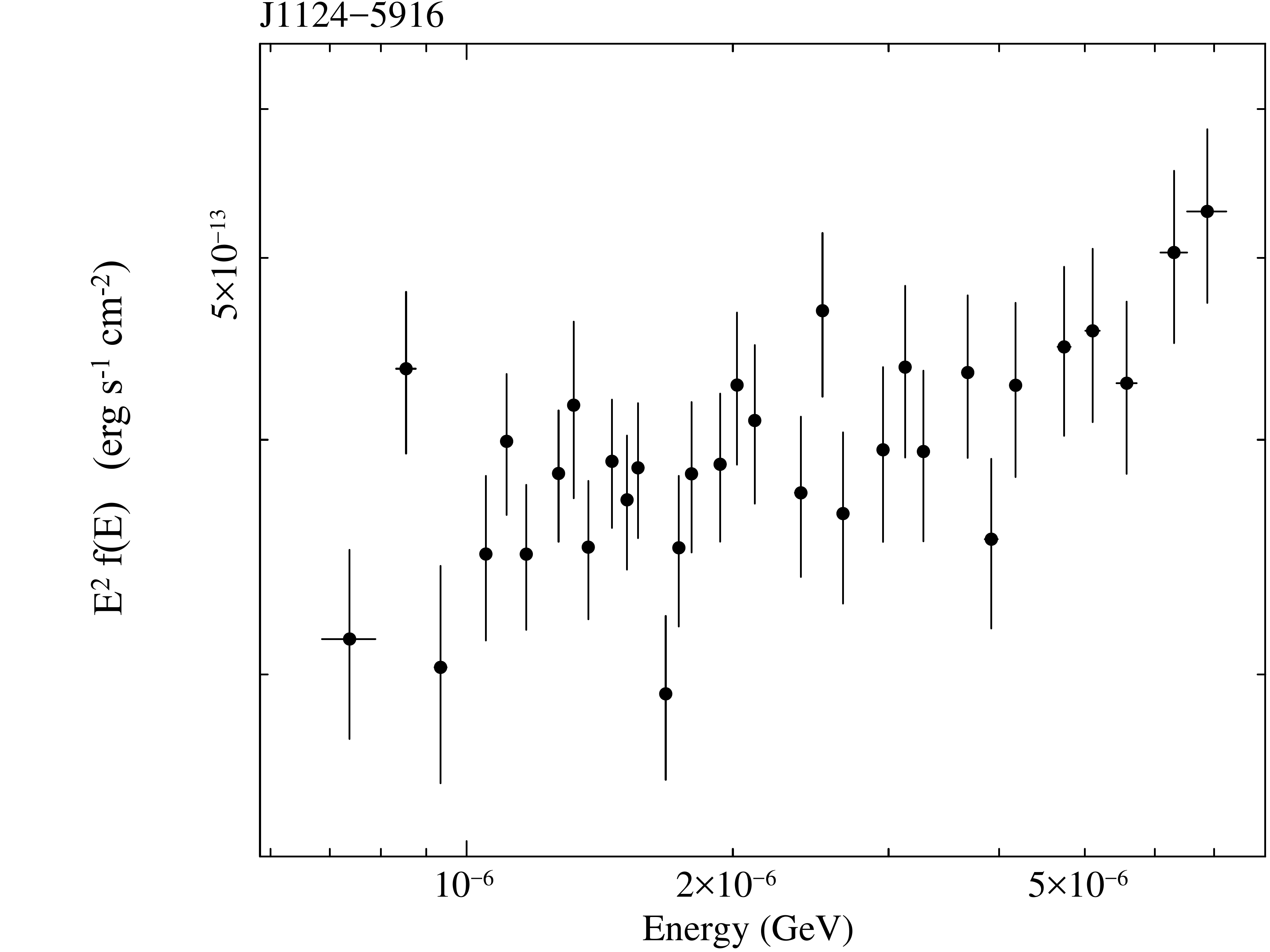}\\
\vspace{0.5cm}
\includegraphics[width=0.508\textwidth]{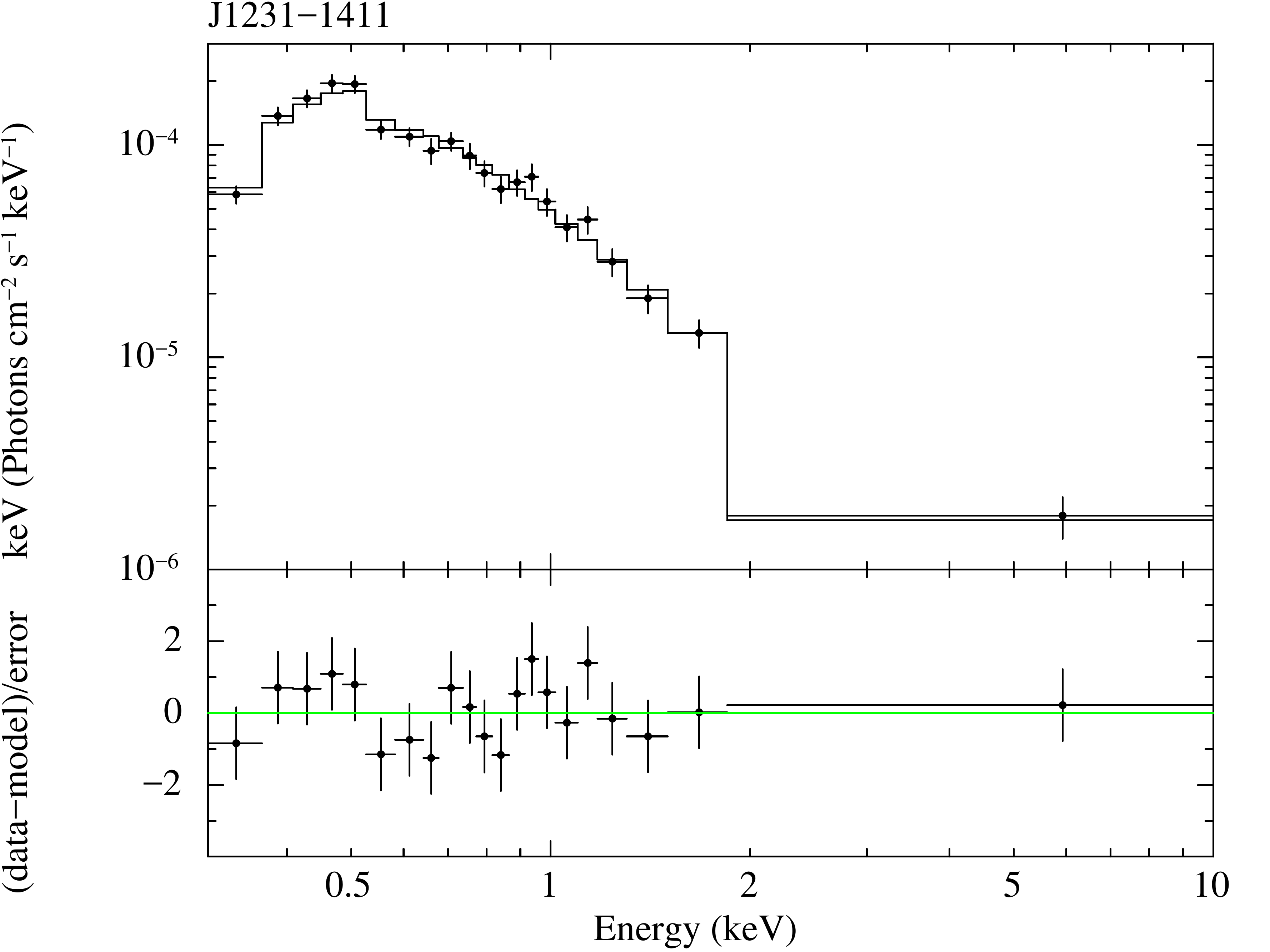}
\hspace{-0.44cm}
\includegraphics[width=0.508\textwidth]{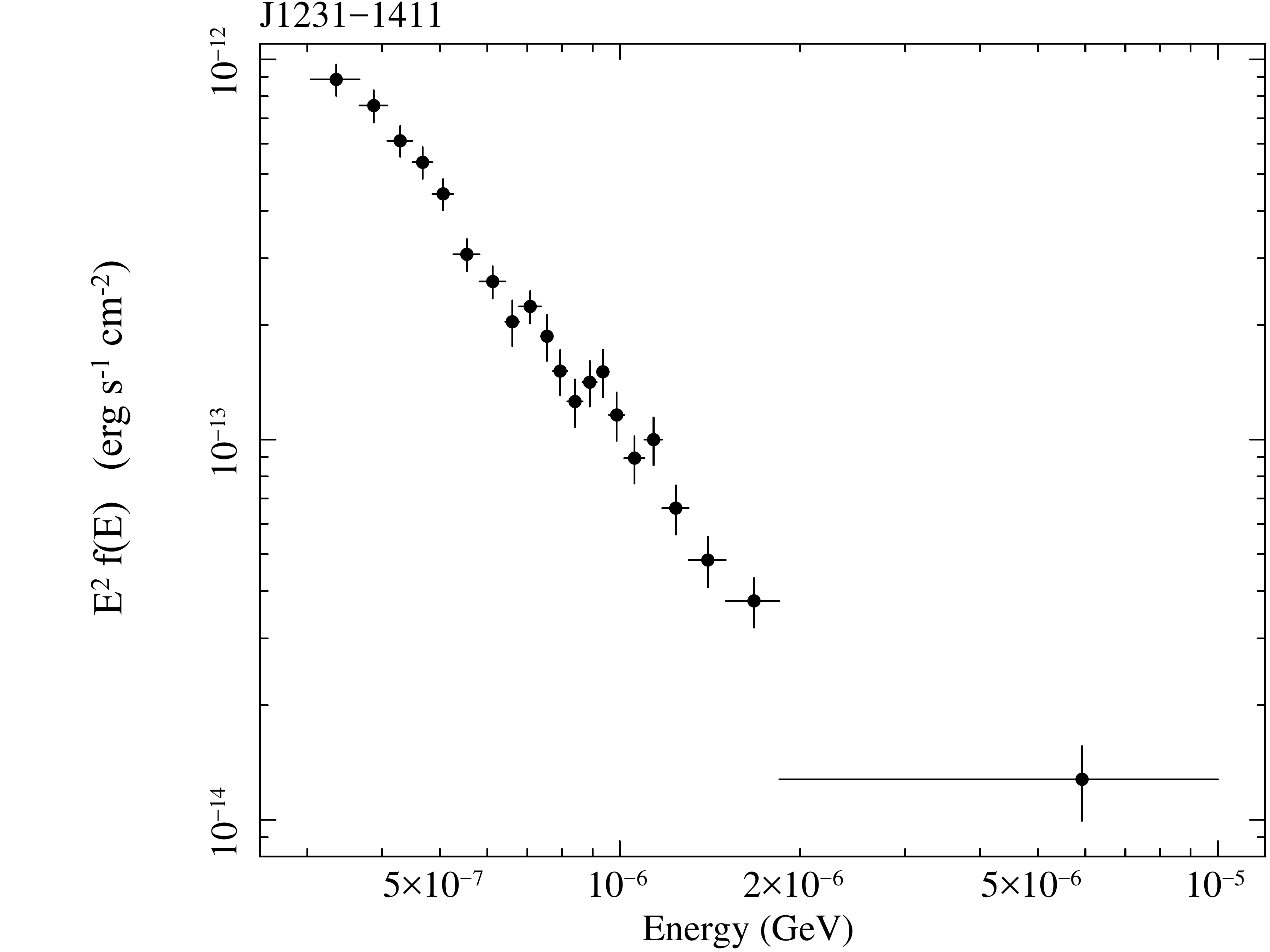}\\
\vspace{0.2cm}
  \contcaption{ }
\label{fig:sedx3}
\end{center}
\end{figure*}

\setcounter{figure}{1}
\begin{figure*}
\begin{center}
\includegraphics[width=0.508\textwidth]{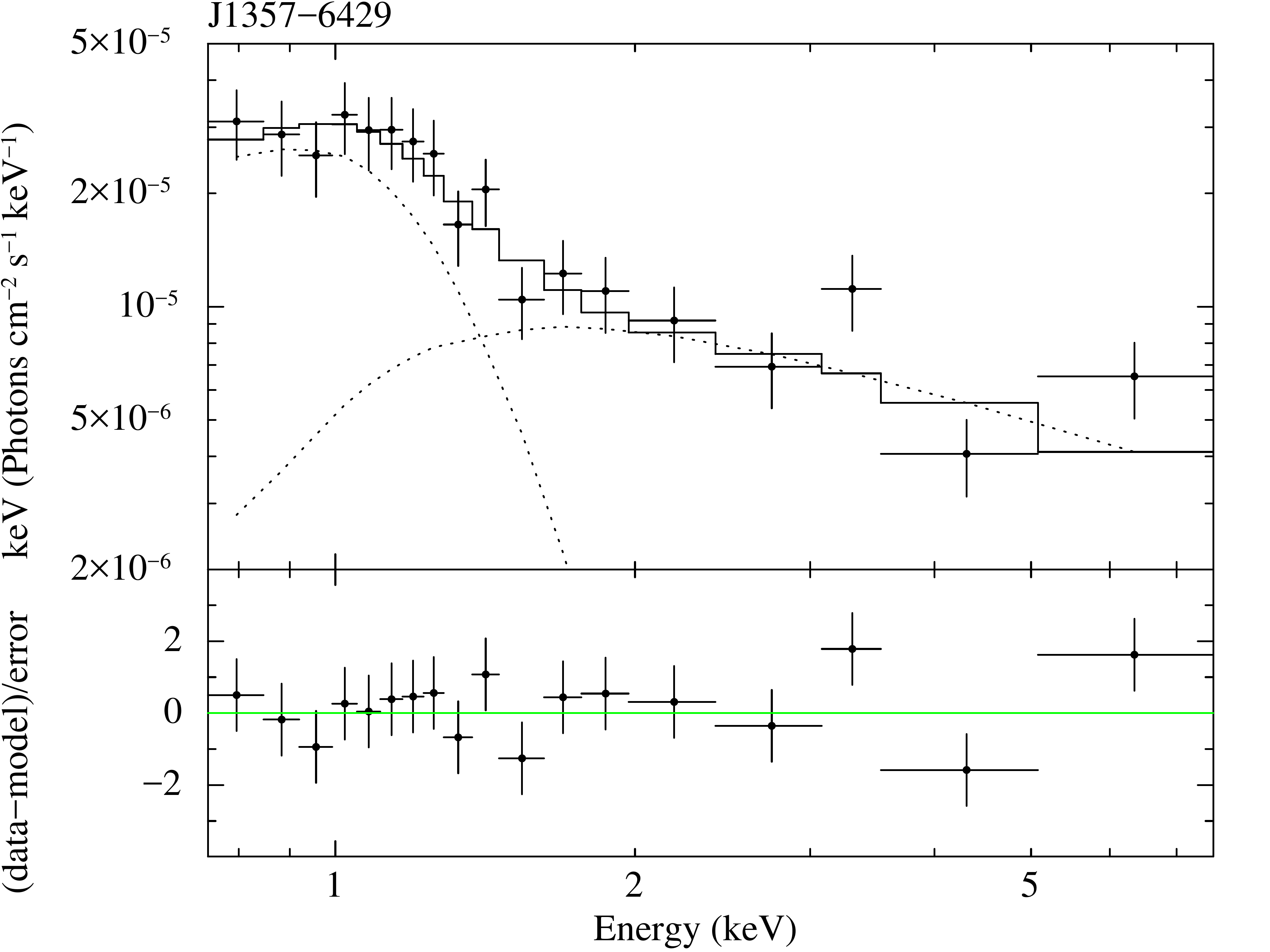}
\hspace{-0.44cm}
\includegraphics[width=0.508\textwidth]{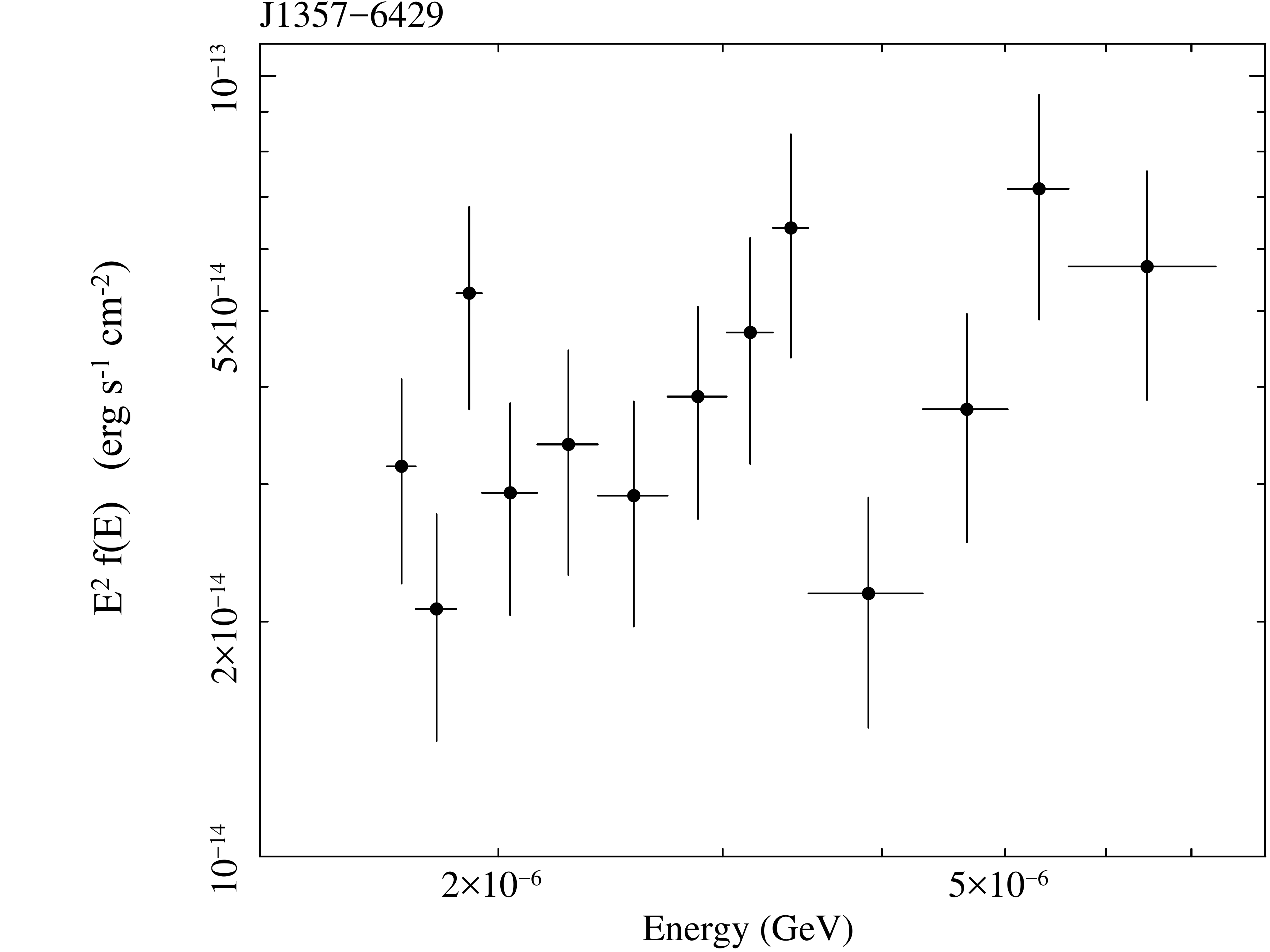}\\
\vspace{0.5cm}
\includegraphics[width=0.508\textwidth]{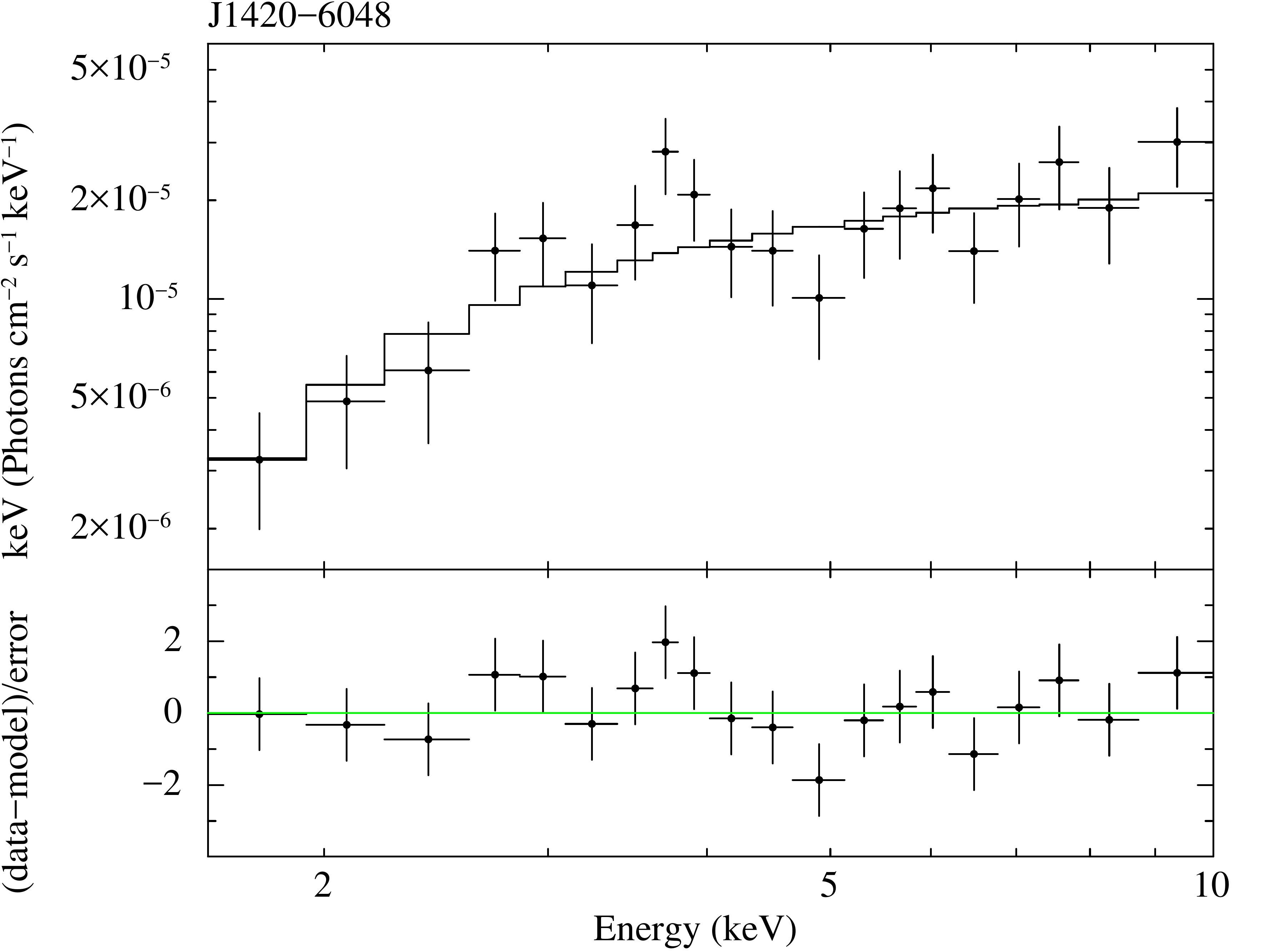}
\hspace{-0.44cm}
\includegraphics[width=0.508\textwidth]{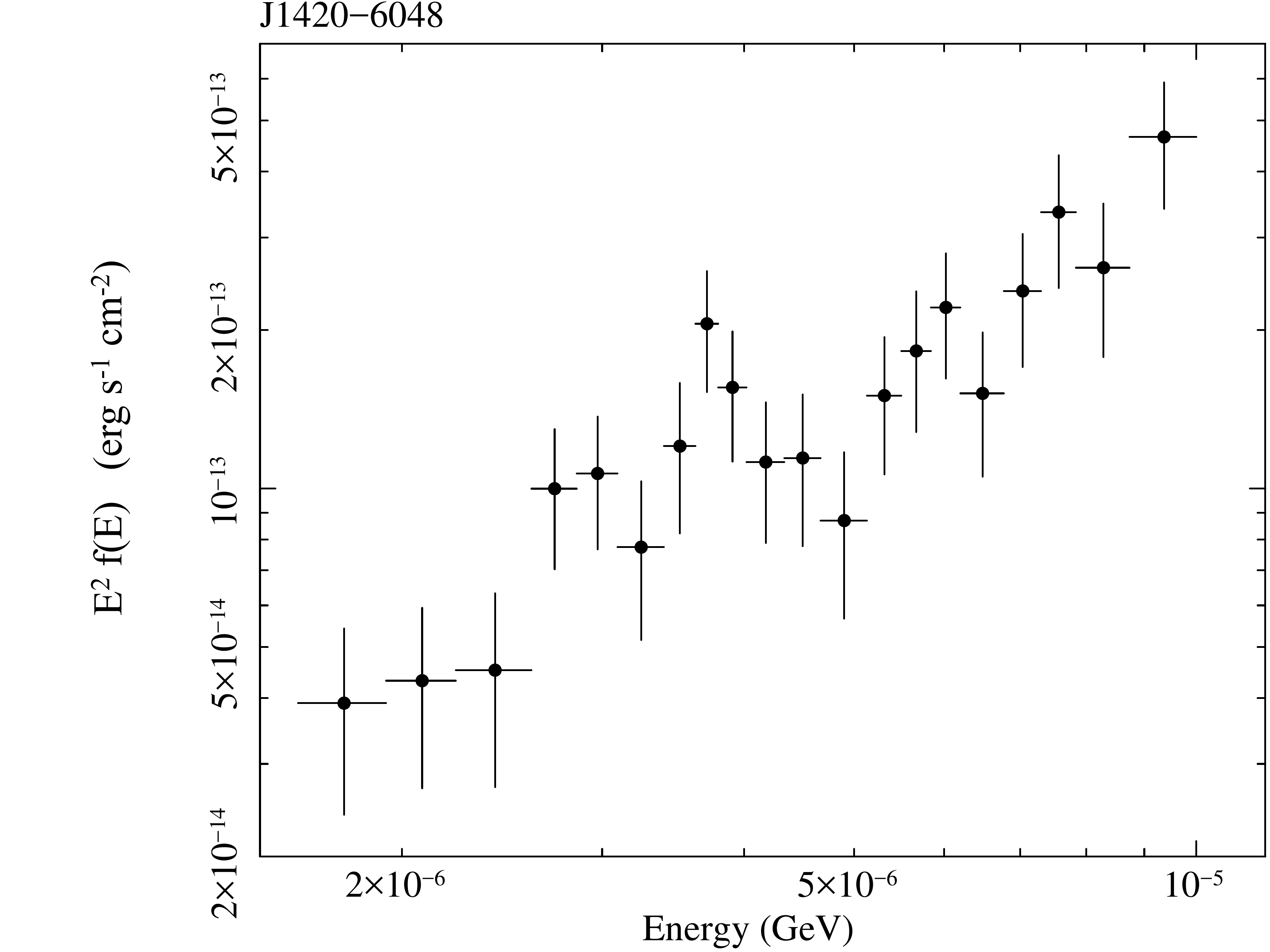}\\
\vspace{0.5cm}
\includegraphics[width=0.508\textwidth]{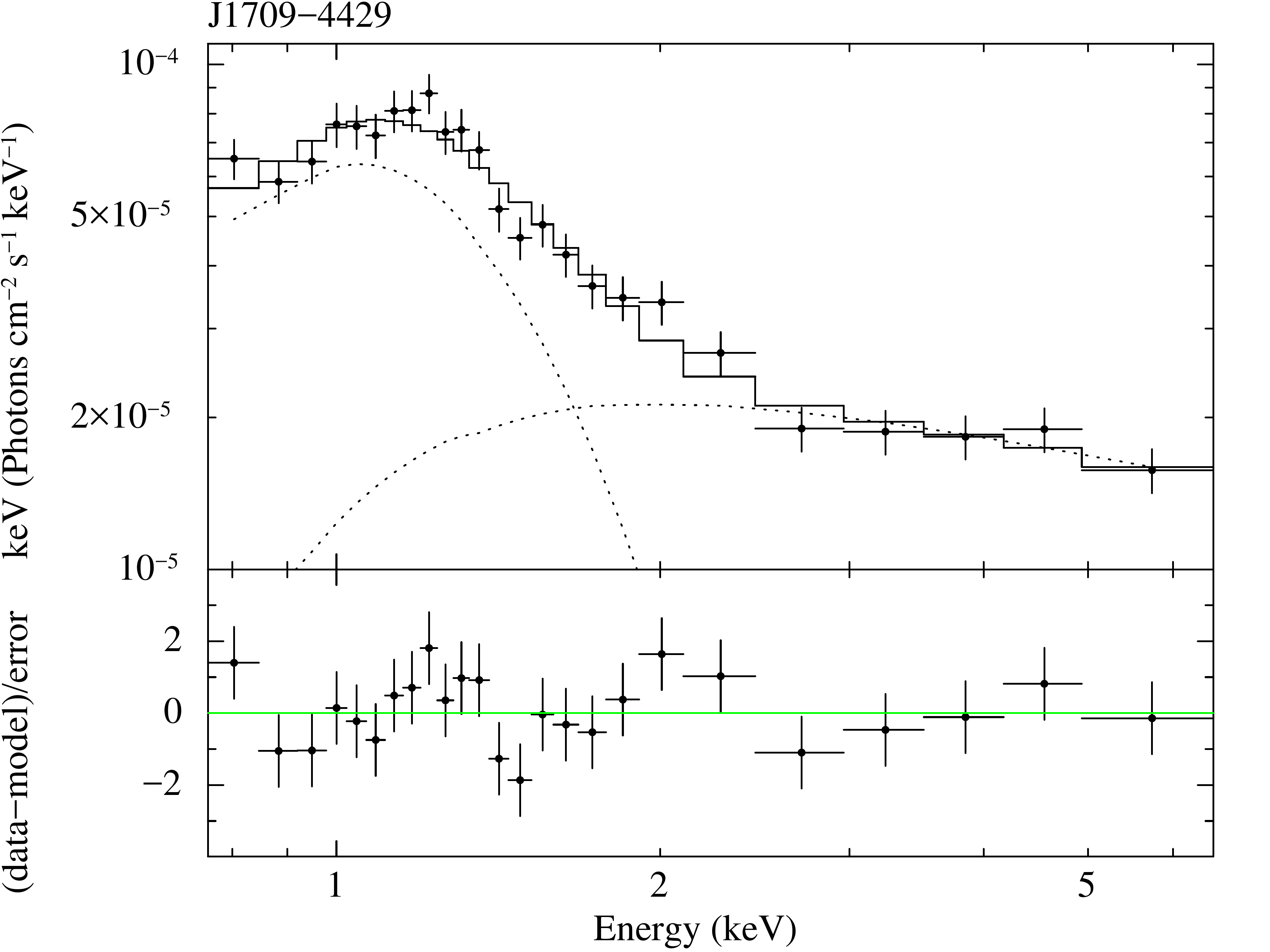}
\hspace{-0.44cm}
\includegraphics[width=0.508\textwidth]{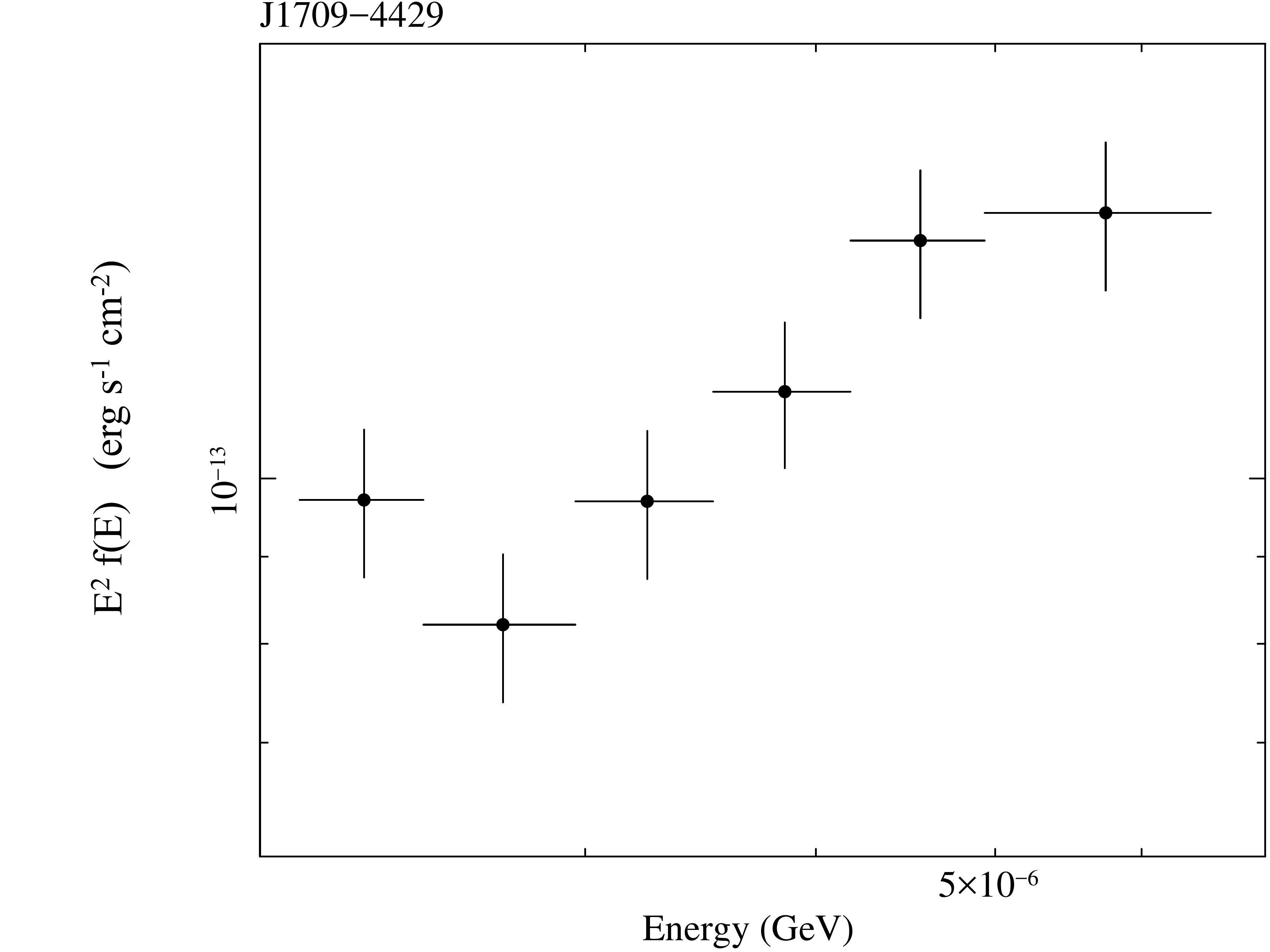}\\
\vspace{0.2cm}
  \contcaption{ }
\label{fig:sedx4}
\end{center}
\end{figure*}

\setcounter{figure}{1}
\begin{figure*}
\begin{center}
\includegraphics[width=0.508\textwidth]{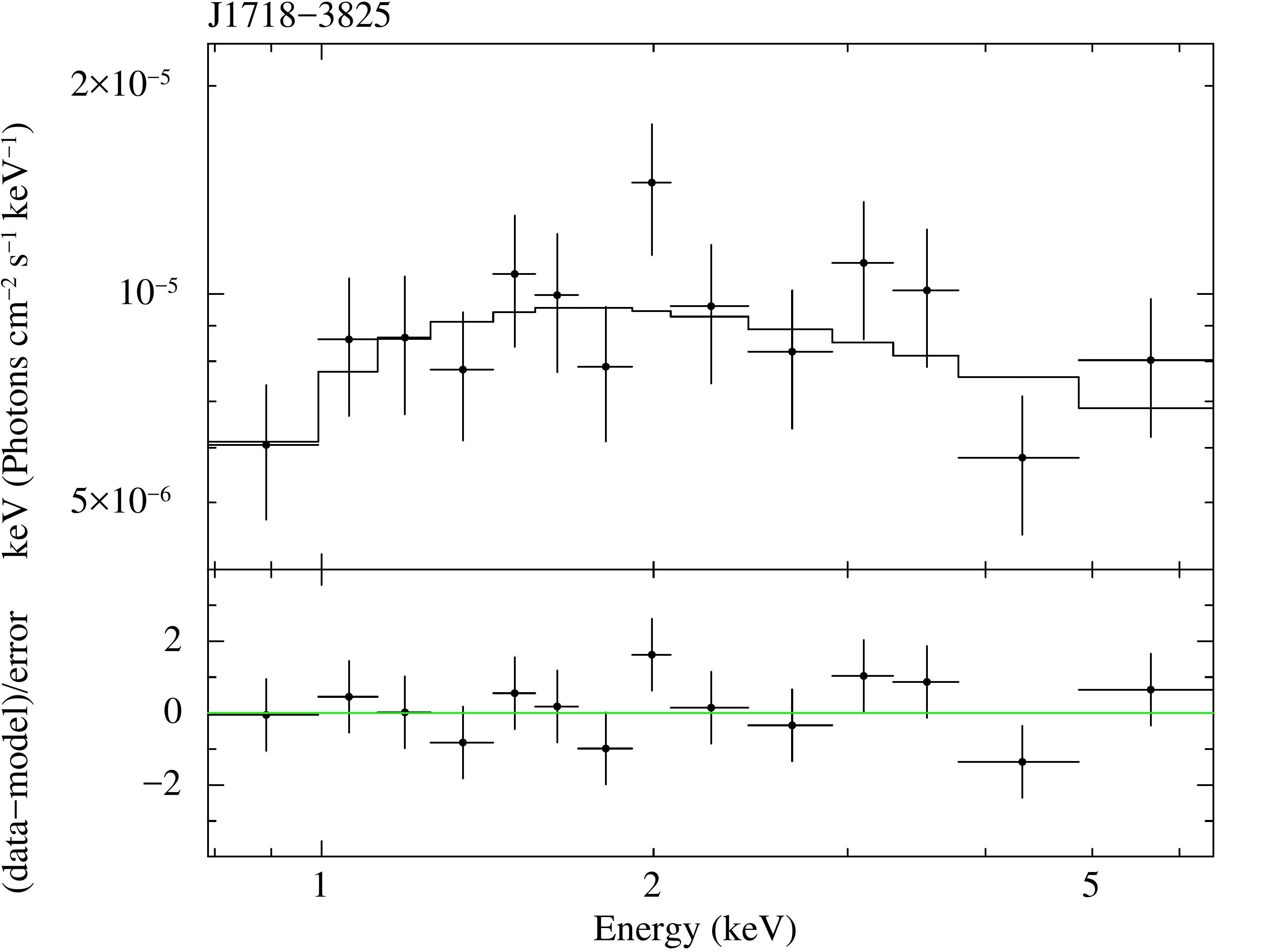}
\hspace{-0.44cm}
\includegraphics[width=0.508\textwidth]{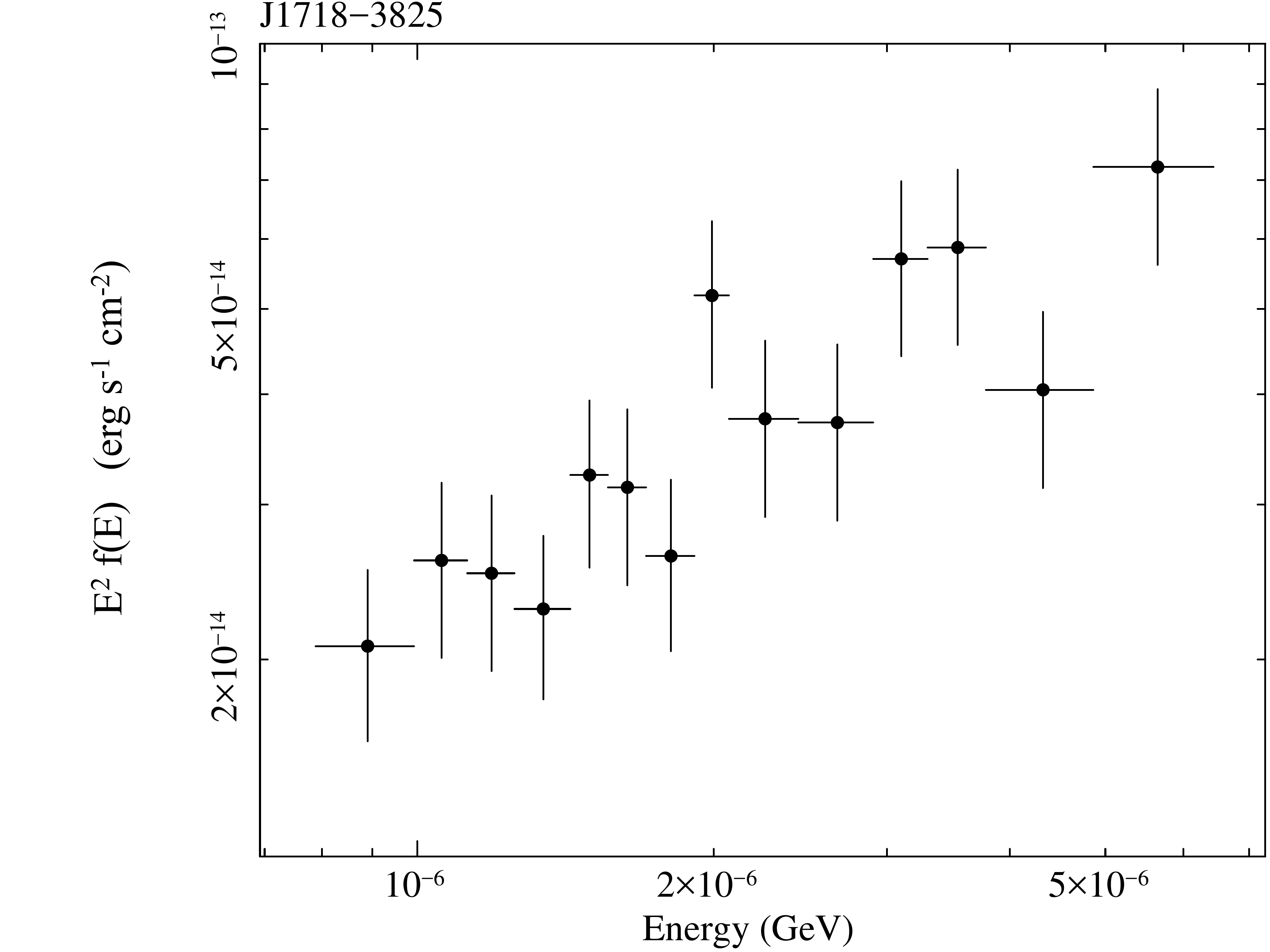}\\
\vspace{0.5cm}
\includegraphics[width=0.508\textwidth]{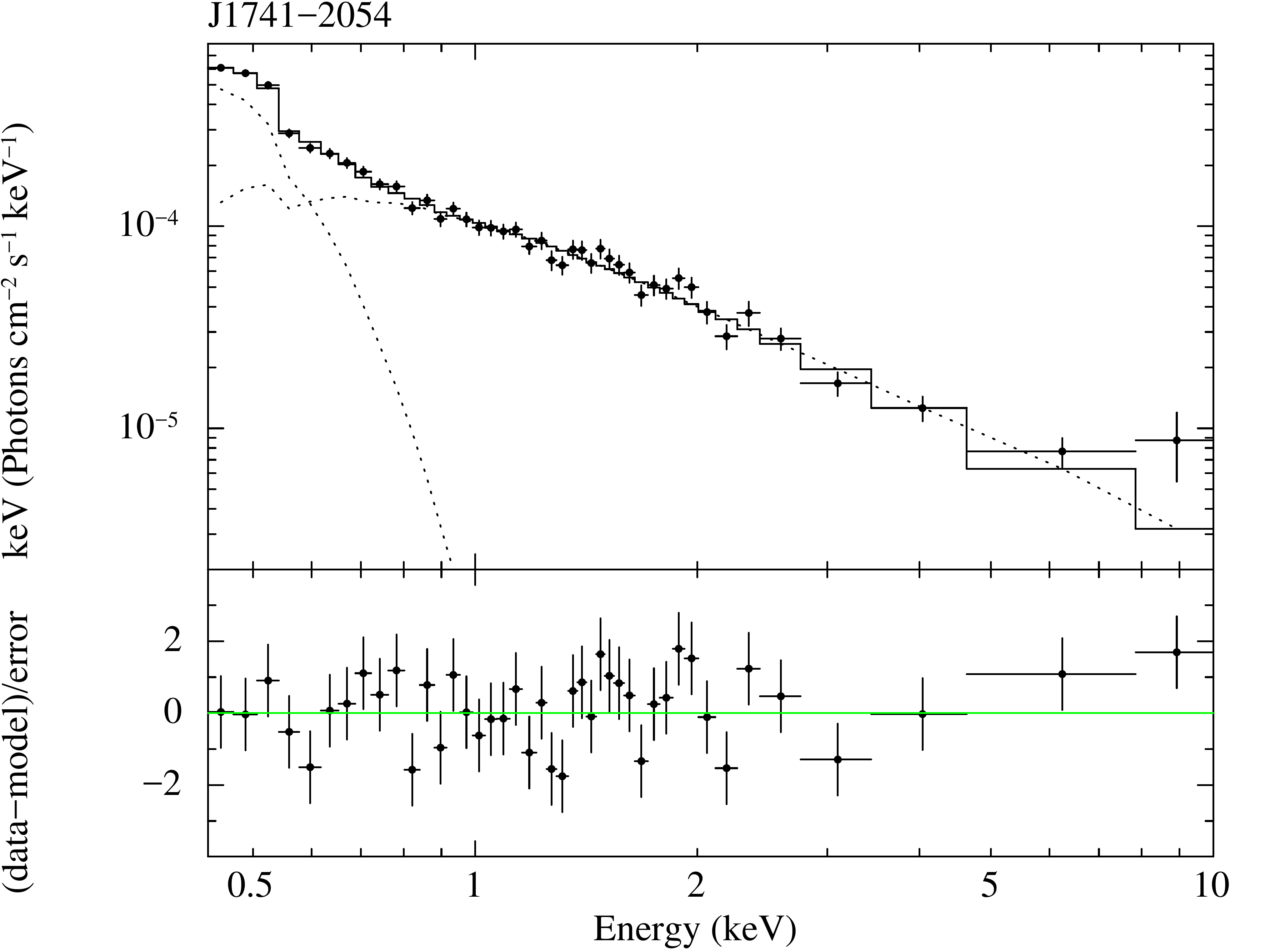}
\hspace{-0.44cm}
\includegraphics[width=0.508\textwidth]{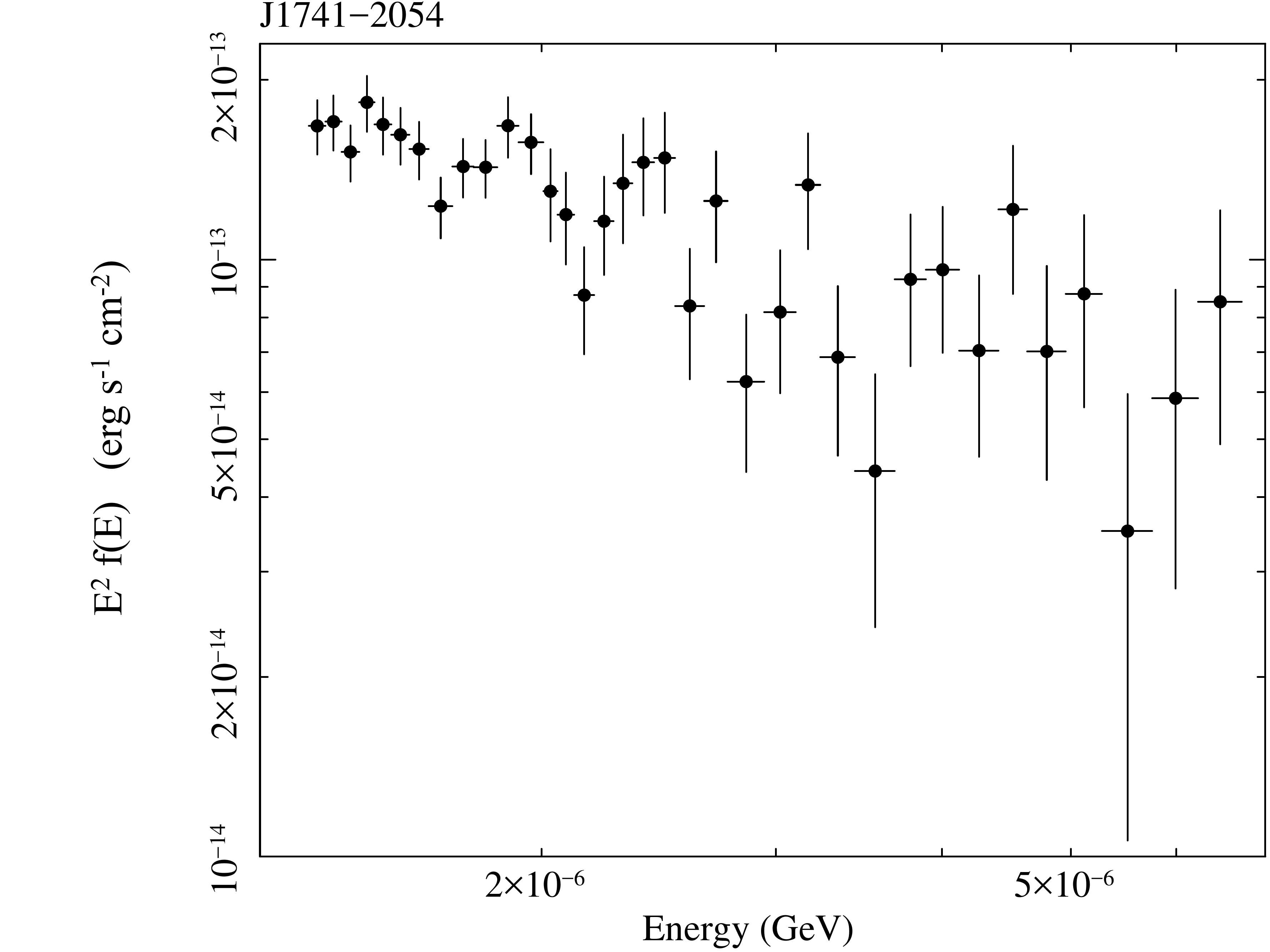}\\
\vspace{0.5cm}
\includegraphics[width=0.508\textwidth]{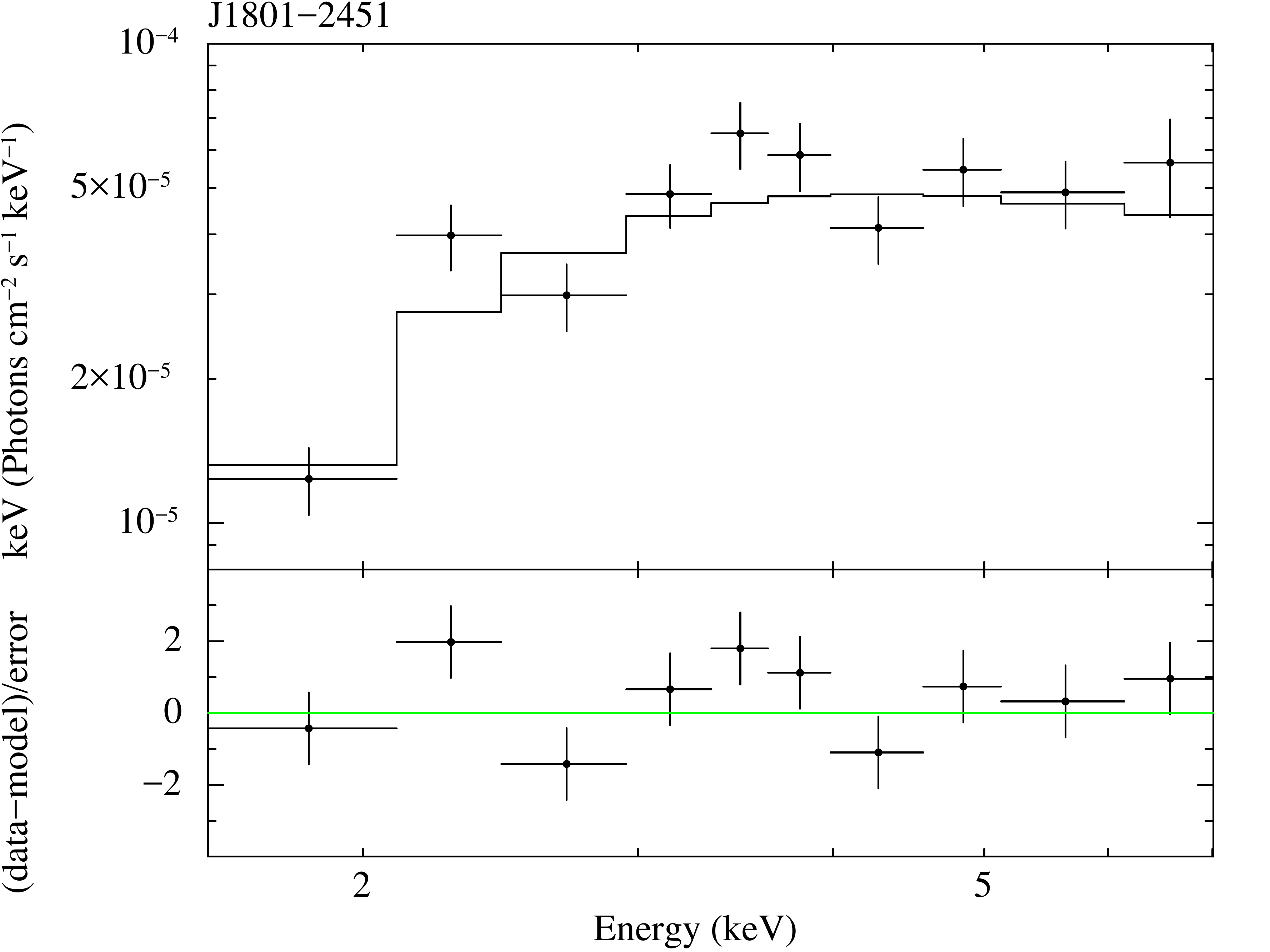}
\hspace{-0.44cm}
\includegraphics[width=0.508\textwidth]{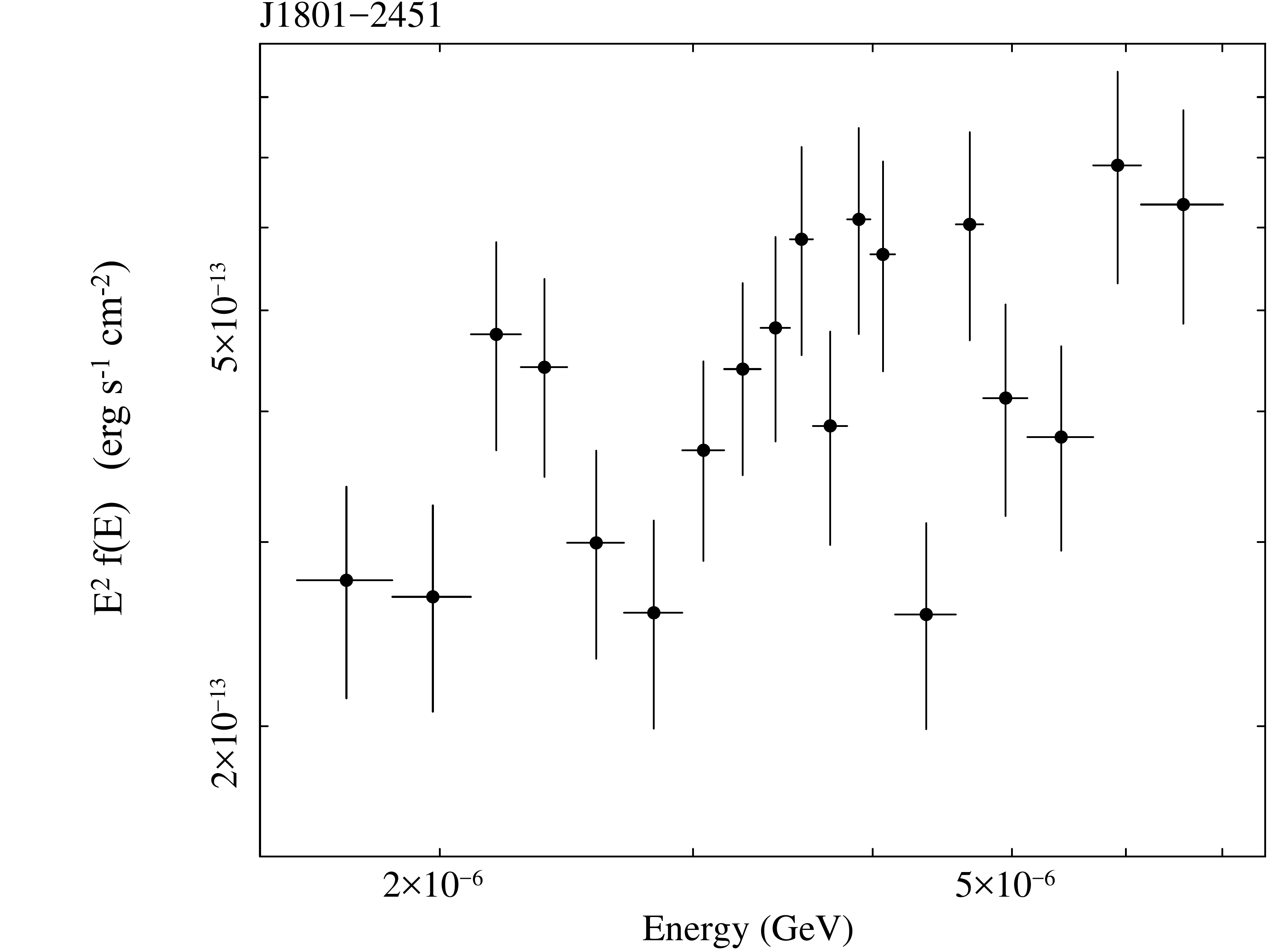}\\
\vspace{0.2cm}
  \contcaption{ }
\label{fig:sedx5}
\end{center}
\end{figure*}

\setcounter{figure}{1}
\begin{figure*}
\begin{center}
\includegraphics[width=0.508\textwidth]{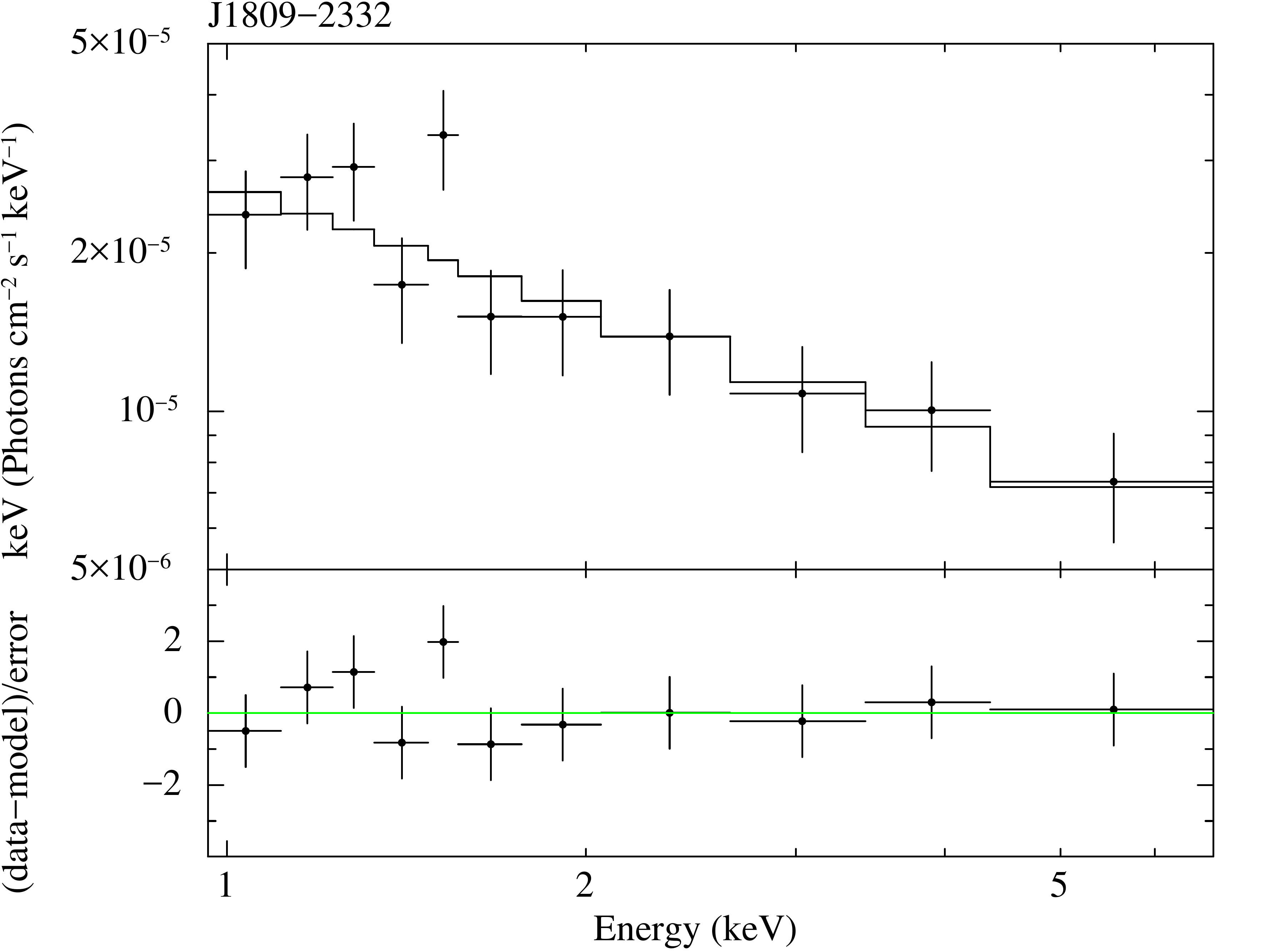}
\hspace{-0.44cm}
\includegraphics[width=0.508\textwidth]{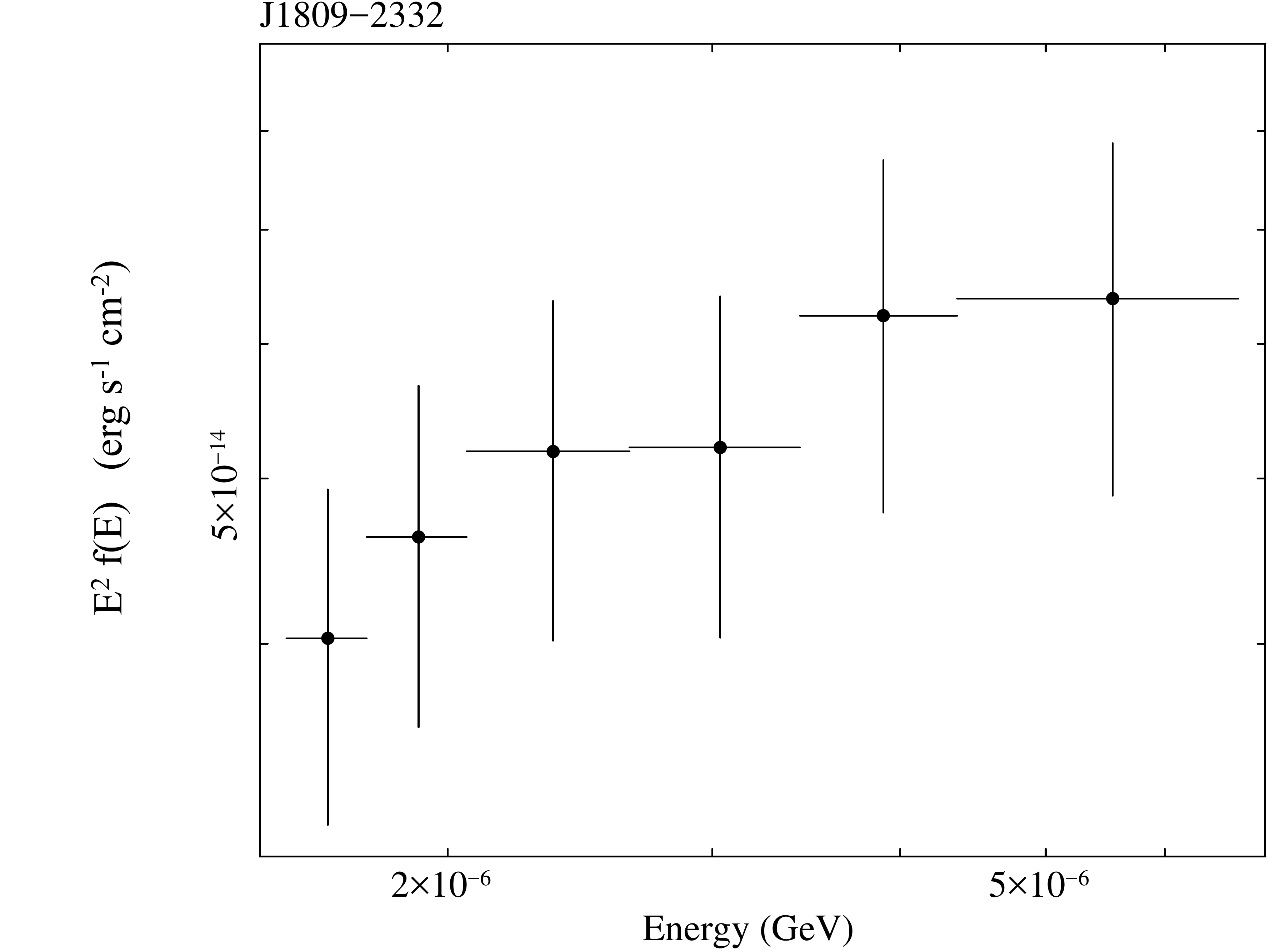}\\
\vspace{0.5cm}
\includegraphics[width=0.508\textwidth]{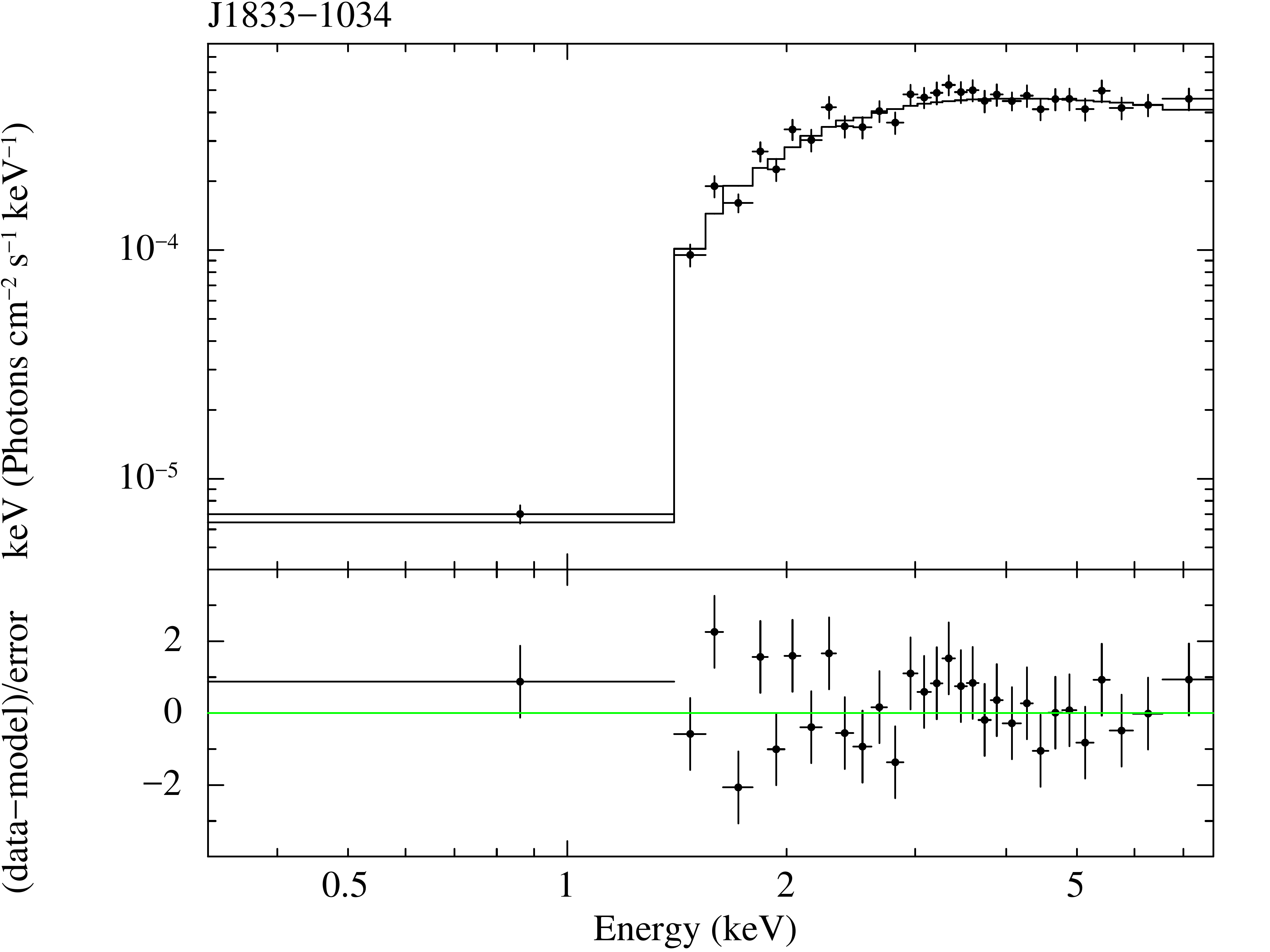}
\hspace{-0.44cm}
\includegraphics[width=0.508\textwidth]{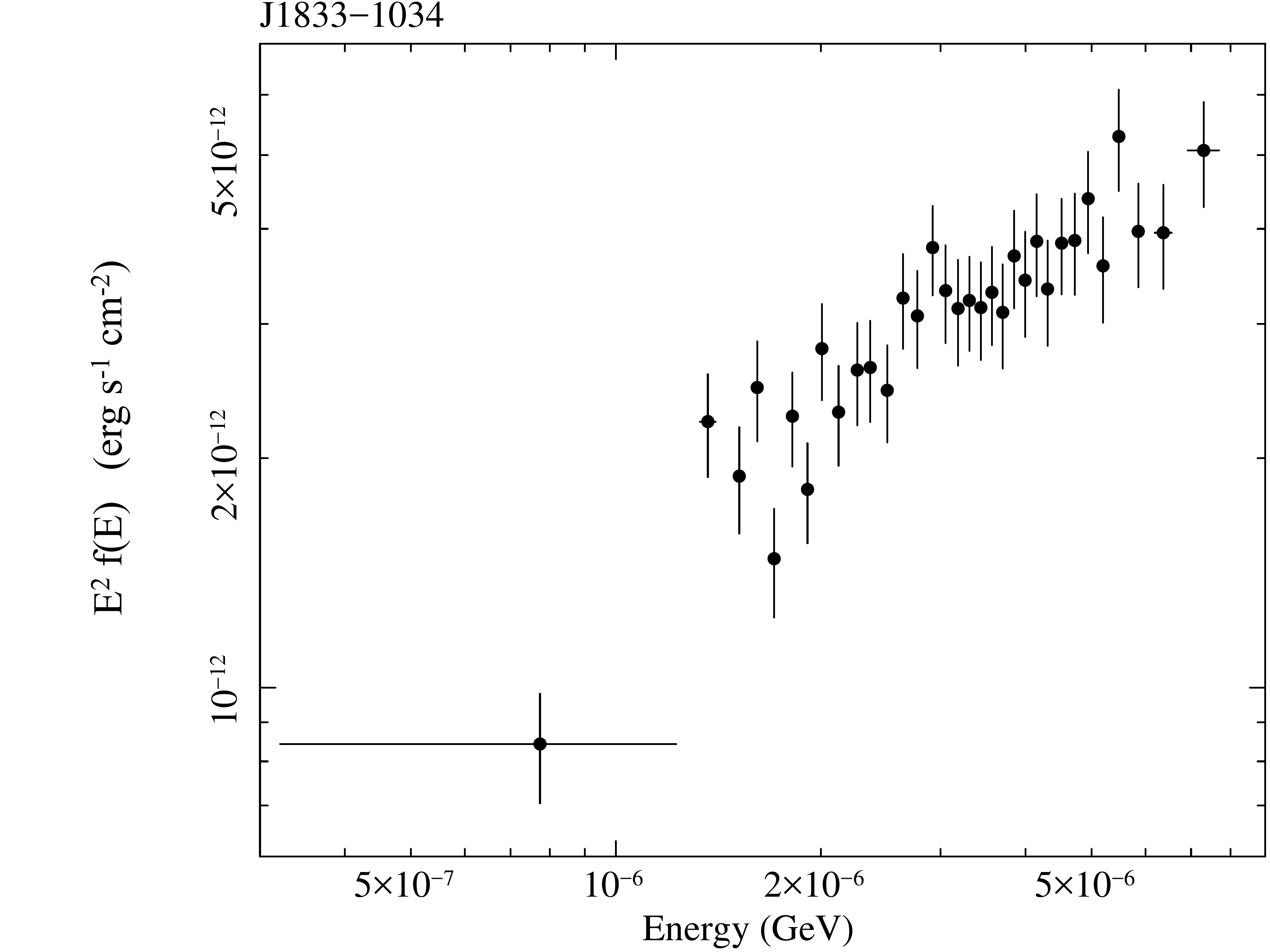}\\
\vspace{0.5cm}
\includegraphics[width=0.508\textwidth]{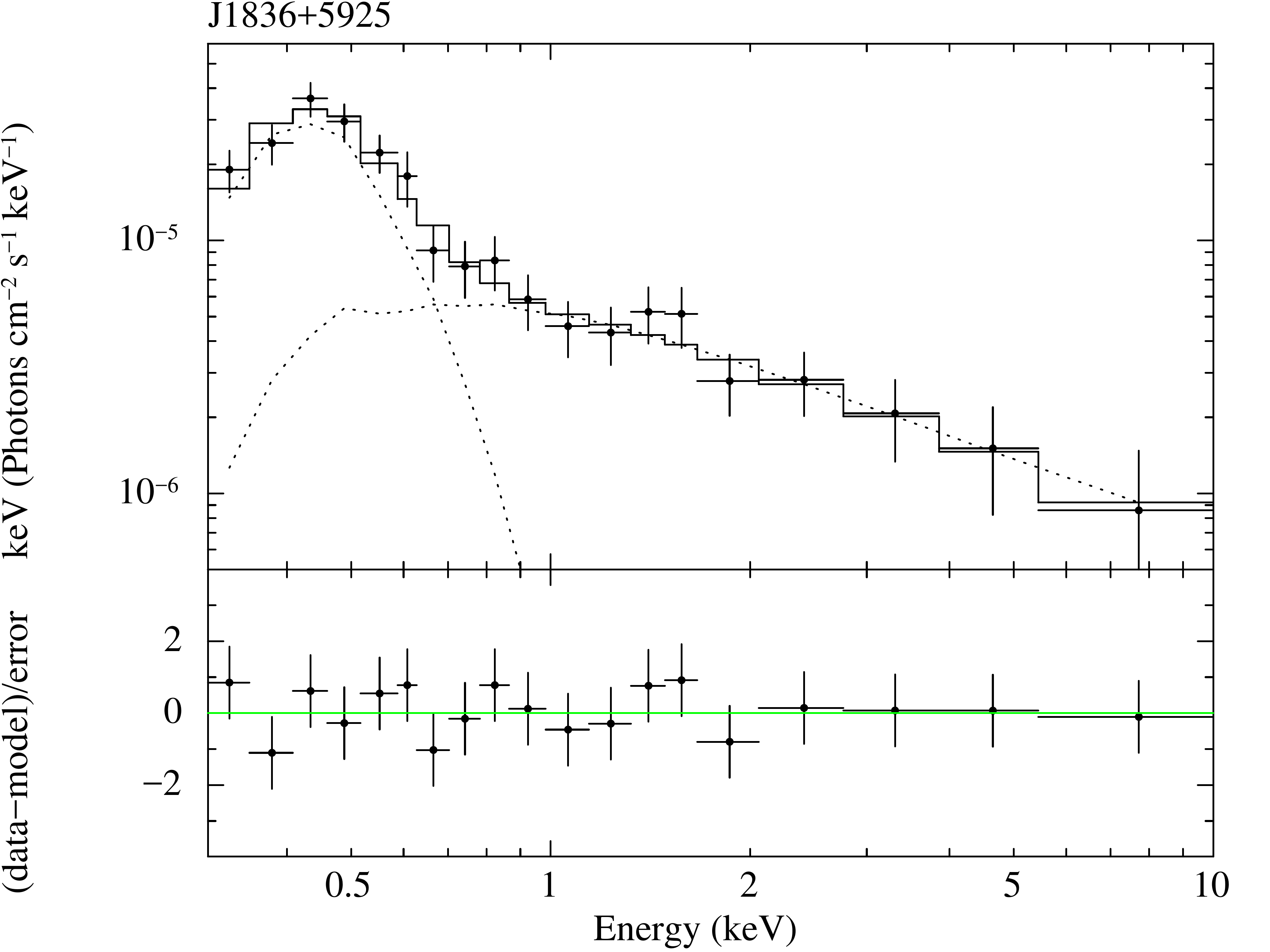}
\hspace{-0.44cm}
\includegraphics[width=0.508\textwidth]{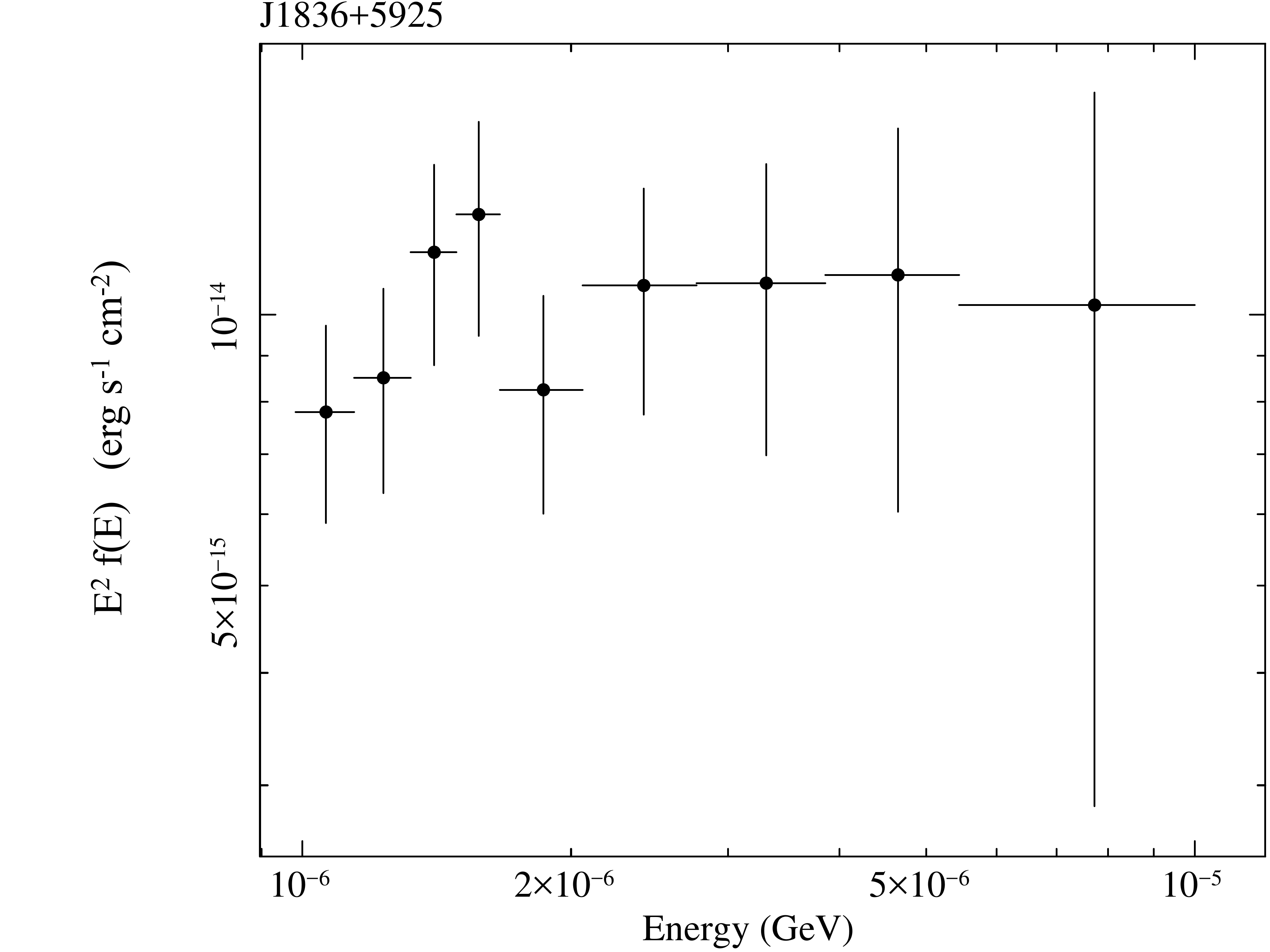}\\
\vspace{0.2cm}
  \contcaption{ }
\label{fig:sedx6}
\end{center}
\end{figure*}

\setcounter{figure}{1}
\begin{figure*}
\begin{center}
\includegraphics[width=0.508\textwidth]{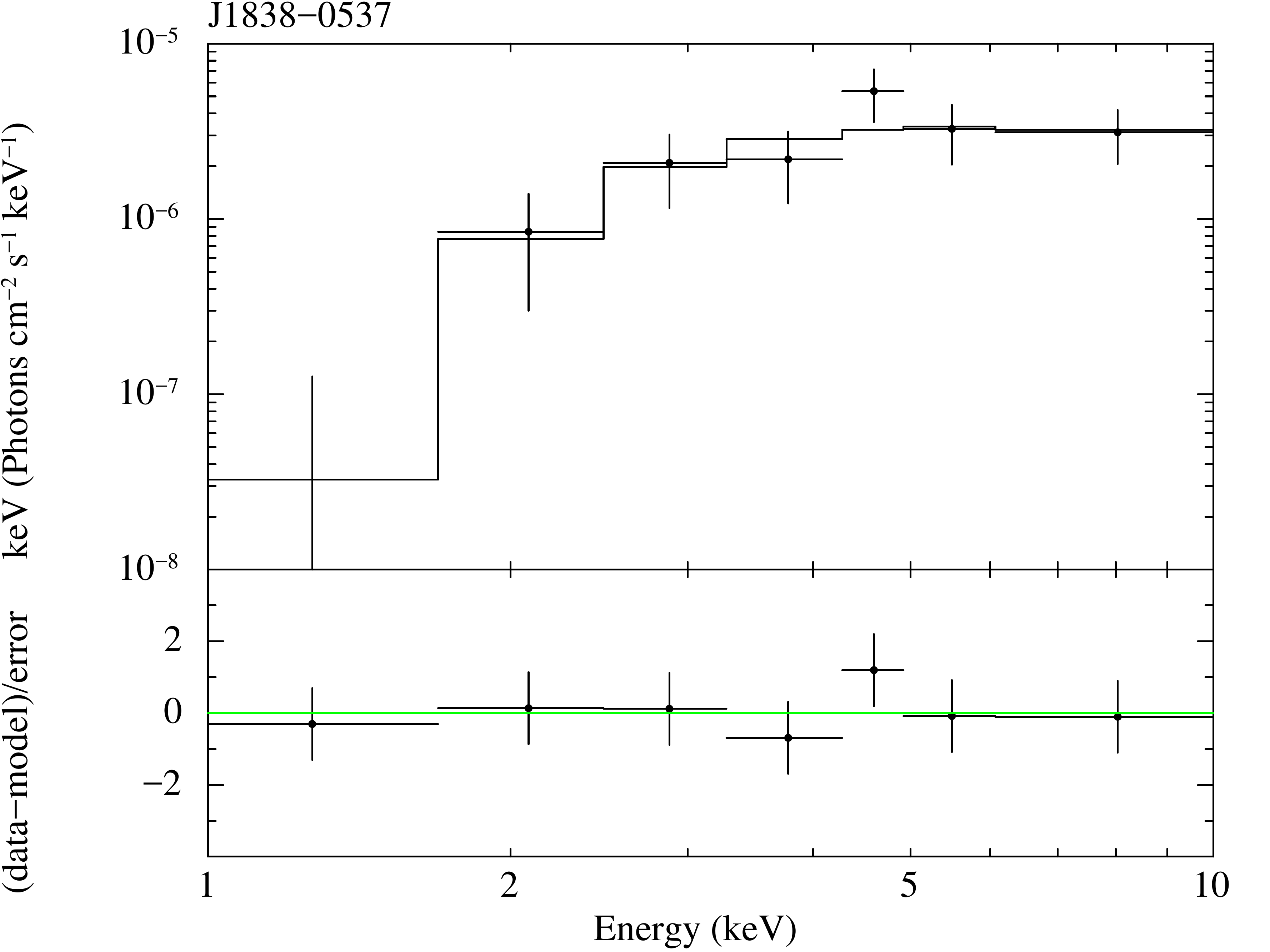}
\hspace{-0.44cm}
\includegraphics[width=0.508\textwidth]{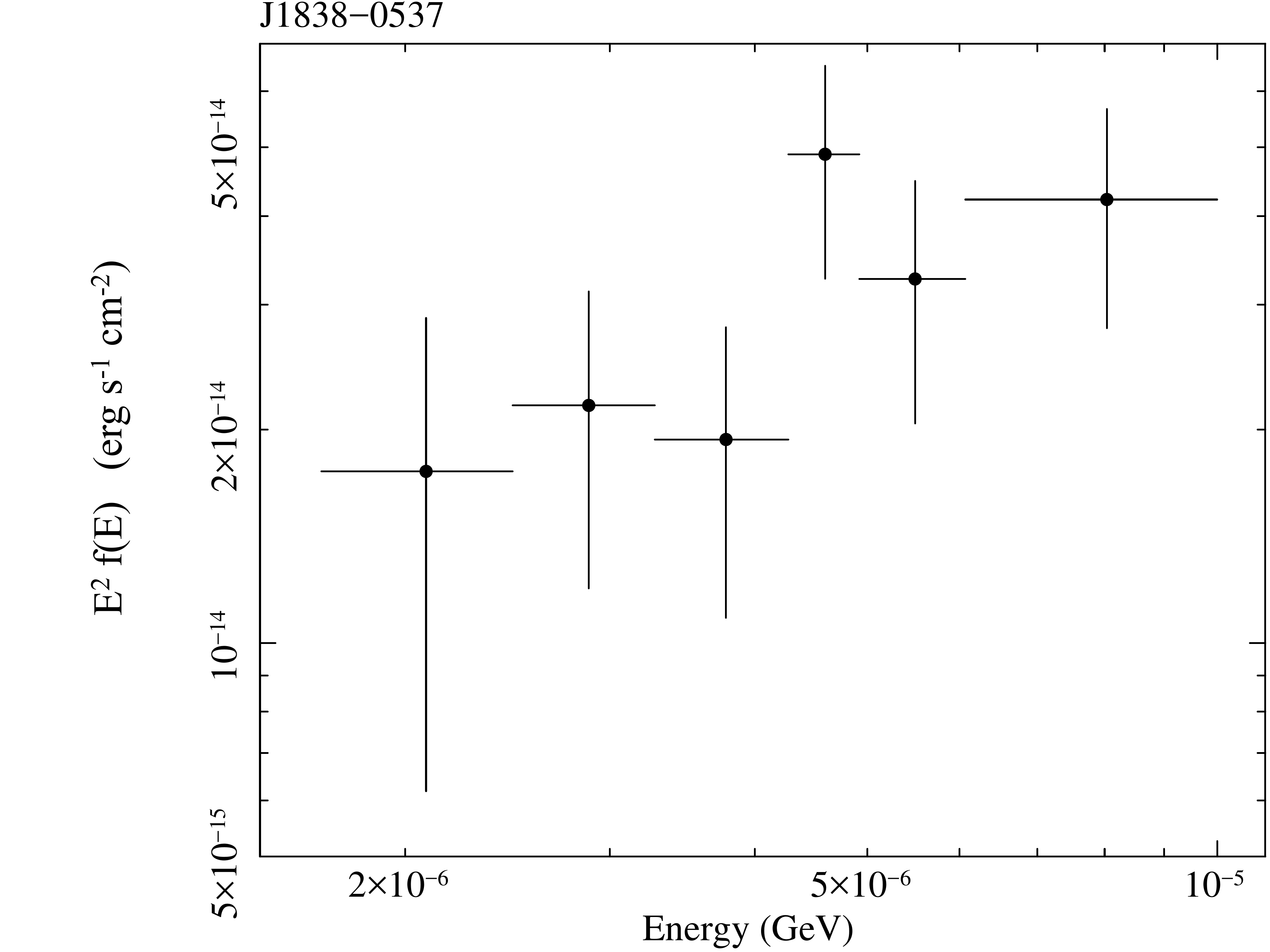}\\
\vspace{0.5cm}
\includegraphics[width=0.508\textwidth]{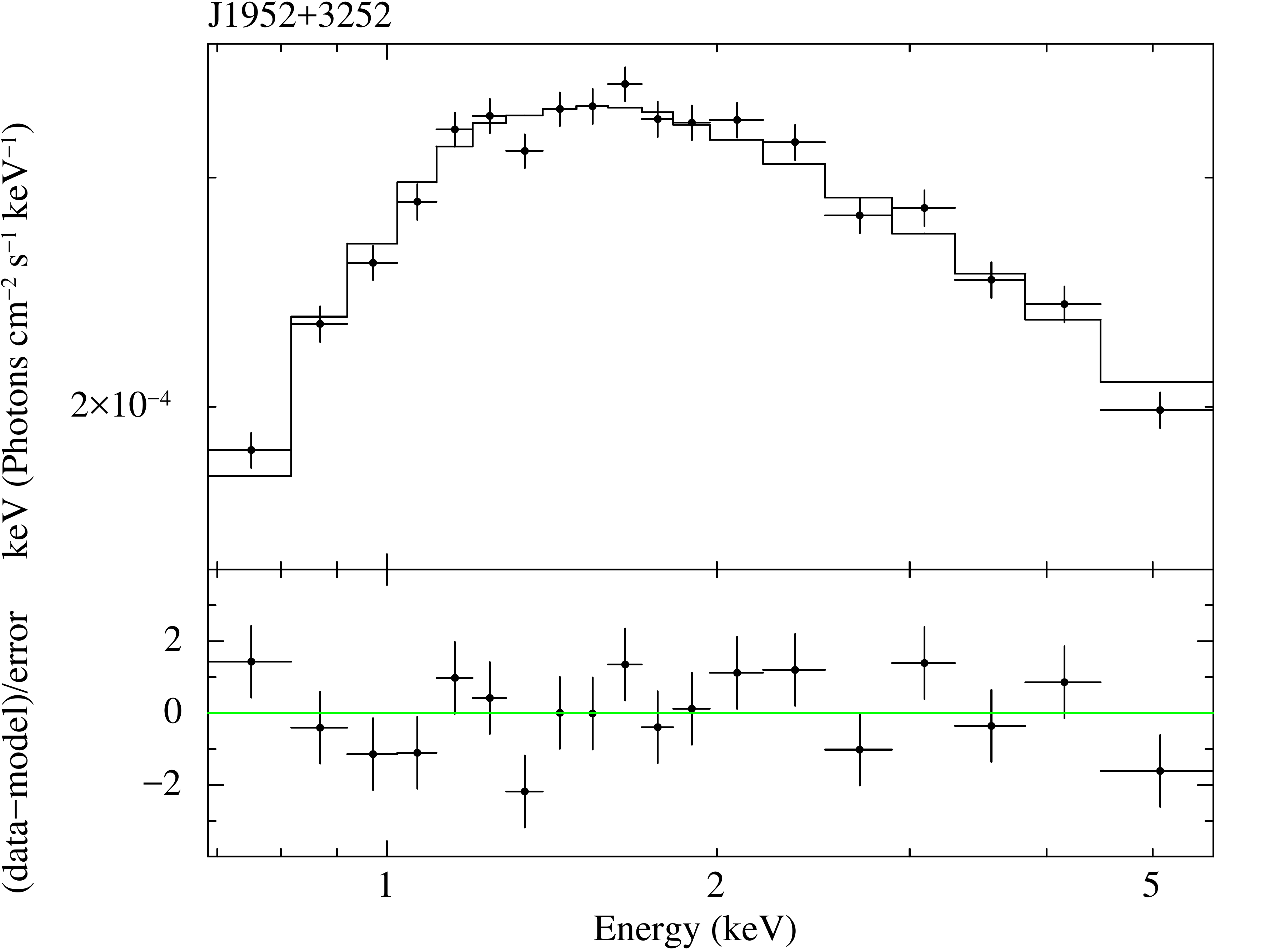}
\hspace{-0.44cm}
\includegraphics[width=0.508\textwidth]{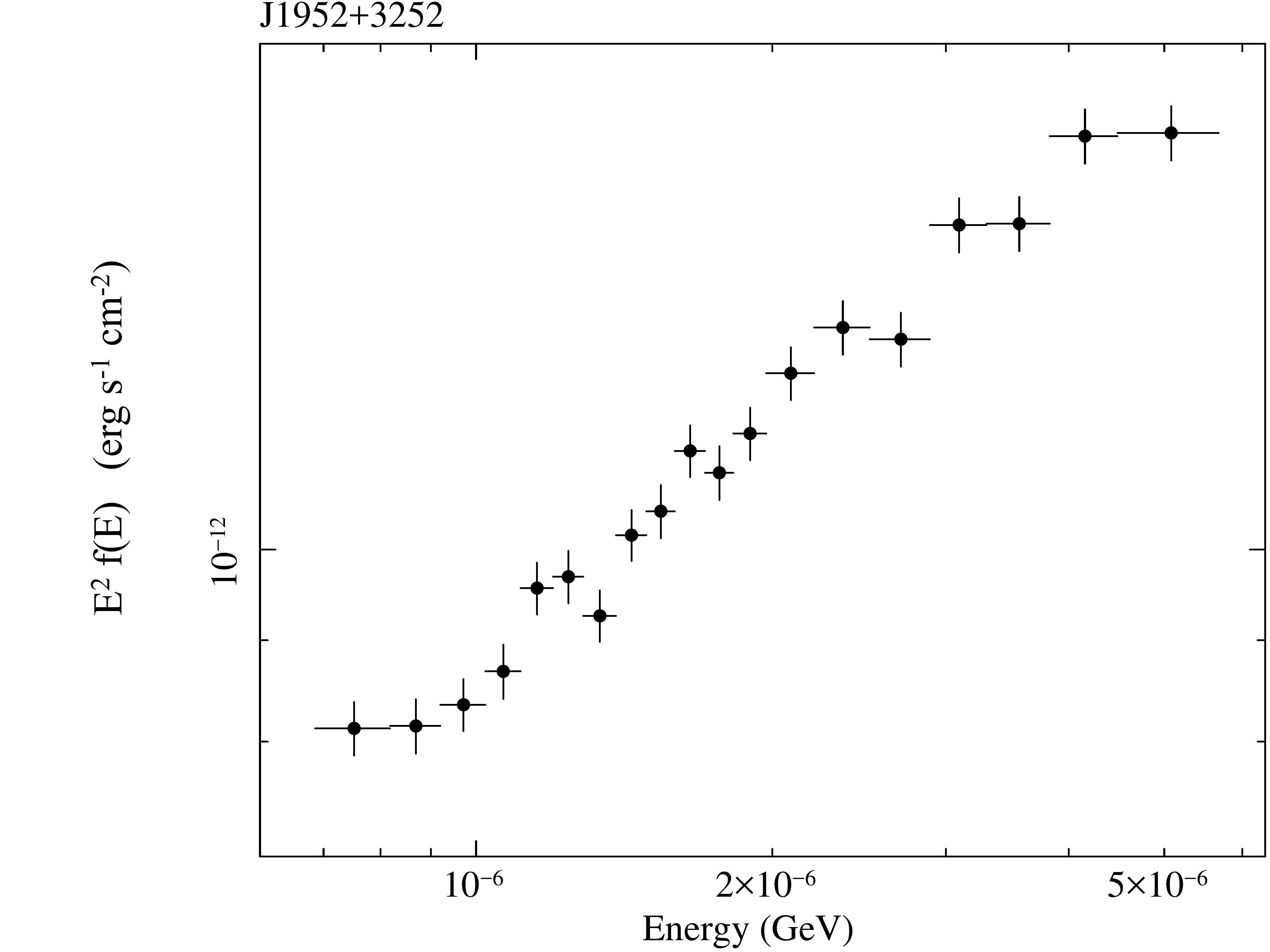}\\
\vspace{0.5cm}
\includegraphics[width=0.508\textwidth]{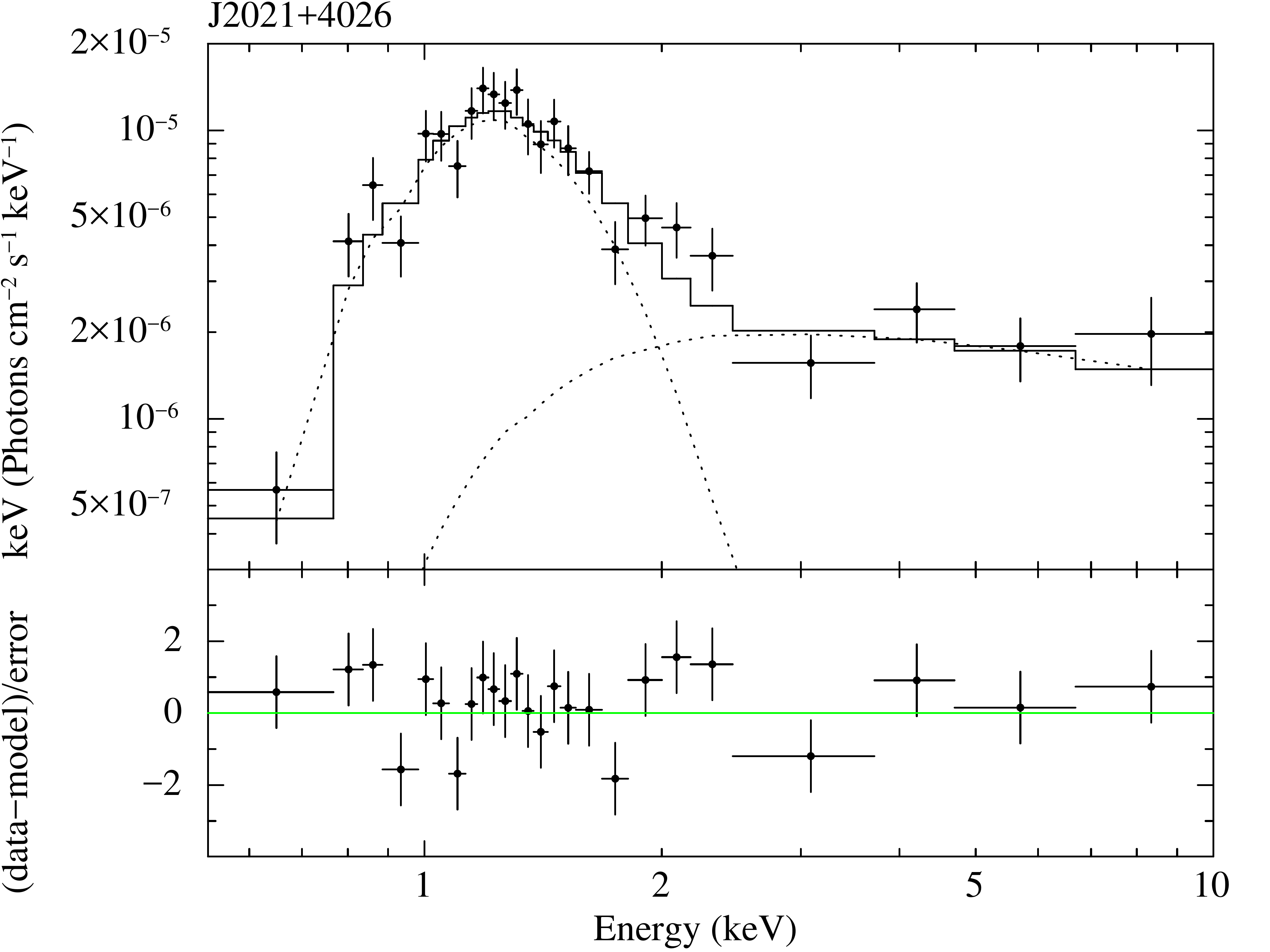}
\hspace{-0.44cm}
\includegraphics[width=0.508\textwidth]{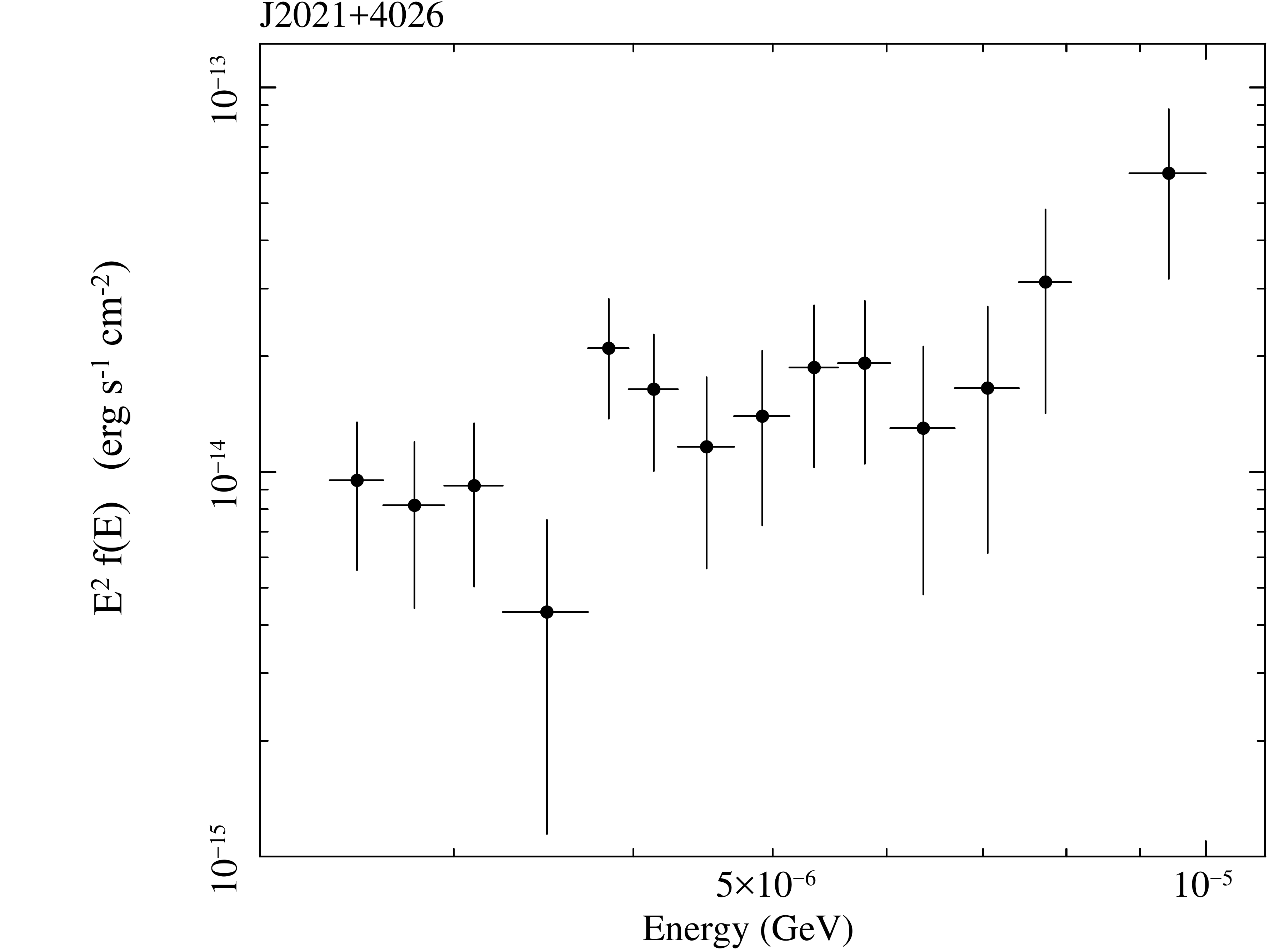}\\
\vspace{0.2cm}
  \contcaption{ }
\label{fig:sedx7}
\end{center}
\end{figure*}

\setcounter{figure}{1}
\begin{figure*}
\begin{center}
\includegraphics[width=0.508\textwidth]{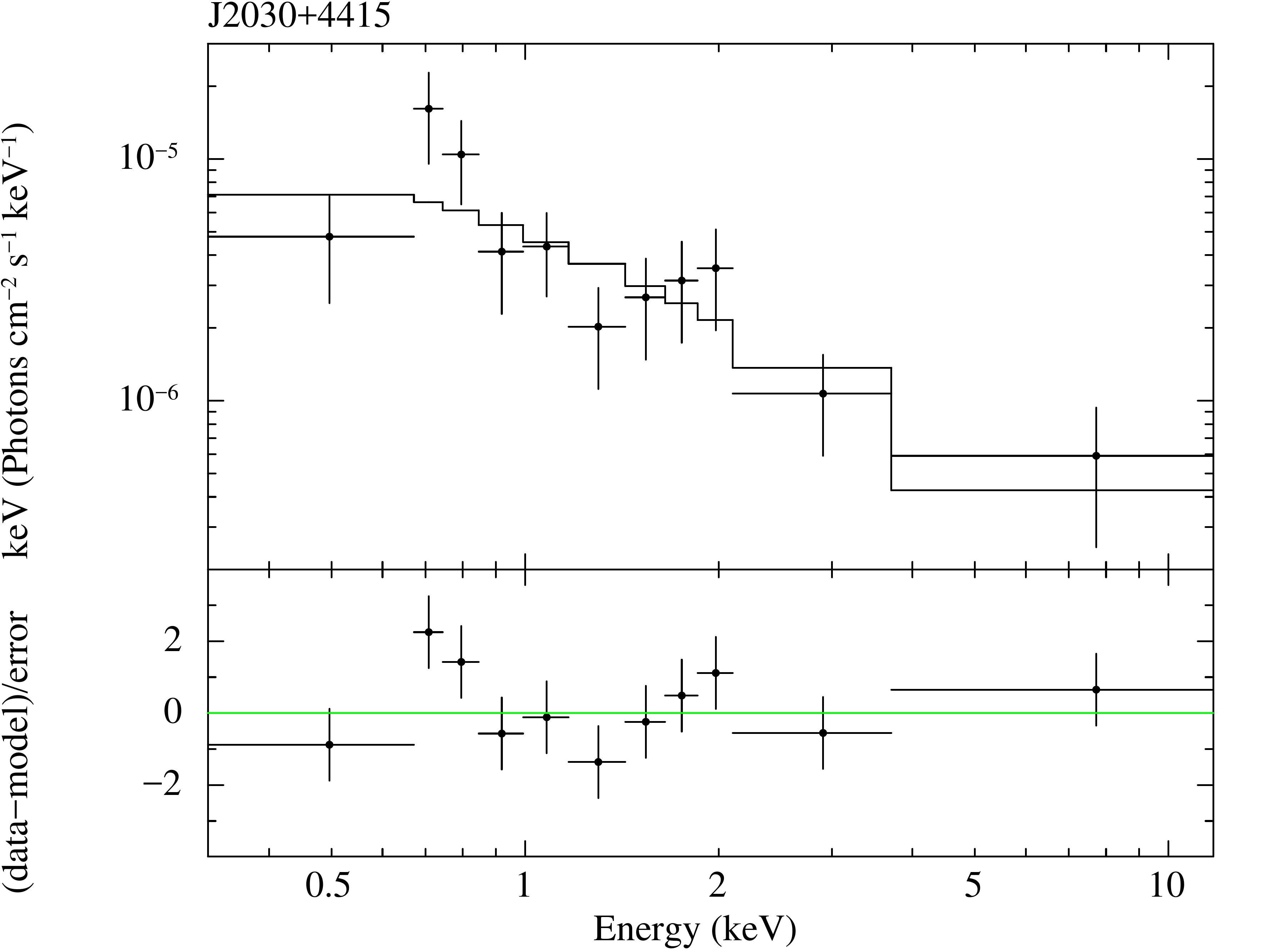}
\hspace{-0.44cm}
\includegraphics[width=0.508\textwidth]{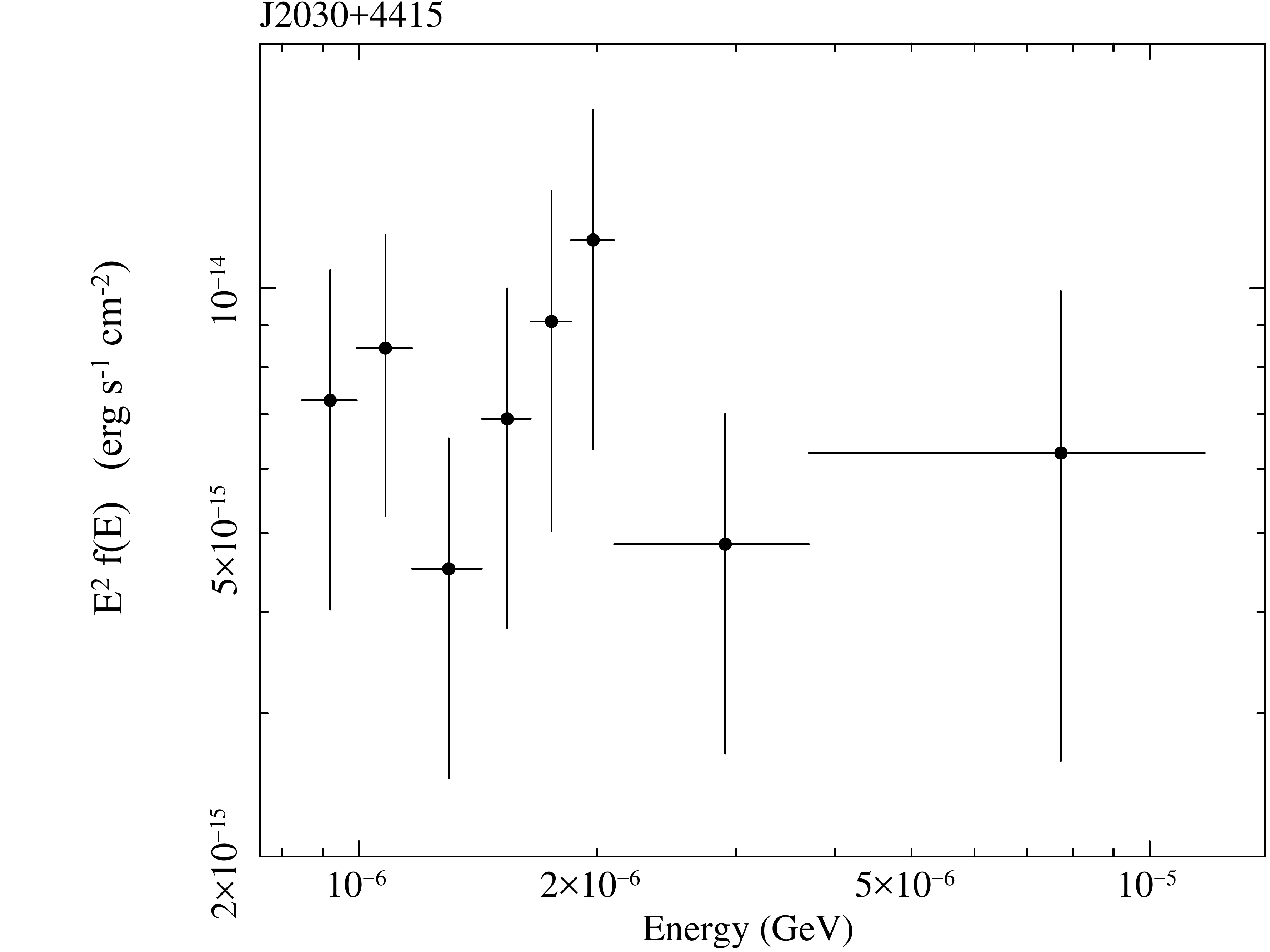}\\
\vspace{0.5cm}
\includegraphics[width=0.508\textwidth]{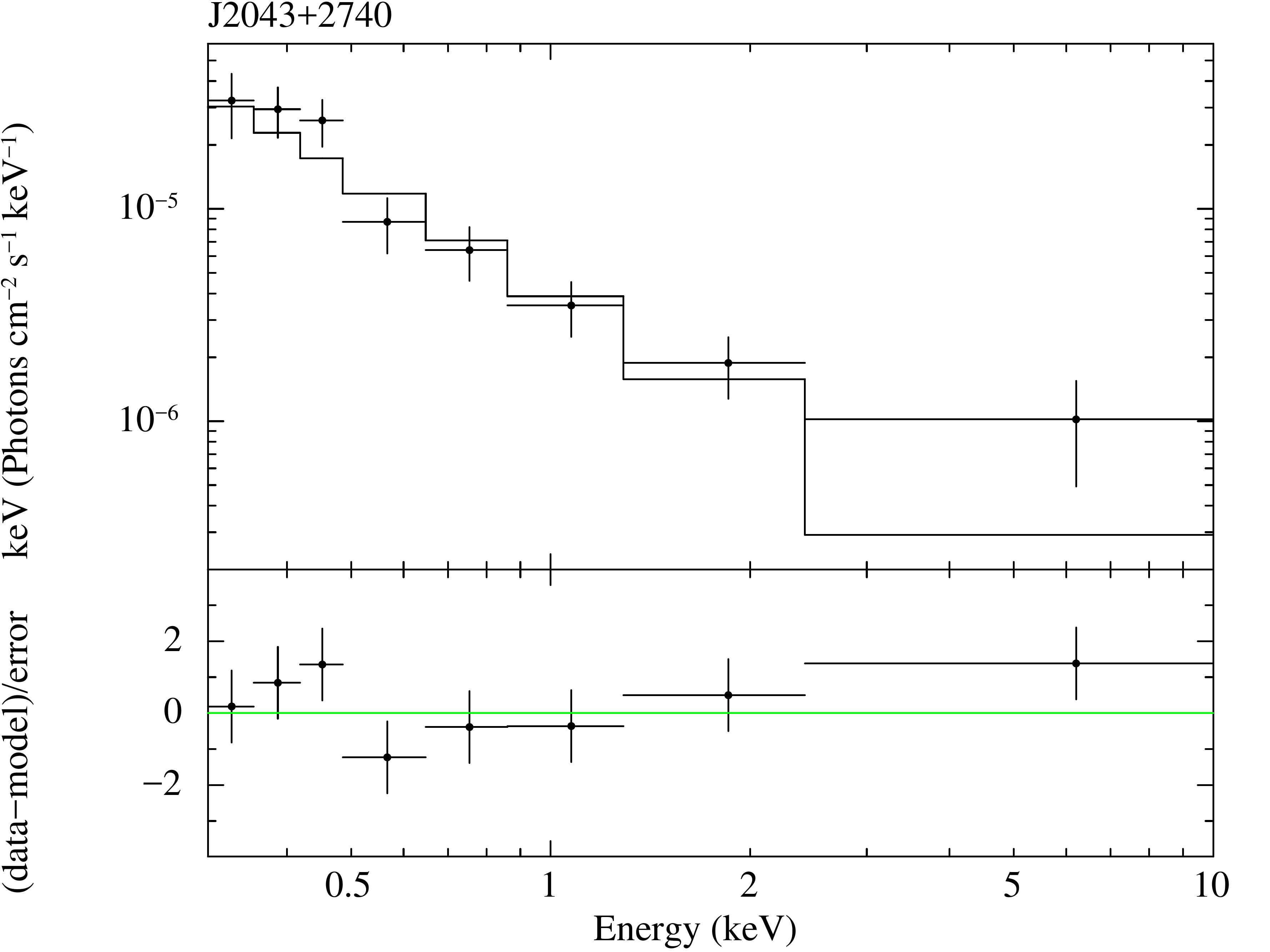}
\hspace{-0.44cm}
\includegraphics[width=0.508\textwidth]{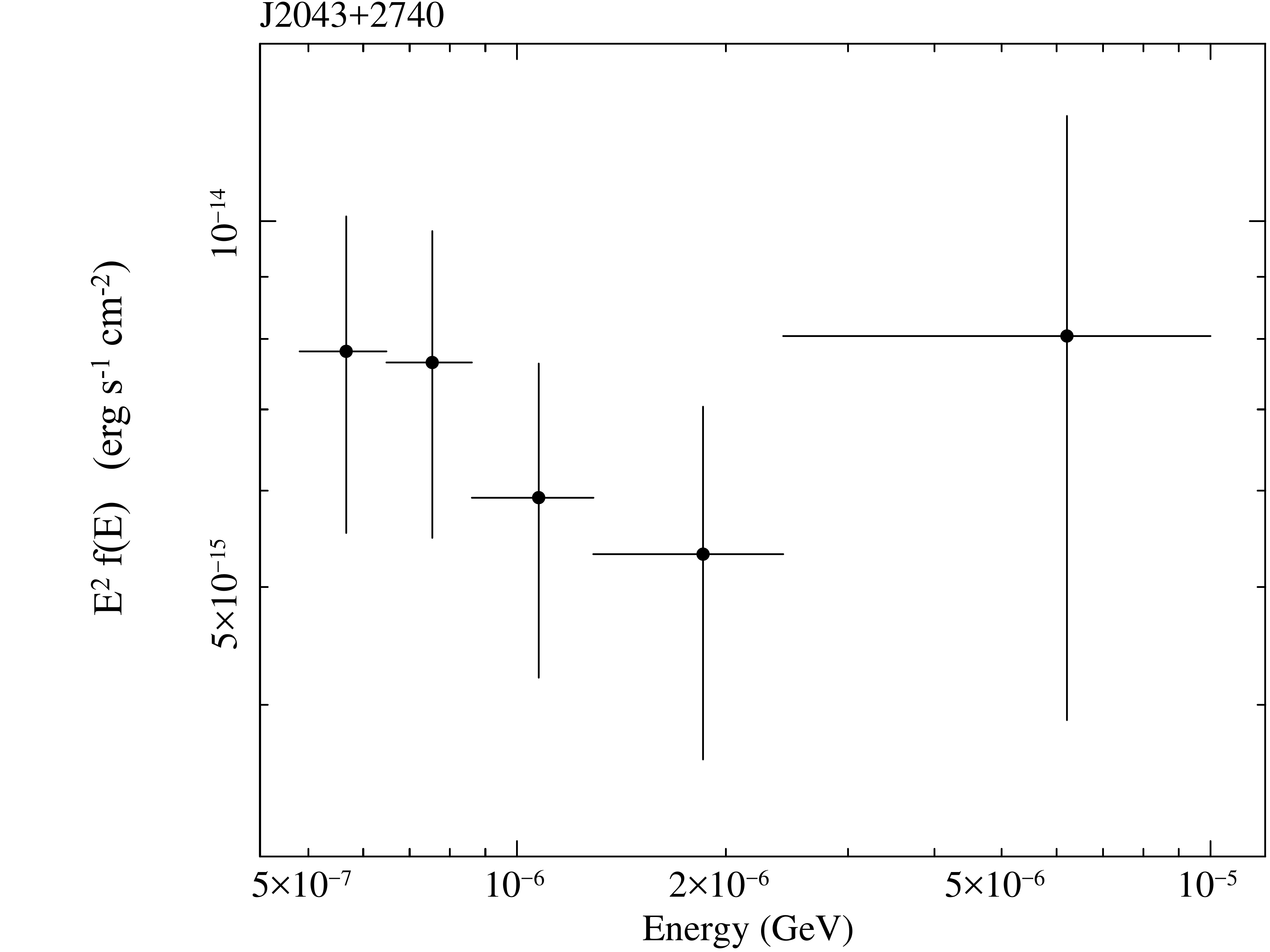}\\
\vspace{0.5cm}
\includegraphics[width=0.508\textwidth]{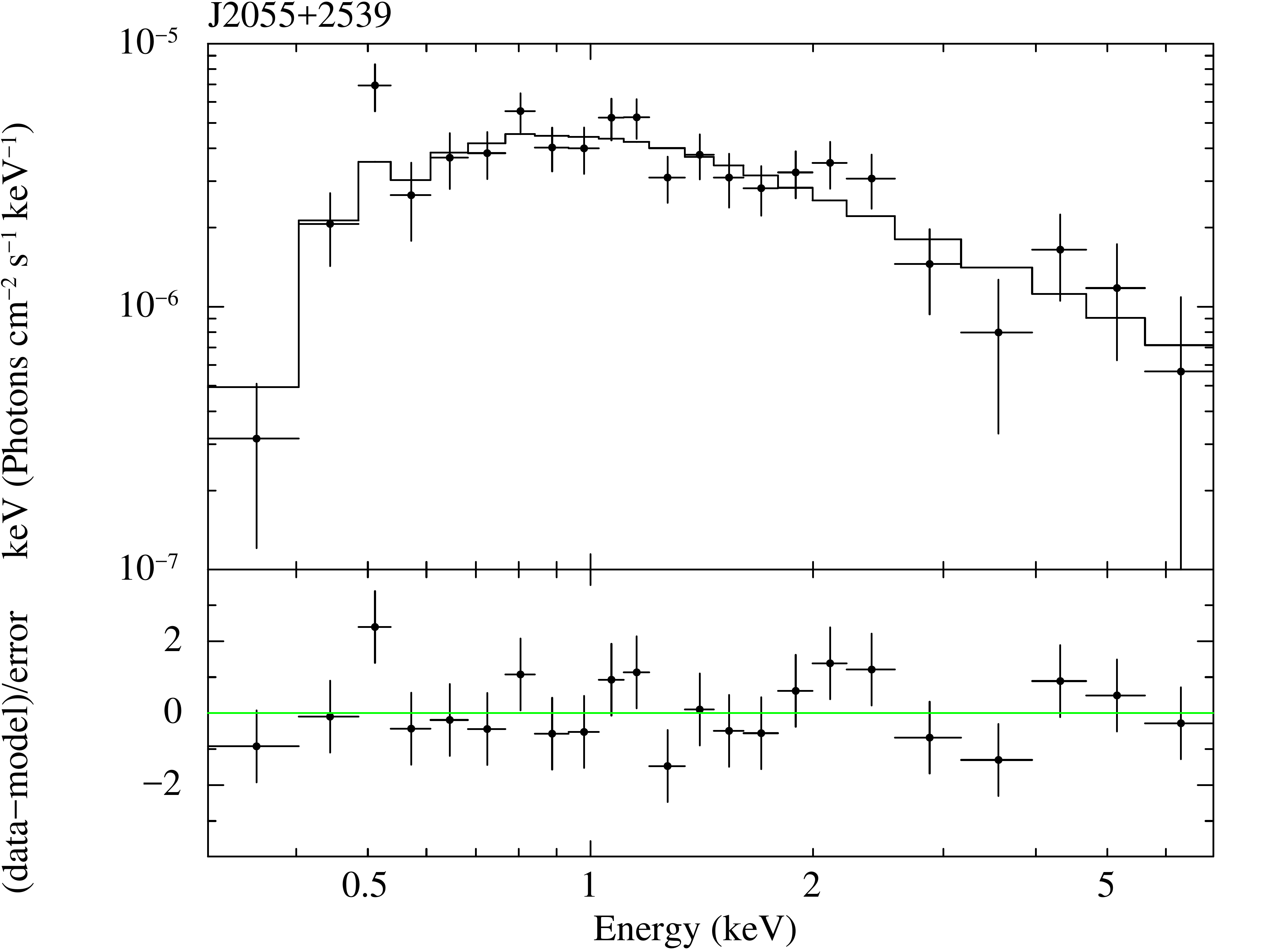}
\hspace{-0.44cm}
\includegraphics[width=0.508\textwidth]{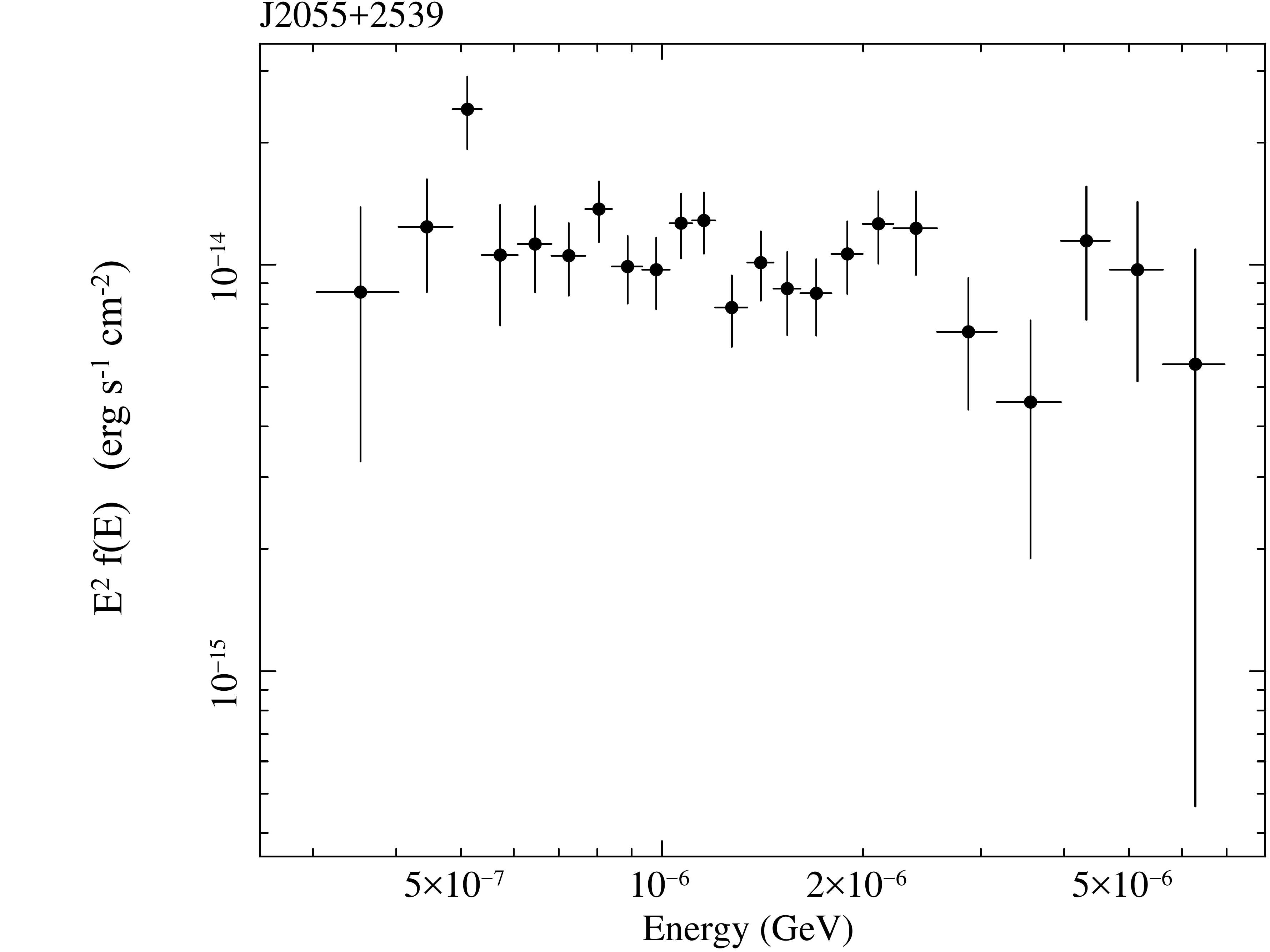}\\
\vspace{0.2cm}
  \contcaption{ }
\label{fig:sedx8}
\end{center}
\end{figure*}

\label{lastpage}
\end{document}